\documentclass[12pt]{iopart}
\usepackage{graphicx}
\usepackage{dcolumn}
\usepackage{bm}
\usepackage{iopams}
\usepackage{color}

\def\be{\begin{equation}}
\def\ee{\end{equation}}
\def\ba{\begin{array}}
\def\ea{\end{array}}
\def\bea{\begin{eqnarray}}
\def\eea{\end{eqnarray}}
\begin{document}
\title[\underline{J. Phys. G: Nucl. Part. Phys. \hspace {5.9cm} T. R. Routray et al. }]
 {Constraints from GW170817 on the bulk viscosity of neutron star 
matter and the $r$-mode instability}
\author{T. R. Routray$^{1,*}$, S. P. Pattnaik$^1$, C. Gonzalez-Boquera$^2$, X. Vi\~nas$^2$, M. Centelles$^2$, B. Behera$^1$}
\address{$^1$School of Physics, Sambalpur University, Jyotivihar-768 019, India.}
\address{$^2$Departament de F\'isica Qu\`antica i Astrof\'isica and Institut de Ci\`encies del Cosmos (ICCUB), 
Facultat de F\'isica, Universitat de Barcelona, Mart\'i i Franqu\`es 1, E-08028 Barcelona, Spain}
$^*$E-mail: trr1@rediffmail.com (corresponding author)\\
\date{\today}

\begin{abstract}
We perform a systematic study of the dependence of the r-mode phenomenology in normal fluid pulsar neutron stars on the symmetry energy slope parameter $L$. An essential ingredient in this study is the bulk viscosity, which is evaluated consistently for several equations of state corresponding to different values of the slope parameter $L$. Direct Urca processes, which are allowed from a critical $L$-value onwards, enhance the bulk viscosity and have large influence on the $r$-mode features, such as the instability boundary and spin-down properties of newborn neutron stars. The magnitude of the changes in the $r$-mode properties induced by the direct Urca processes are driven by the $L$-value of the equation of state and the mass of the pulsar. The study has been done by using a family of equations of state of $\beta$-equilibrated neutron star matter obtained with the finite range simple effective interaction, which provides realistic results for nuclear matter and finite nuclei properties. These equations of state predict the same properties in symmetric nuclear matter and have the same value of the symmetry energy parameter, $E_s(\rho_0)$, but differ in the slope parameter $L$. The range chosen for the variation of $L$ is decided from the tidal deformability data extracted from the GW170817 event and the maximum mass constraint.
\end{abstract}

\noindent {PACS: 21.10.Dr, 21.60.-n, 23.60.+e., 24.10.Jv.}

\noindent{\it Keywords}: Neutron star matter; Tidal deformability;  Bulk viscosity; Direct Urca; $r$-mode instability.
   
\bigskip

\section{Introduction}

The equation of state (EOS) of asymmetric nuclear matter (ANM) and, in particular, the density
 dependence of the nuclear symmetry energy, $E_s(\rho)$, in the supra-saturation density 
regime, have remained elusive in spite of several efforts made in finding an answer to this problem. 
The LIGO-Virgo collaboration's analysis of the data recorded
 in the GW170817 event of a merger of two neutron stars (NS) in a binary system \cite{Abbott2017,AbbottAPJ2017,Abbott2018,Abbott2019} has provided
new constraints in terms of the tidal deformability. The analysis of these constraints can be of crucial relevance in the ongoing efforts in 
disentangling the underlying uncertainties associated with the equation of 
state of isospin-rich dense nuclear matter to a reasonable extent.
 Reliable information on the mass-weighted tidal 
deformability parameter, $\tilde{\Lambda}$, and the possible ranges for the masses of the
 two NSs participating in the merger, are of great value for the nuclear 
astrophysics community. Progress has already been made in this context, where, for instance, Krastev and 
Li \cite{Krastev2019} have analyzed the density dependence of the symmetry energy, 
$E_s(\rho)$, with the MDI interaction including the constraint derived on 
$\tilde{\Lambda}$ coming from the data of the GW170817 event in addition to the earlier findings 
from diffusion studies in heavy-ion collisions (HIC) involving radioactive ion beams 
\cite{Chen2005,Li2005,Tsang2004}. These authors have concluded that the
results for the tidal deformability consistent with the constraint derived from the GW170817 
event can be reproduced with the stiff-to-soft and soft-to-stiff combinations of $E_s(\rho)$.
 That is, if $E_s(\rho)$ is stiff in the low-density region below the saturation density, 
then it requires a soft behaviour in the high-density region, which we refer as 
stiff-to-soft symmetry energy, and vice-versa for the soft-to-stiff case. As a consequence this analysis provides upper and lower limits for the slope of the symmetry energy compatible with the data extracted from the GW170817 event. Qualitatively, an analogous conclusion on the possible
 density dependence of the $E_s(\rho)$ also follows from the studies on $\pi^+\,\pi^-$ 
production and elliptic flow in HIC favouring a neither stiff nor very soft behaviour
 for the $E_s(\rho)$ \cite{Xiao2009,Kohley2012,Hong2014}.

Another feature of interest from the nuclear EOS point of view is the damping mechanism 
for the oscillations of matter in astrophysical objects, such as $r$-modes in pulsar NSs, just after the merger of two NSs etc. It has been shown in the 
simulation studies \cite{Alford2019} that the bulk viscosity is the dominant
dissipating mechanism for establishing thermal equilibrium in the post-merger scenario. Alford et al.
\cite{AMS2012,Alford2019a} have made a critical analysis of the influence of the bulk viscosity in the context of the $r$-mode instability boundary and have shown that at high temperatures $T\geq 10^{10}$K the bulk viscosity dominates the gravitational instability. However, at very high temperatures $T>10^{11}$K the $\beta$-decay rates match with the frequency of the stellar instability, which is of the order of $10^{3}-10^{4}$ Hz, the bulk viscosity shows a resonating behaviour, which becomes maximum and thereafter decreases rapidly. Our aim in this work is to study of the influence of the nuclear EOS, in particular, its $L$ dependence, on the spin-down feature of newly born hot NSs of normal $n+p+e+\nu$ composition. Therefore, we shall be using the bulk viscosity in its high frequency limit, which is relevant for the study. In this context, in Ref.\cite{Vidana2012} it has been shown, by considering several model interactions having widely varying $L$ values, that the high temperature boundary, 
which corresponds to the intermediate 
temperature branch of Alford et al. \cite{AMS2012}, shifts to the low temperature region reducing the area under the instability boundary for the EOSs having larger $L$-values.

The main contributions to the bulk viscosity 
come from the Urca processes in which nucleons emit leptons (l), namely, electrons (e) or 
muons ($\mu$), which are absorbed by other nucleons. The Urca reactions are of two types, 
the direct Urca (DU) and the modified Urca (MU) processes. A direct Urca process comprises two reactions,
\begin{equation}
n\longrightarrow p+l+\bar{\nu}_{l} \hspace{2cm} \mathrm{and}\hspace{2cm}  p + l\longrightarrow n+\nu_{l}, 
\label{eq.1}
\end{equation}
where, $n (p)$ is the neutron (proton), $l=e$ or $\mu$ is the lepton and $\nu_{l} (\bar{\nu_{l}})$ 
is the associated neutrino (anti-neutrino). The occurrence of the direct Urca process is, however, subject to the 
 constraint
\begin{equation}\label{eq.bi}
k_{n} \leq k_{p} + k_{l},
\label{eq.2}
\end{equation}
where, $k_{i}$ ($i=n, p, e, \mu$) are the respective momenta in $\hbar$ units.
If the direct Urca process is not allowed by the momentum conservation, 
given by Eq.~(\ref{eq.2}), 
then the bulk viscosity is determined by the modified Urca processes,
\begin{equation}
n+N\longrightarrow p+N+l+\bar{\nu_{l}} \hspace{1cm} \mathrm{and}\hspace{1cm} p+N+l \longrightarrow n+N+\nu_{l},
\label{eq.3}
\end{equation}
where $N$ is an additional nucleon required to conserve the momentum of the particles in the 
reactions. Depending on the type of nucleon, we have two branches, the neutron branch if $N$ is 
 a neutron and the proton branch if $N$ is a proton. In all
these Urca processes the neutrino that escapes the volume of the star takes away the energy, thereby 
reducing the perturbation and moving the system towards equilibrium.
The crucial role of the EOS of ANM becomes evident since	
it determines the $\beta$-equilibrium condition, 
\begin{equation}
\mu_{n}-\mu_{p}=\mu_{e}=\mu_{\mu},
\label{eq.4}
\end{equation}
with $\mu_{i}, i=n,p,e,\mu$, being the respective chemical potentials in the $n+p+e+\mu$ 
matter, referred to as neutron star matter (NSM).
 The $\beta$-equilibrium condition can be expressed in terms of the EOS of ANM as
\begin{equation}
\mu_{n}-\mu_{p} = 2\frac{\partial e(\rho,\beta)}{\partial \beta}=\mu_{e}=\mu_{\mu},
\label{eq.5}
\end{equation}
where, $e(\rho,\beta)$ is the energy per particle in ANM at a total density 
$\rho=\rho_{n}+\rho_{p}$ having an isospin asymmetry 
$\beta=\frac{\rho_{n}-\rho_{p}}{\rho_{n}+\rho_{p}}$, $\rho_{n}$ and $\rho_{p}$ being 
the neutron and proton densities, respectively. In the studies of neutron star phenomenology, 
it is a common practice to solve the $\beta$-stability condition using the quadratic approximation 
(also known as parabolic approximation (PA)) of the energy per particle in ANM 
\cite {Lattimer2001,Lattimer2007,Xu2009,Moustakidis2015}. Under this 
approximation, 
$e(\rho,\beta)$ is expressed as 
\begin{equation}
e(\rho,\beta)=e(\rho)+\beta^{2}E_{s}(\rho),
\label{eq.6}
\end{equation}
where, $e(\rho)$ is the energy per particle in symmetric nuclear matter (SNM) and 
$E_{s}(\rho)$ is the symmetry energy. In the conventional way, $E_s(\rho)$ is defined as 
the coefficient of the $\beta^2$ term in the Taylor series expansion of energy per particle, $e(\rho,\beta)$, of ANM about $\beta$=0. 
 Another common generic definition of $E_s(\rho)$, which we shall adopt here, and which is valid 
in both asymptotic limits of the isospin asymmetry, is the one expressed in terms of the difference 
between the energy per particle in pure neutron matter (PNM) and symmetric nuclear matter, 
\begin{equation}
E_s(\rho)= e^{PNM}(\rho)-e(\rho),
\label{eq.6a}
\end{equation}
where, $e^{PNM}(\rho)$ is the energy per particle in PNM. The $\beta$-stability condition in 
Eq.~(\ref{eq.5}) under the parabolic approximation of the energy given in Eq.~(\ref{eq.6}) becomes,
\begin{equation}
\mu_{n}-\mu_{p}=4\beta E_{s}(\rho)=\mu_{e}=\mu_{\mu}.
\label{eq.7}
\end{equation}
The $\beta$-stability condition, together with the charge neutrality of NSM, i.e., 
$\rho_{p}=\rho_{e}+\rho_{\mu}$, $\rho_{e}$ and $\rho_{\mu}$ being the electron and muon 
densities, determines the equilibrium particle fractions, $Y_{i}=\frac{\rho_{i}}{\rho}$, 
$i=n,p,e,\mu$, in NSM. 
 From Eq.~(\ref{eq.7}) it is evident that the symmetry energy is 
crucial for determining the particle fractions at each density value
in NSM that, together
 with condition (\ref{eq.2}), decide whether the direct Urca is a 
likely process at that density or not. 
The density for which Eq.(\ref{eq.2}) gets satisfied
is the threshold density for the direct Urca reaction, which becomes
 the dominant mechanism contributing to the bulk viscosity in NSM. 
In this work we shall perform our study using the finite-range simple effective interaction (SEI). This interaction provides realistic descriptions for nuclear, neutron and
spin-polarized matter \cite{trr2015} and for the global properties of finite nuclei \cite{trr2013,trr2016}.
Our objective in the present work is two-fold. We shall first use the mass-weighted 
tidal deformability parameter $\tilde{\Lambda}$ data of the GW170817 
event to estimate, within our chosen model, the limiting values of the slope parameter $L(\rho_0)$ for the stiffness of the nuclear symmetry energy $E_s(\rho)$. The $L$ values extracted from many 
studies of different nature \cite{Li2008,Tsang2012,Li2013,Horowitz2014,Lattimer2014,
Baldo2016,Lattimer2013,Margueron2018} lie in a very wide range between approximately 
20 and 100 MeV. The choice of the slope parameter $L$ from the GW constraint seems a  
reasonable option, as far as in this case the extracted $L$-value is not determined by strong 
interaction probes.
Within the allowed range of $L$ values permitted in our model from the GW170817 data, we analyze the bulk viscosity in NSM
and its influence on the $r$-mode instability boundary. The intensity of the gravitational waves emitted under the $r$-mode instability and its detectability will be examined in terms of the mode saturation amplitude. We also examine the correlation between the intensity of the GWs under the $r$-mode instability and the tidal deformability of 
isolated pulsar neutron stars
taking into account the information extracted from the  GW170817 observational data.

In Section~\ref{sec1.2}, the formulation of the tidal deformability parameter is discussed. 
The expressions for the bulk viscosity 
under direct and modified Urca processes 
 are outlined. In the same section, the features of the finite-range SEI 
and the determination of its parameters \cite{trr1998,trr2013,trr97,trr2015} are briefly recalled.
 In Section~\ref{sec1.3}, the limiting values for the slope parameter, $L(\rho_0)$, are ascertained in the SEI model 
using the constraint on $\tilde{\Lambda}$ obtained from the GW170817 event and the $\sim$2$M_\odot$ maximum mass limit. The appearance of direct Urca in the NS of given mass and its influence on the bulk viscosity in the high frequency limit is examined as a function of the slope parameter $L$ within the range of values allowed by the SEI model. 
The study has been done for 1.4$M_{\odot}$ and 1.8$M_{\odot}$ NSs, with particular emphasis on the ability of the PA of the energy in ANM to reproduce the exact results. 
The impact of the $L$-dependence of the EOS on
the $r$-mode instability boundary and on the spin-down phenomenology of newborn NS is examined. Finally, 
 Section~\ref{sec1.4} contains a brief summary of this work. 

%

\section{Formalism}\label{sec1.2}

In this section we provide a brief formulation of the tidal deformability of a NS binary system and then outline 
the expressions for the bulk viscosity under the direct and modified Urca processes.
\subsection{Tidal deformability of a neutron star binary system}

In a binary system composed of two neutron stars, each component star induces a
 perturbing gravitational tidal field on its companion, leading to a mass-quadrupole 
deformation in each star. 
To linear order, the tidal deformation of each component of the binary system is 
described by the so-called tidal deformability $\Lambda$, which is defined as 
the ratio between the 
induced quadrupole moment and the external tidal field~\cite{Flanagan08, Hinderer08}.
For a single neutron star, the tidal deformability can be written in terms of the dimensionless tidal 
Love number $k_2$, and the mass $M$ and radius $R$ of the NS~\cite{Flanagan08, Hinderer08, Hinderer2010},
\begin{equation}
 \Lambda= \frac{2}{3}k_2 \left(\frac{R c^2}{G M} \right)^5,
 \label{eq:lambda}
\end{equation}
 where, $G$ is the gravitational constant and $c$ the speed of light.
The mass and radius of a NS are determined by the solution of the TOV 
equations. The Love number $k_2$ is defined as
\begin{eqnarray}
 k_2&=& \frac{8C^5}{5} \left(1-2C \right)^2 \left[ 2+2C (y-1) -y\right]
 \times \left\{ 2C \left[6-3y + 3C(5y-8) \right] \right.\nonumber \\
 &+&  4 C^3 \left[ 13-11y+C (3 y -2) + 2 C^2 (1+y)\right]\nonumber\\
 &+&3 \left.(1-2C)^2 \left[2-y+2C(y-1) \right] \mathrm{ln}(1-2C)\right\}^{-1} ,
\end{eqnarray}
where, 
\begin{equation}
C=\frac{GM}{Rc^2}
\end{equation} 
is the compactness parameter of the star and 
\begin{equation}
y=\frac{R \bar{\beta}(R)}{\bar{H} (R)}.
\end{equation}

The functions $\bar{\beta}(r^{\prime})$ and $\bar{H}(r^{\prime})$ are obtained by solving 
the following set of coupled differential equations~\cite{Hinderer08, Hinderer2010}:
\begin{eqnarray}
 \frac{d \bar{H}(r^{\prime})}{dr^{\prime}} &=& \bar{\beta} \\
 \frac{d\bar{\beta}(r^{\prime})}{d r^{\prime}} &=& \frac{2G}{c^2} \left(1-\frac{2Gm}{r^{\prime} c^2} \right)^{-1} 
\bar{H} 
 \left\{ -2 \pi \left[ 5 H + 9P + \frac{d H}{d P} 
(H+P) \right] + \frac{3 c^2}{r^{\prime 2} G} \right.\nonumber \\
 &+&\left. \frac{2G}{c^2} \left(1-\frac{2Gm}{r^{\prime} c^2} \right)^{-1} \left( \frac{m}{r^{\prime 2}} 
+ 4 \pi r^{\prime} P\right)^2 \right\}\nonumber \\
 &+& \frac{2 \bar{\beta}}{r^{\prime}} \left(1-\frac{2Gm}{r^{\prime} c^2} \right)^{-1} \left\{-1+\frac{Gm}{r^{\prime} c^2} 
+ \frac{2 \pi r^{\prime 2} G}{c^2} \left( H -P\right) \right\},
\label{eq.14a}
\end{eqnarray}
where, $m=m(r^{\prime})$ is the mass enclosed inside a radius $r^{\prime}$, and $H$ and $P$ are the corresponding 
energy density and pressure. 
One solves this system, together with the TOV 
equations, integrating outwards and considering as boundary conditions
$\bar{H}(r^{\prime})= a_0 r^{\prime 2}$ and $\bar{\beta}(r^{\prime})= 2 a_0 r^{\prime}$
as $r^{\prime} \rightarrow 0$. The constant $a_0$ is arbitrary, as it cancels in the expression for 
the Love number~\cite{Hinderer2010}.

For a binary NS system, the mass-weighted tidal deformability $\tilde{\Lambda}$, 
 defined as
\begin{equation}\label{eq:wLambda}
 \tilde{\Lambda} = \frac{16}{13} \frac{(M_1 + 12M_2)M_1^4 \Lambda_1 +(M_2 + 12M_1)M_2^4 
\Lambda_2 }{(M_1+M_2)^5},
\end{equation}
takes into account the contribution from the tidal effects to the phase evolution of the 
gravitational wave spectrum of the inspiraling NS binary. 
In the definition of $\tilde{\Lambda}$ in Eq.~(\ref{eq:wLambda}), $\Lambda_1$ and $\Lambda_2$,
refer to the tidal deformabilities of each neutron star in the system and $M_1$ and $M_2$ are their corresponding masses. 
The definition of $\tilde{\Lambda}$ fulfills $\tilde{\Lambda}=\Lambda_1=\Lambda_2$, 
when, $M_1=M_2$.

\subsection{Bulk viscosity under Urca processes}

The system of neutrons, protons, electrons and muons that fulfils the $\beta$-equilibrium 
condition given by Eq.~(\ref{eq.4}) is said to be in fully thermodynamical 
equilibrium. A deviation of the particle number density $\rho_{i}$, 
$i=n, p, e$, $\mu$, 
from its equilibrium value due to any perturbation mechanism implies
non-zero differences in the instantaneous chemical potentials $\mu_{i}$,
\begin{equation}
\eta_{e}= \mu_{n}-\mu_{p}-\mu_{e} \quad,  \eta_{\mu}= \mu_{n}-\mu_{p}-\mu_{\mu}.
\label{eq.8}
\end{equation}
This will cause an asymmetry in the direct and inverse rates, $\Gamma_{l}$ and 
$\bar{\Gamma_{l}}$ ($l=e$ or $\mu$), of the direct Urca reaction, which under a linear approximation 
can be written as \cite{Haensel2000}  
\begin{equation}
\Gamma_{l} - \bar{\Gamma_{l}} = -\lambda_{l}\eta_{l},
\label{eq.9}
\end{equation}
where $\lambda_{l}$ is the asymmetry coefficient for a given type of lepton. The neutrinos and antineutrinos emitted in the non-equilibrium direct Urca reactions dissipate the energy, which 
damps out the perturbation and thereby contribute to the bulk viscosity $\zeta^{DU}$. The expression for $\zeta_l^{DU}$, $l=e,\mu$, is obtained as \cite{Sawyer1989,Haensel1992,AMS2012},
\begin{equation}
\zeta_l^{DU}=\frac{|\lambda_l| C_l^2}{\omega_p^2+\frac{4 \lambda_l^2 B_l^2}{\rho}} ,
\label{eq.9a}
\end{equation}
where, 
\begin{equation}
C_l=\rho \frac{\partial \eta_l}{\partial \rho},
\label{eq.9b}
\end{equation}
and
\begin{equation}
B_l=\frac{\partial \eta_l}{\partial Y_P}.
\label{eq.9c}
\end{equation}
All the quantities in Eq.~(\ref{eq.9a}), i.e., $\lambda_l$, $C_l$ and $B_l$ given by Eqs.~(\ref{eq.9b}) and (\ref{eq.9c}), are to be evaluated at beta-equilibrium condition. The contribution of the second term in the denominator of Eq.~(\ref{eq.9a}) is very small in comparison to the $r$-mode frequency $\omega_p\sim10^4$, for $T\leq10^{10}$K and can be neglected in the present context of the study \cite{ Sawyer1989,Haensel1992}. The contributions from the direct Urca processes of the two types of leptons is additive. Thus the
bulk viscosity of NSM containing electron and muon becomes, $\zeta^{DU}=\zeta_{e}^{DU}+\zeta_{\mu}^{DU}$. Each leptonic contribution, 
$\zeta_{l}^{DU}$, under the high frequency limit, is now expressed as \cite{Haensel1992,Haensel2000}
\begin{equation}
\zeta_{l}^{DU}=\frac{\left|\lambda_{l}\right|}{\omega_{p}^{2}}C^{2}_{l} ,
\label{eq.10}
\end{equation}
with $\omega_{p}$ being the frequency of the perturbation. Since our aim is to study the bulk viscosity by varying the asymmetry stiffness of the EOS, we express $C_{l}$ defined in Eq.~(\ref{eq.9b}), with the help of Eq.(\ref{eq.5}), as
\begin{equation}
C_{l}= - \rho \frac{\partial^{2} e(\rho, Y_{p})}{\partial \rho \partial Y_{p}} - \frac{c^{2}p^{2}_{fl}}{3\mu_{l}}
\label {eq.11}
\end{equation}
where, $e(\rho, Y_{p})$ is the energy per nucleon in the NSM at density $\rho$ and equilibrium
 proton fraction $Y_{p}$, which is related to the isospin asymmetry parameter as 
$\beta=(1-2Y_{p})$ and $p_{fl}$ being the Fermi momentum of the lepton. The asymmetry coefficient $\lambda_{l}$ is computed by evaluating 
the direct and inverse rates, $\Gamma_{l}$ and $\bar{\Gamma_{l}}$, respectively, 
which allows to write the bulk viscosity under direct Urca as 
\cite{Haensel2000}
\begin{equation} 
\zeta_{l}^{DU} = \frac{17G_{F}^2\cos^2 \theta_{c}(1+3g^{2}_{A})C_{l}^{2}}{240 \pi \hbar^{10}c^{3} \omega_{p}^{2}} m^{*}_{n} m^{*}_{p} m^{*}_{e} 
(k_{B}T)^{4}\Theta_{npl},
\label {eq.12} 
\end{equation}
where, the index $l$=$\mu, e$ refers to both types of leptons, $G_{F}$ 
is the Fermi coupling constant, $\theta_{c}$ is the Cabibbo angle, $g_{A}$ is the axial
 vector normalization constant. In (\ref{eq.12}) $m^{*}_{n},\  m^{*}_{p},\  m^{*}_{e}$ are the neutron, proton and electron 
effective masses, and $\Theta_{npl}$ is the step function which is 1 if direct Urca is on and 0 otherwise.

Using a similar formulation, the bulk viscosity for a modified Urca 
 process Eq.~(\ref{eq.3}) has been obtained in Ref.~\cite{Haensel2001} for 
the neutron and proton branches and for both types of leptons, $e$ and $\mu$, respectively.
These bulk viscosities read 
\begin{equation}
\zeta_{ne}^{MU}=\frac {367 G_{F}^{2} g_{A}^{2} m^{*3}_{n} m^{*}_{p}k_{fp}C_{e}^{2}} {1512 \pi^{3} \hbar^{9} c^{8} \omega_{p}^{2}}
 \left(\frac{f^{\pi}}{m_{\pi}}\right)^{4}(k_{B}T)^{6}\alpha_{n} \beta_{n}
\label{eq.13}
\end{equation}
\begin{equation}
\zeta_{pe}^{MU}=\zeta_{ne}^{MU}\bigg(\frac{m^{*}_{p}}{m^{*}_{n}}\bigg)^{2} 
\frac{(3k_{fp}+k_{fe}-k_{fn})^{2}}{8k_{fp}k_{fe}}\Theta_{pe}
\label{eq.14}
\end{equation}
\begin{equation}
\zeta_{n\mu}^{MU}=\zeta_{ne}^{MU}\bigg(\frac{k_{f\mu}}{k_{fe}}\bigg)
\bigg(\frac{C_{\mu}}{C_{e}}\bigg)^{2}
\label{eq.15}
\end{equation}
\begin{equation}
\zeta_{p\mu}^{MU}=\zeta_{ne}^{MU}
\bigg(\frac{C_{\mu}m^{*}_{p}}{C_{e}m^{*}_{n}}\bigg)^{2} \frac{(3k_{fp}+k_{f\mu}-k_{fn})^{2}}
{8k_{fp}k_{f\mu}}\bigg(\frac{k_{f\mu}}{k_{fe}}\bigg)\Theta_{p\mu} ,
\label{eq.16}
\end{equation}
where, $\zeta_{ne(pe)}^{MU}$ and $\zeta_{n\mu(p\mu)}^{MU}$ correspond to the neutron (proton) branches 
of the electronic and muonic MU contributions to the bulk viscosity, $f^{\pi}$ is the pion-nucleon
 coupling constant ($f^{\pi} \approx 1$), $m_{\pi}$ is the pion mass, 
$\alpha_{N}$ accounts for the momentum dependence of the squared matrix element in the Born
 approximation and $\beta_{N}$ contains various corrections (N refers to n(p) for neutron(proton) branch).
The step functions entering in these equations 
are 
$\Theta_{pe}(\Theta_{p\mu})$=1 if $k_{fn}< (3k_{fp}+k_{fe(\mu)})$, and 0 
otherwise \cite{Yakovlev1995, Kaminker2016}. 
To evaluate the bulk viscosity at a given density 
under direct and modified Urca processes given by Eq.~(\ref{eq.12}) and Eqs.~(\ref{eq.13})-(\ref{eq.16}), 
respectively, it is required to know $C_{l}$, $m^{*}_{n}$ and $m^{*}_{p}$, which can be 
calculated using the EOS of ANM. In the present calculation we will use the EOS resulting 
from the finite range simple effective interaction (SEI) introduced 
in Refs.~\cite{trr1998,trr2013}.
\subsection{The fitting procedure of SEI}
 The SEI interaction used in this work for describing the nuclear part of the NSM 
reads
\begin{eqnarray}
v_{eff}({\bf r},{\bf R})&=&t_0 (1+x_0P_{\sigma})\delta({\bf r}) \nonumber \\
&&+\frac{t_3}{6}(1+x_3 P_{\sigma})\left(\frac{\rho({\bf R})}
{1+b\rho({\bf R})}\right)^{\gamma} \delta({\bf r}) \nonumber \\
&& + \left(W+BP_{\sigma}-HP_{\tau}-MP_{\sigma}P_{\tau}\right)e^{-r^2/\alpha^2},
\label{eq.17}
\end{eqnarray}
where, ${\bf r}$ and ${\bf R}$ are the relative and center of mass coordinates,
respectively. The SEI in Eq.~(\ref{eq.17}) has in total 12 parameters,
 namely,
$b$, $t_0$, $x_0$, $t_3$, $x_3$, $\gamma$, $\alpha$,
$W$, $B$, $H$ and $M$ plus the spin-orbit strength parameter $W_0$, which enters
in the description of finite nuclei. The SEI interaction is similar in 
form to the Skyrme force, where the gradient terms are replaced by the 
single finite-range contribution. A similar analogy can be drawn with the 
Gogny interaction, where one of the two finite range terms is replaced by the 
zero-range $t_0$-term. One more difference in this context is that the density-dependent
term of SEI contains the factor (1+b$\rho$)$^\gamma$ in the denominator, 
where the parameter $b$ is fixed to prevent the supra-luminous behavior in nuclear matter
at high densities \cite{trr97}.

The formulation of nuclear matter and neutron matter using SEI has been 
discussed at length in Refs.~\cite{trr2013,trr2015}, but for the sake of convenience of the 
reader we report in Appendix~A the expressions of the energy density 
of asymmetric nuclear matter, as well as its symmetric nuclear matter and pure neutron
matter limits, the neutron and proton single-particle potentials and the respective effective 
masses in ANM obtained with SEI. 
We shall now outline, in brief, the determination of the parameters 
involved in the study of nuclear and neutron matter. The study of asymmetric 
nuclear matter involves altogether nine parameters, 
namely, $\gamma$, $b$, $\varepsilon_{0}^{l}$, $\varepsilon_{0}^{ul}$,
$\varepsilon_{\gamma}^{l}$,$\varepsilon_{\gamma}^{ul}$, $\varepsilon_{ex}^{l}$,
$\varepsilon_{ex}^{ul}$ and $\alpha$, whose connection to the interaction 
parameters is given in the Appendix A. 
However, symmetric nuclear matter requires only the following three combinations
of the strength parameters in the like (l) and unlike (ul) channels
\begin{eqnarray}
\left(\frac{\varepsilon_{0}^{l}+\varepsilon_{0}^{ul}}{2}\right)=\varepsilon_0,
\left(\frac{\varepsilon_{\gamma}^{l}+\varepsilon_{\gamma}^{ul}}{2}\right)=\varepsilon_{\gamma},
\left(\frac{\varepsilon_{ex}^{l}+\varepsilon_{ex}^{ul}}{2}\right)=\varepsilon_{ex},
\label{eq.18}
\end{eqnarray}
together with $\gamma$, $b$ and $\alpha$, 
i.e., altogether six parameters.
For a given value of the exponent
$\gamma$, which determines the stiffness in SNM, and assuming 
standard values for the nucleon mass, saturation density, $\rho_0$, and 
binding energy per particle at saturation, $e(\rho_0)$, 
the remaining five parameters $\varepsilon_{0}$, $\varepsilon_{\gamma}$,
$\varepsilon_{ex}$, $b$ and $\alpha$ 
of SNM are determined in the following way. 
The range $\alpha$ and the exchange strength $\varepsilon_{ex}$ are determined 
simultaneously by adopting an optimization procedure \cite{trr1998}, which uses the condition 
that the nuclear mean field in SNM at saturation density 
vanishes for a kinetic energy of the nucleon of 300 MeV, a result extracted from optical model analysis
of nucleon-nucleus data \cite{ber1988,gale87,gale90,cser92}. 
The parameter $b$ is determined as mentioned before. 
The two remaining parameters,
namely $\varepsilon_{\gamma}$ and $\varepsilon_{0}$, are obtained from the saturation
conditions. The stiffness parameter $\gamma$ is kept as a free parameter 
and its allowed values are chosen in such a way that the corresponding pressure-density relation 
in symmetric matter lies within the region compatible 
with the analysis of flow data in heavy-ion collision experiments 
\cite{Danielewicz2002}. It is verified that the upper limit for $\gamma$ for which the pressure-density 
relation is obeyed is $\gamma$=1, which corresponds to the nuclear matter incompressibility 
$K(\rho_0)$=283 MeV. Therefore, we can study the 
nuclear matter properties by assuming different values of $\gamma$ up to 
the limit 
$\gamma$=1. 

In order to study the asymmetric nuclear matter we need to know how the strength 
parameters
$\varepsilon_{ex}$, $\varepsilon_{\gamma}$ and $\varepsilon_{0}$ 
of Eq.~(\ref{eq.18})
split into
the like and unlike components. The splitting of $\varepsilon_{ex}$ into
$\varepsilon_{ex}^{l}$ and $\varepsilon_{ex}^{ul}$ is decided from the 
condition that the entropy density in PNM should not exceed 
that of the SNM. This prescribes that the critical value for 
the splitting of the exchange strength parameter to be $\varepsilon_{ex}^{l}=2\varepsilon_{ex}/3$ \cite{trr2011}. 
The splitting of the remaining two strength parameters 
$\varepsilon_{\gamma}$ and $\varepsilon_{0}$,
is obtained from the values of the symmetry energy parameter $E_s(\rho_0)$ and 
its derivative $E_s^{'}(\rho_0)$ = $\rho_0 \frac{dE_s(\rho_0)}{d\rho_0}$ 
at saturation density $\rho_0$. For a given symmetry energy 
$E_s(\rho_0)$ within its accepted range of values \cite{dutra12}, we can produce different density dependence of symmetry energy 
by assigning arbitrary values to $E_s^{'}(\rho_0)$. The slope parameter in each case will be $L(\rho_0)=3E_s^{'}(\rho_0)$. 
In the study where the variation of $L(\rho_0)$ is not an explicit requirement, the value of $E_s^{'}(\rho_0)$ is fixed 
from the condition that asymmetric nucleonic contribution in NSM, i.e.,  
$S^{NSM}$$(\rho)$=$[H(\rho,Y_p)-H(\rho, Y_p=1/2)]$ is maximum, with
$Y_p$ being the equilibrium proton fraction in charge neutral $\beta$-stable $n+p+e+\mu$ matter. The characteristic 
$E_s^{'}(\rho_0)$ value thus obtained 
predicts a density dependence of the symmetry energy which is neither stiff
nor very soft \cite{trr2007}. With the parameters 
determined in this way, the SEI is able to reproduce 
the trends of
the EOS and the momentum dependence of the mean field properties 
with similar quality as predicted by microscopic calculations
\cite{Sammarruca2010,Wiringa1988,trr2009,trr2011}. It is also found from finite nuclei studies using SEI that there is a 
particular value of $E_s(\rho_0)$ associated with a given incompressibility (given $\gamma$),  
 for which the binding energies of 161 even-even spherical
nuclei and the charge radii of 86 even-even spherical nuclei are predicted with minimum root mean square ({\it rms}) deviations 
 from the experimental data when $E_s^{'}(\rho_0)$ takes it characteristic value 
\cite{trr2013,trr2015,trr2016}. 

\section{Results and Discussion}\label{sec1.3}

\subsection{Symmetry energy, mass-radius relation and tidal deformability}

\begin{figure}[ht]
\vspace{1.5cm}
	\begin{center}
		\includegraphics[width=0.7\columnwidth,angle=0]{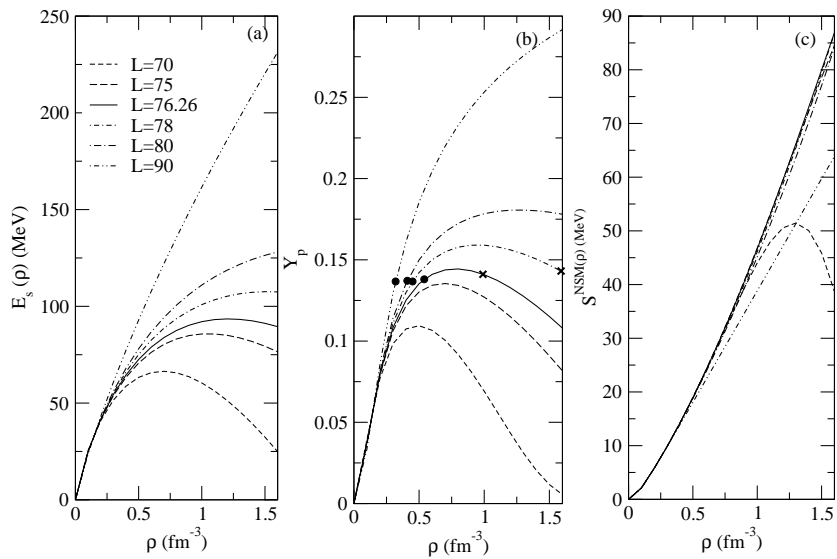}
		\end{center}
	\label{fig:esym} 
	\caption{The symmetry energy $E_s(\rho)$ in panel (a), equilibrium proton fraction $Y_p$ in NSM in panel (b) and asymmetric nucleonic contribution $S^{NSM}(\rho)=[H(\rho,Y_p)-H(\rho,Y_p=1/2)]$ in panel (c) as functions of the density $\rho$ for the EOSs having 
	$\gamma$=1/2 and different $L$-values, in the range 70 MeV $ \leq L \leq 90 $ MeV. In panel (b) the dots and crosses on the curves indicate the beginning and end of the direct Urca process, respectively. The continuous curves in the three panels correspond 
	to the characteristic $L$ value.\label{Fig1}}
	\end{figure}

 In this work we use the EOS based on the SEI with a Gaussian form factor and characterized by $\gamma=\frac{1}{2}$ (corresponding to $K(\rho_0)=246$ MeV)
and symmetry energy $E_{s}(\rho_{0})$=35 MeV. This choice is motivated by our previous
studies of nuclear matter and finite nuclei performed in Refs.~\cite{trr2013,trr2015,trr2016}.
The values of the parameters of this interaction, as well as the corresponding nuclear
matter properties at saturation, are given in Table~\ref{Table1}.
The behaviour of the symmetry energy as a function of the density, calculated using the generic expression (\ref{eq.6a})
with the aforementioned SEI parametrization, is shown in panel 
(a) of Figure~\ref{Fig1} for different values of the slope parameter $L(\rho_0)$.
 The equilibrium proton fractions $Y_p(\rho)$, obtained from the
solution of charge neutral $\beta$-stable $n+p+e+\mu$ matter~(\ref{eq.4}) using the exact analytical expression for  
$\mu_n-\mu_p$, are displayed in panel (b) of Figure~\ref{Fig1}. The exact analytical expression for $\mu_n-\mu_p$ is obtained from the expression in equation (\ref{eq.A.5}) as explained in Appendix A. The asymmetric 
nucleonic contribution in NSM, $S^{NSM}(\rho)$ is shown in panel (c). The stiffest behavior
of  $S^{NSM}(\rho)$ is found for $E^{'}_{s}(\rho_{0})$=25.42 MeV, which corresponds to a 
characteristic value of the slope parameter $L(\rho_0)$=76.26 MeV.
\begin{table}[ht]
\caption{Values of the nine parameters of ANM for the EOS of SEI corresponding to $\gamma=1/2$ together with 
their nuclear matter saturation properties.\label{Table1}}
\renewcommand{\tabcolsep}{0.05cm}
\renewcommand{\arraystretch}{1.2}
\begin{tabular}{|c|c|c|c|c|c|c|c|c|c|c|c|}\hline
\hline
$\gamma$ & $b$& $\alpha$ & $\varepsilon_{ex}$ & $\varepsilon_{ex}^{l}$ & $\varepsilon_{0}$ &
$\varepsilon_{0}^{l}$ & $\varepsilon_{\gamma}$& $\varepsilon_{\gamma}^{l}$ \\
& $\mathrm{fm^3}$ & $\mathrm{fm}$ & $\mathrm{MeV}$ & $\mathrm{MeV}$ & $\mathrm{MeV}$ & $\mathrm{MeV}$ & $\mathrm{MeV}$ & $\mathrm{MeV}$ \\
\hline
$\frac{1}{2}$& 0.5914 & 0.7597 &-94.4614  &-62.9743 &-78.7832 &-45.8788 &77.5068 &57.76866 \\\hline
\multicolumn{9}{|c|}{Nuclear matter properties at saturation density} \\
\hline
\multicolumn{1}{|c|}{$\gamma$}&\multicolumn{1}{|c|}{$\rho_0$ ($\mathrm{fm}^{-3}$)} & \multicolumn{2}{c|}{$e (\rho_0) $ (MeV)}
& \multicolumn{1}{c|}{$K (\rho_0)$ (MeV)} & \multicolumn{1}{c|}{$\frac{m^*}{m}(\rho_0,k_{f_0})$}
& \multicolumn{1}{c|}{$E_s (\rho_0)$ (MeV)} & \multicolumn{2}{c|}{$L (\rho_0)$ (MeV)} \\
\hline
\multicolumn{1}{|c|}{$\frac{1}{2}$}& \multicolumn{1}{|c|}{0.1571} & \multicolumn{2}{c|}{-16.0} & \multicolumn{1}{c|}{245.6}
& \multicolumn{1}{c|}{0.7111} & \multicolumn{1}{c|}{35.0} & \multicolumn{2}{c|}{76.26} \\\hline
\end{tabular}
\end{table}
%

\begin{figure}[ht]
\vspace{1.5cm}
	\begin{center}
		\includegraphics[width=0.7\columnwidth,angle=0]{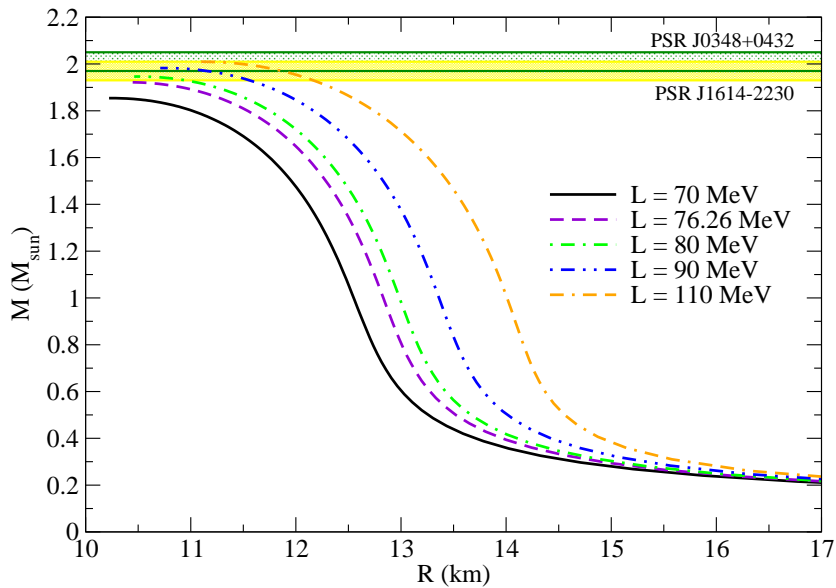}
		\end{center}
	\caption{Mass-radius relation obtained with the SEI $\gamma=1/2$ EOS for $L=70, 76.26,80,90$ and $110$ MeV. The 
	horizontal bands correspond to the observations of highly massive NS of $M=1.97\pm 0.04 M_\odot$ in the pulsar
	PSR J1614-2230 (yellow band) and of $M=2.01 \pm 0.04 M_\odot$ in the pulsar PSR J0348+0432 (green band)~\cite{Demorest2010, Antoniadis2013}.}
	\label{fig:MR_SEI}
\end{figure}

 In Figure~\ref{fig:MR_SEI} we plot the mass-radius relation for neutron stars obtained 
by solving the TOV equations for the different sets of the
SEI EOS with $\gamma=1/2$ and $E_{s}(\rho_{0})$=35 MeV but with different values of
 the slope parameter $L$, namely $L=70, 76.26, 80,90$ and $110$ MeV.
Constraints given by the observation of highly massive NSs are also
drawn~\cite{Demorest2010, Antoniadis2013}. From this plot we see that SEI
interactions of $\gamma=1/2$ with a relatively small value of the slope parameter $L$, as for example
 the $L=70$ MeV set, predict a maximum NS mass of 1.85 $M_\odot$, which does not reach the NS mass
constraint. However, the interactions with the characteristic $L$ value
(76.26 MeV in the used SEI) or larger are compatible with the $\sim$ 2 $M_\odot$ constraint. In the context of the maximum mass constraint, it is worthwhile to mention that the maximum mass of 1.97$\pm$0.04 $M_{\odot}$ of the PSR J1614-2230, measured by Demorest et al \cite{Demorest2010} was subsequently ascertained by more precise measurements to be 1.928$\pm$0.017 $M_\odot$ by Fonseca et al. \cite{Fonseca2016} and 1.908$\pm$0.016 $M_\odot$ by Arzoumanian et al. \cite{Arzoumanian2018} in two different analysis. The most recent maximum NS mass measured for the PSR J0740+6620 by Cromartie et al. \cite{Cromartie2019} is $2.14^{+0.10}_{-0.09}$ $M_\odot$. In view of the observed uncertainty in the predictions of maximum masses from the data taken over different periods by different groups of scientists, as in case of the PSR J1614-2230, EOSs capable of predicting $\sim$ 2 $M_\odot$ can be admitted for the studies associated with the features of NS \cite{Rezzola2018}. 
In solving the TOV equations to obtain the mass-radius relationship and the tidal deformability
(see below), we have used the BPS-BBP EOS \cite{Baym1971,Baym1971b} up to a density 0.07468 fm$^{-3}$ and 
the SEI EOS with $\gamma=1/2$ and $E_{s}(\rho_{0})$=35 MeV for higher densities.

\begin{figure}[ht]
\vspace{1.5cm}
	\begin{center}
		\includegraphics[width=0.7\columnwidth,angle=0]{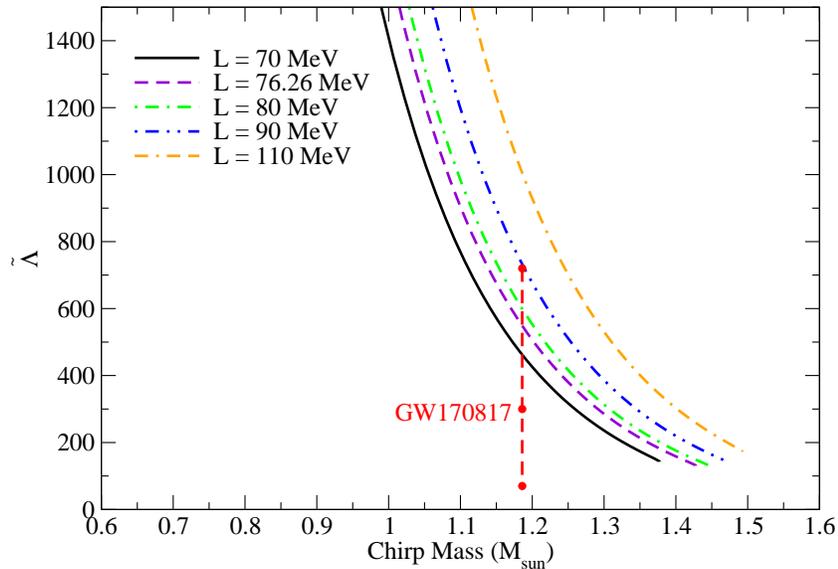}
		\end{center}
	\caption{Mass-weighted tidal deformability as a function of the chirp mass of a binary NS system calculated 
	using the SEI $\gamma=1/2$ EOS for $L=70, 76.26,80,90$ and $110$ MeV. The latest 
	constraints coming from the GW170817 detection are also included~\cite{Abbott2019}.
\label{fig:weighted_adim_lambda_ChirpMass}}
\end{figure}

The recent event GW170817 \cite{Abbott2017,AbbottAPJ2017,Abbott2018,Abbott2019} accounting for the detection of GWs coming from the merger of a NS 
binary system has allowed the LIGO and Virgo collaboration to 
obtain constraints on the mass-weighted 
tidal deformability $\tilde{\Lambda}$ and on the chirp mass $\mathcal{M}$, 
which for a binary NS system conformed of two stars of masses $M_1$ and $M_2$ 
is defined as 
\begin{equation}
 \mathcal{M}= \frac{(M_1 M_2)^{3/5}}{(M_1+M_2)^{1/5}}.
\end{equation}
In the initial data analysis of GW170817 by the LIGO and Virgo collaboration, values of 
$\tilde{\Lambda} \leq 800$ and $\mathcal{M} = 1.188^{+0.004}_{-0.005} M_\odot$ were 
reported~\cite{Abbott2017}. Moreover, they estimated the masses of the two NSs to be in the 
range $M_1 \in (1.36,1.60) M_\odot$ and  $M_2 \in (1.17,1.36) M_\odot$. In a recent 
 reanalysis of the data~\cite{Abbott2019},
the values have been further constrained to $\tilde{\Lambda} =300^{+420}_{-230}$, 
$\mathcal{M} = 1.186^{+0.001}_{-0.001} M_\odot$,
$M_1 \in (1.36,1.60) M_\odot$ and  $M_2 \in (1.16,1.36) M_\odot$. The large error bars in mass-weighted tidal deformability is due to the large uncertainty associated with the data recorded in the GW170817 event. 
We plot in Figure~\ref{fig:weighted_adim_lambda_ChirpMass} the results of
the mass-weighted tidal deformability $\tilde{\Lambda}$ against the chirp mass $\mathcal{M}$ of a 
 NS binary system for our EOSs with $\gamma$=1/2 and $L$=70, 76.26, 80, 90 and 100 MeV. In the same figure the range of $\tilde{\Lambda}$ extracted from the analysis of the data of the GW170817 event in Ref.~\cite{Abbott2019} with mass ratio of $\eta=M_1 M_2 / (M_1+M_2)^2=0.250$ (i.e., 
a binary system with NSs of equal masses $M_1=M_2$) is shown.
We, thus, find an upper bound of $L\sim90$ MeV for the value of the slope parameter of the symmetry energy, as 
the results from the EOSs with $L=70$, 76.26, 80 and 90 MeV fit inside the observational constraint  
of the GW170817 data, whereas the predictions of the EOSs with $L > 90$ MeV turn out to be above the observed band.
Within this range of the slope parameter $L$, the radii predicted for canonical neutron stars of
$1.4 M_\odot$ conform to the limit 10.5 km$\leq R_{1.4} \leq$13.3 km
of the binary components concluded by the LIGO-Virgo collaboration in their recent analysis \cite{Abbott2018}.

\begin{figure}[ht]
\vspace{1.5cm}
        \begin{center}
                \includegraphics[width=1\columnwidth,angle=0]{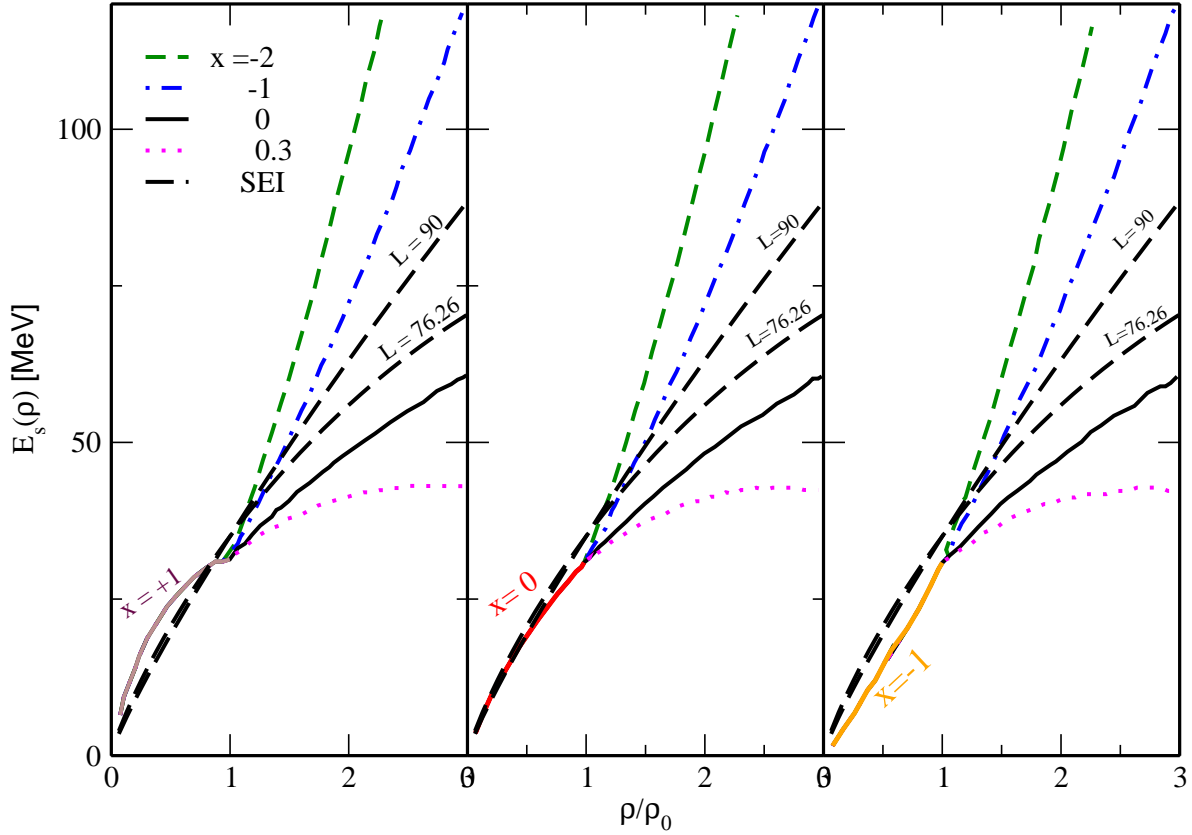}
                \end{center}
\caption{Density dependence of the symmetry energy for the EOSs of SEI corresponding
to $\gamma$=1/2 and $L$=76.26 and 90 MeV are compared with the stiff-to-soft and
soft-to-stiff symmetry energies of Ref.\cite{Krastev2019}. For details, see the text.}\label{Krastev.eps}
\end{figure}

 In Ref.~\cite{Krastev2019}  Krastev and Li have discussed the phenomenology of
the tidal deformability parameter and the radius of 1.4 $M_\odot$ NS in terms of the
density content of the symmetry energy. They find that both soft-to-stiff and
stiff-to-soft combinations of the density dependence of $E_s(\rho)$ can reproduce
the constraints for $\Lambda_{1.4}$ and $R_{1.4}$ within their prescribed range of
uncertainty. By showing this, they claim that in order to derive some definite
conclusion on the density dependence of $E_s(\rho)$, more accuracy in the measurement of the
 astrophysical constraints is needed. The density dependence of the the symmetry energy
$E_s(\rho)$ for the EOSs of ANM used in this work with slope parameter
in the range between $L$= 76.26 and 90 MeV are compared with the soft-to-stiff and 
stiff-to-soft results of \cite{Krastev2019} and shown in Fig.~\ref{Krastev.eps}. 
The high density behavior of the symmetry energy used in the present work lies within the 
curves for $x$=0 and $x$=-1 of Krastev and Li, which correspond to $L$=62 and 107 MeV, respectively.
\subsection{Bulk viscosity of neutron-star matter}

\begin{table}
\caption{
%
The density range for occurrence of direct Urca in NSM, $\rho_{e(\mu)}^{DU}$ in $fm^{-3}$ for 
electrons (muons) as allowed under the considered EOS. Also shown are the central density $\rho_c$ in $fm^{-3}$, the radius $R$ in $km$ and the distance over which 
direct Urca process occurs in the volume of NS $R_{DU}$ in $km$.
The column containing a single value for $\rho_{e(\mu)}^{DU}$ 
denotes the switch-on density for the direct Urca process, which continues upto the central density.
These results are given in 1.4 $M_{\odot}$ and 1.8 $M_{\odot}$ NSs for the EOSs of ANM having 
$L$=70, 76.26, 80 and 90 MeV under the exact and PA calculations.
}
\renewcommand{\tabcolsep}{0.18cm}
\renewcommand{\arraystretch}{1.2}
\begin{tabular}{|c|c|c|c|c|c|c|c|c|c|}
\hline
\multicolumn{4}{|c}{} & \multicolumn{3}{|c|}{(1.4 $M_{\odot}$)} & \multicolumn{3}{c|}{(1.8 $M_{\odot}$)}\\
\hline
$L$     &&  $\rho_{e}^{DU}$ & $\rho_{\mu}^{DU}$
                                 &$\rho_c$  & R   &$R_{DU}$&  $\rho_c  $	&   R     &  $R_{DU}$   \\
$MeV$     &&  $fm^{-3}$ & $fm^{-3}$				
                                 &$fm^{-3}$ & km  & km &  $fm^{-3}$ 	&   km     & km   \\												
\hline
 70.00   &Exact & x  				&	x	    & 0.54	&	12.167&x    			&0.92		&11.054&x													\\
			   &PA    & x  				&	x	    & 0.54	&	12.195&x    			&0.92		&11.070&	x												\\\hline
 76.26   &Exact & 0.54-0.99 &	x	    &	0.50	&	12.483&x    			&0.78		&11.583&6.194													\\
				 &PA    & 0.52-1.03	&	x	    & 0.50	&	12.512&x    			&0.78		&11.600&6.454												\\\hline
 80.00   &Exact & 0.41			&	0.59	& 0.48  &	12.655&4.844			&0.74		&11.788&7.754													\\
					&PA    & 0.40			&	0.57  & 0.48	&	12.684&5.184			&0.74		&11.805&7.874													\\\hline
 90.00   &Exact & 0.32			&	0.41	&	0.45  &	13.023&6.994			&0.67		&12.224&8.754													\\
				 &PA    & 0.31			&	0.40  & 0.45	&	13.052&7.254			&0.67		&12.240&8.874													\\\hline
\hline
\end{tabular}
\label{table2}
\end{table}
\begin{figure}[ht]
\vspace{1.5 cm}
	\begin{center}
		\includegraphics[width=0.9\columnwidth,angle=0]{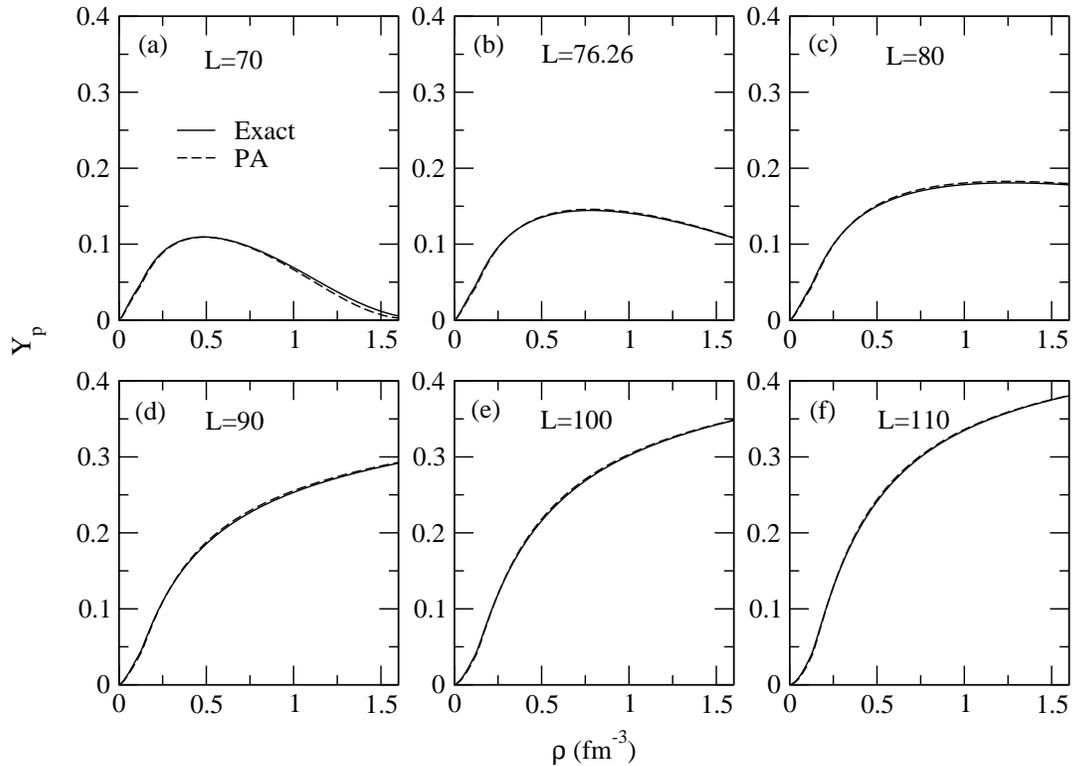}
		\end{center}
	\caption{Comparison of the equilibrium proton fraction in
	NSM obtained under the exact and PA for the different EOSs of ANM 
	corresponding to $L$=70, 76.26, 80, 90, 100 and 110 MeV shown in individual 
	panels as a function of the density.\label{fig:yp_exact_compare}}
\end{figure}
We shall now study the behaviour of the bulk viscosity $\zeta$, when the slope parameter
$L$ varies within the limits constrained in the present model from the tidal deformability data of the GW170817 event and the 
$\sim$2$M_\odot$ maximum mass limit. Our study is 
performed 
using both the exact
 expression and the PA of the energy in ANM. 
The proton fraction, $Y_{p}$, in $\beta$-equilibrated NSM is obtained by solving Eq.~(\ref{eq.5}) for the exact expression of the energy and Eq.~(\ref{eq.7}) for the PA. 
The equilibrium proton fraction $Y_{p}$ thus obtained 
under the exact and the PA for the EOSs having different values of the slope parameter 
$L$=70, 76.26, 80, 90, 100 and 110 MeV 
are compared in the different panels of Figure ~\ref{fig:yp_exact_compare}. 
\begin{figure}[ht]
\vspace{1.5cm} 
	\begin{center}
		\includegraphics[width=0.7\columnwidth,angle=0]{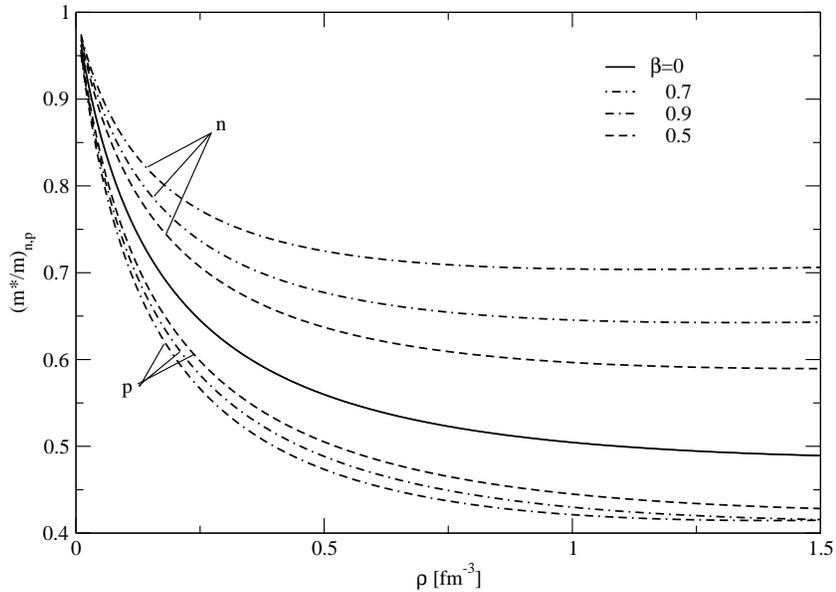}
		\end{center}
	\caption{Neutron and proton effective masses 
	as a function of density for values of isospin asymmetry $\beta$=0.5, 0.7, 0.9. The central continuous curve is the result for SNM.}
	\label{fig:efm12new}
\end{figure}
It can be seen that there is an overall good agreement in the predictions of $Y_{p}$ 
between the PA and the exact expression of the EOS, for all values of $L$. However, for low values of $L$, a small-scale difference 
between the exact and PA results may be noted from panels (a) and (b) of Figure ~\ref{fig:yp_exact_compare}. 
We shall examine in the following applications the physical consequences, if any, of this small 
difference in the equilibrium proton fractions resulting from the exact and PA of the energy, in particular, for the 
EOSs in the close vicinity of the direct Urca threshold. It may be noted that the $\beta$-equilibrium proton fraction $Y_{p}$ can be considered as a topological mapping of the density dependence of symmetry energy. This can be seen from the corresponding curves in the two panels (a) and (b) of Figure~\ref{Fig1} for different $L$.
%
As the neutron and proton effective masses, $(m^*/m)_n$ and $(m^*/m)_p$,
in NSM are also required in the calculation of the bulk viscosity $\zeta$, 
we display in Figure~\ref{fig:efm12new} the results for 
$(m^*/m)_n$ and $(m^*/m)_p$, calculated from 
Eq.~(\ref{eq.A.6}) of Appendix A, as a function of the density for some 
representative values of the asymmetry $\beta$. It is to be noted here that the 
 neutron and proton effective masses 
 are determined by only $\epsilon^l_{ex}$, $\epsilon^{ul}_{ex}$ and 
 the range $\alpha$ of the form factor of the interaction,
which take the same value for all the parametrizations with different slope of the 
symmetry energy $L$ for given $\gamma$ and $E_{s}(\rho_{0})$.

The quantity $C_{l}$ in Eq.~(\ref{eq.11}) is evaluated for 
 the set of EOSs having slope parameter $L$=70, 76.26, 80, 90, 100 and 110 MeV
using the exact expression of the energy per
particle in ANM as well as its PA.  
The expression of $C_l$ for the exact analytical expression of the energy in ANM 
 reads 
\begin{equation}
C_l=-\frac{\rho}{2} 
\left[
\left(\frac{\partial \mu_p}{\partial \rho_n}-\frac{\partial \mu_n}{\partial \rho_n}\right)\left(1+\beta\right)
+
\left(\frac{\partial \mu_p}{\partial \rho_p}-\frac{\partial \mu_n}{\partial \rho_p}\right)\left(1-\beta\right)
\right] 
-\frac{c^2 p_{fl}^2}{3 \mu_l},
\label{eq.A.9}
\end{equation}
while for the PA of the energy is given by
\begin{equation}
C_l=4 \beta \rho \frac{d E_{s}}{d \rho}-\frac{c^2 p_{fl}^2}{3 \mu_l}.
\label{eq.A.10}
\end{equation}
The density range 
for the occurrence of the direct Urca process, which is determined from the condition~(\ref{eq.2}) in NSM, 
is given in Table~\ref{table2} for each of the EOSs. 
The direct Urca process does not occur at any density
for $L$=70 MeV. For the EOS corresponding to the characteristic $E'_{s}(\rho_{0})$ value ($L$=76.26 MeV), 
the direct Urca predicted by the exact and PA calculations start at slightly different values of the density 
and also are 
switched-off at certain different densities, as it can be seen from Table~\ref{table2}. 

\begin{figure}[ht]
\vspace{1.5cm}
	\begin{center}
		\includegraphics[width=0.9\columnwidth,angle=0]{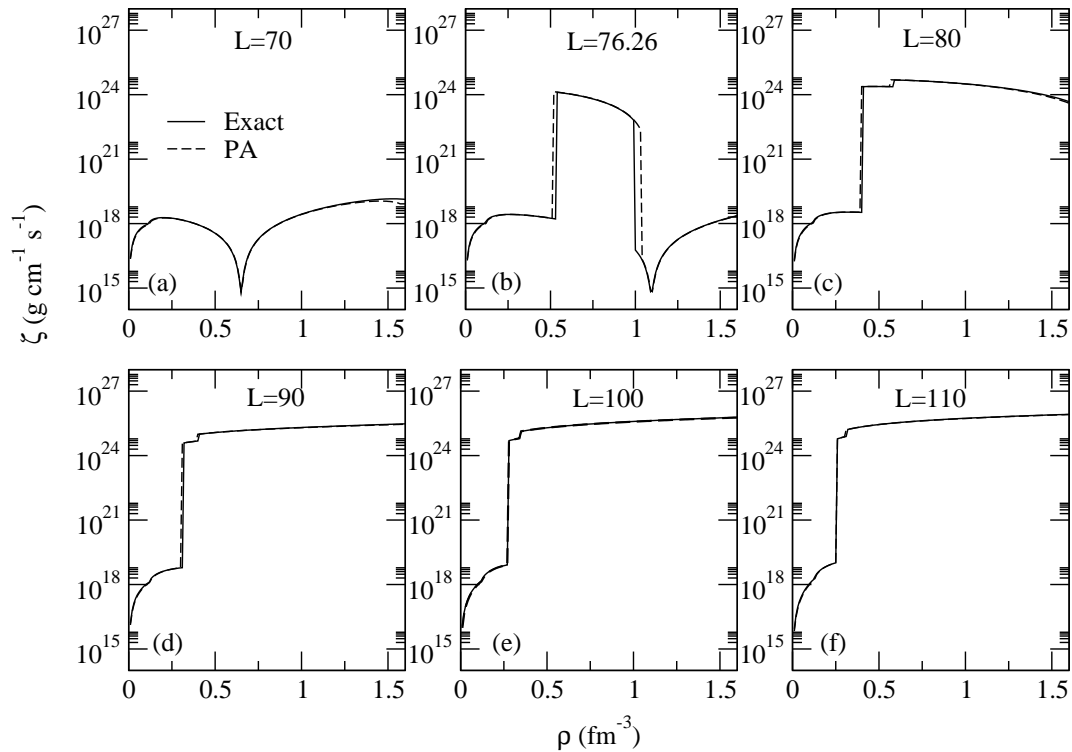}
		\end{center}
	\caption{Comparison of the bulk viscosity $\zeta$ in NSM 
	obtained under the exact and PA calculations for the different EOSs of ANM corresponding to $L$=70, 76.26, 80, 90 and 110 MeV 
shown in individual panels as a function of density. These results are obtained using $\omega_p$=10$^{4}$ and T=10$^9$K.}
\label{fig:xi}
\end{figure}

The total bulk 
viscosity 
is provided by the sum of the contributions of the modified and direct Urca processes 
$\zeta=\zeta^{MU}+\zeta^{DU}$.
As the bulk viscosity is additive
for the different species of leptons, the total modified Urca contribution is $\zeta^{MU}=\zeta_{ne}^{MU}+\zeta_{pe}^{MU}+ 
\zeta_{n\mu}^{MU}+\zeta_{p\mu}^{MU}$ and the total direct Urca contribution is $\zeta^{DU}=\zeta_{e}^{DU} +\zeta_{\mu}^{DU}$.
The individual contributions to the total modified Urca contribution coming from electrons and muons of the neutron and proton branches are computed from Eqs.~(\ref{eq.13})-(\ref{eq.16}).  
Similarly, the two direct Urca contributions to the bulk viscosity are calculated from 
Eq.~(\ref{eq.12}) for both kinds of leptons $l=e,\mu$. 
These different contributions to the bulk viscosity are computed for the representative
values of $\omega_{p}$=10$^{4}$ and T=10$^9$K. The results for the total $\zeta$, thus obtained in the high-frequency limit, are shown in the different panels of Figure ~\ref{fig:xi} as a function of the density for the set of EOSs corresponding to $L$=70, 76.26, 80, 90, 100 and 110 MeV.
From Table 2 we see that for the EOS of $L$=70 MeV, there is only MU contribution 
and no DU processes are allowed. However, for $L$=76.26 MeV the bulk viscosity curve predicted by the exact (PA) calculations shows 
a sudden jump to a higher value at the onset of direct Urca for electrons at $\rho$=0.54 
fm$^{-3}$ ($\rho$=0.52 fm$^{-3}$) and abruptly falls at $\rho$=0.99 fm$^{-3}$ ($\rho$=1.03 fm$^{-3}$), 
where the direct Urca process gets switched-off. 
 Direct Urca processes involving muons do not occur at any density 
in this case of the characteristic EOS.
For the EOSs corresponding to the higher values of $L$=80, 90, 100 and 110 MeV, the direct Urca process under 
the exact and PA calculations
starts almost at same density, marked by a sudden rise in the respective bulk viscosity curves. 
From this density onward, 
both DU and MU processes contribute to $\zeta$
along the 
whole range of considered densities.
The value of this direct Urca onset density decreases with increasing $L$.
The direct Urca process including muons also occurs for 
these higher values of the slope parameter above the critical $L$,
%
which are marked by a smaller sudden rise in the respective curves of the bulk viscosity.
It is found that for EOSs with $\gamma$=1/2 and different slope 
parameter $L$, at and around its characteristic value (in this case 
$L$=76.26 MeV), the densities for which the direct Urca process is 
switched on and off in $\zeta$ are slightly different under the predictions of the exact and PA calculations. 
The small difference
in the predicted values of the densities for switch-on and switch-off of the direct Urca process is prominent for the characteristic 
value of the slope parameter in the present case. This difference between the exact and PA results decreases as $L$ increases becoming 
almost negligible for $L$ $\geq$ 80 MeV. This is due to the small differences in the predictions of the proton fractions 
obtained under the exact and PA for the EOSs having a slope parameter value around the characteristic $L$.
\begin{figure}[ht]
\vspace{1.5cm}
	\begin{center}
		\includegraphics[width=0.9\columnwidth,angle=0]{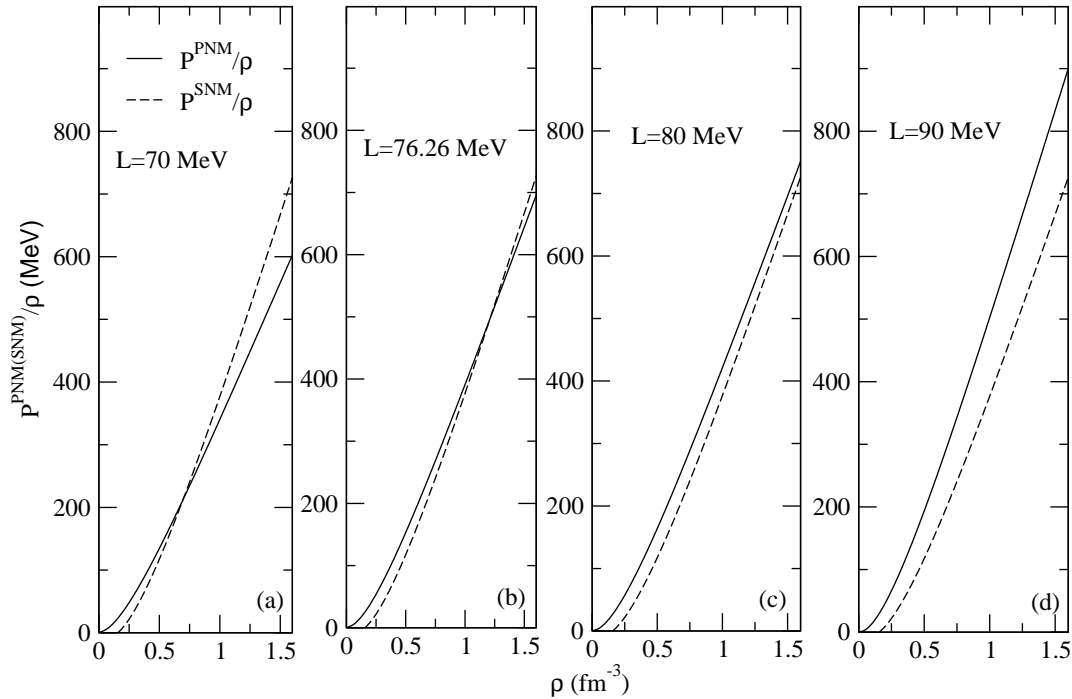}
		\end{center}
	\caption{Pressure per particle in PNM, $P^{PNM}(\rho)$/$\rho$, and in SNM, $P^{SNM}(\rho)$/$\rho$, as a function of density for the EOSs 	corresponding to $L$=70, 76.26, 80 and 90 MeV shown in four panels (a), (b), (c) and (d), respectively.}
	\label{fig:pressure}
\end{figure}
\begin{figure}[ht]
\vspace{1.5 cm}
	\begin{center}
		\includegraphics[width=0.9\columnwidth,angle=0]{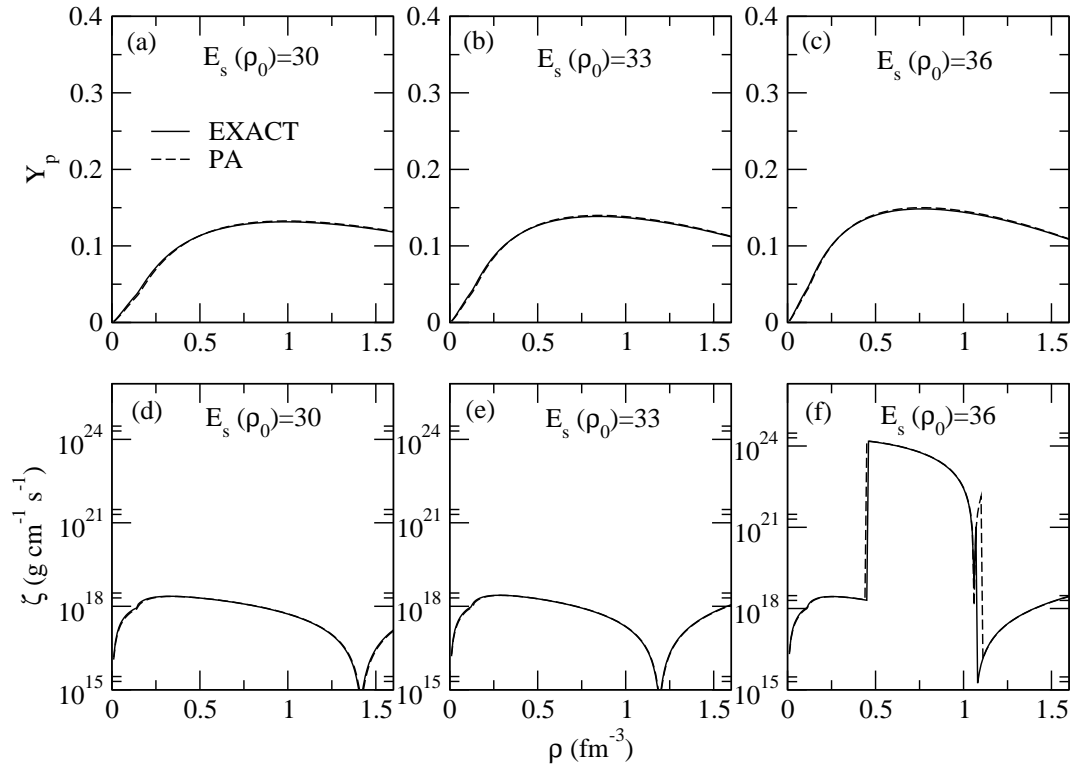}
		\end{center}
	\caption{Equilibrium proton fraction as a function of density, in the upper panels, for the three EOSs of ANM 
having $E_s(\rho_0)$=30, 33 and 36 MeV and for respective characteristic $L$ values. In the lower panels, the corresponding 
bulk viscosity $\zeta$ as a function of $\rho$ for exact and PA are shown.} 
\label{fig:xii}
\end{figure}
%

We note the presence of dips in the curves of the bulk viscosity $\zeta$ displayed in the panels (a) and (b) of 
Figure ~\ref{fig:xi}. 
These dips can be understood by analyzing the behaviour of the function $C_l^{2}$ in the expression for $\zeta$ 
in Eq.~(\ref{eq.12}). The function $C_l$ for the exact and PA cases is given in Eqs.~(\ref{eq.A.9}) and (\ref{eq.A.10}),
 respectively. Eq. (\ref{eq.A.10}) can be used to understand the occurrence of dips in the $\zeta$ curves 
in the case of low $L$ values. The nuclear part, which is the first term in the right 
hand side of the Eq.~(\ref{eq.A.10}), contains the derivative of the symmetry energy with respect to density. Now, if we use the 
expression for $E_s$ given in Eq.~(\ref{eq.6a}), the particle density times the density derivative 
of the symmetry energy results into the difference between the pressures per particle in PNM and SNM, i.e., 
$\rho dE_s(\rho)/d\rho$=[$P^{PNM}(\rho)-P^{SNM}(\rho)$]/$\rho$, where $P^{PNM}(\rho)$ and $P^{SNM}(\rho)$ are the pressures in 
PNM and SNM, respectively. We display the density dependence of $P^{PNM}(\rho)$/$\rho$ and 
$P^{SNM}(\rho)$/$\rho$ for the EOSs corresponding to $L$=70, 76.26, 80 and 90 MeV in four different panels of Figure~\ref{fig:pressure}. 
It can be seen that for a lower value of the slope parameter $L$, the pressure in SNM crosses-over to the pressure 
in PNM at a certain density, which shifts to the higher values with an increase in $L$. At the cross-over point, the function 
$C_l$ makes a change from positive value to negative one and increases negative-wise with increase in density. 
As $C_l$ appears with a quadratic power in the expression of $\zeta$ in Eq.~(\ref{eq.12}), the function $\zeta$ becomes
minimum at the density corresponding to the cross-over point. Since the density at this point 
takes a higher value when $L$ increases, the dips in the 
figures in panels (a) and (b) of Figure ~\ref{fig:pressure} are shifted to higher densities. 
However, if the value of the slope parameter increases further, the density dependence of $P^{PNM}(\rho)$/$\rho$ 
becomes stiffer than that of $P^{SNM}(\rho)$/$\rho$ and their difference increases when the density increases, 
as it can be seen from panel (d) of Figure~\ref{fig:pressure} for $L$=90 MeV. Due to this behaviour, there are no dips  
in the curves of $\zeta$ in Figure ~\ref{fig:xi} for the EOSs of $L$=90, 100 and 110 MeV. 

 The appearance of a dip in the bulk viscosity $\zeta$ is a general feature that can be 
observed in all EOSs of ANM with relatively low $L$ value. In order to show this, we have considered different EOSs in 
ANM with $\gamma$=1/2 but assuming different values of the symmetry energy, namely, $E_s(\rho_0)$=30, 33 and 
36 MeV. The characteristic $E'_s(\rho_0)$ values are 22.23 ($L$=66.69), 24.14 ($L$=72.42) and 26.1 ($L$=78.30) MeV, respectively. The comparison of the equilibrium proton 
fractions of the three characteristic EOSs is shown in the upper three panels of Figure ~\ref{fig:xii}, for both 
the exact and PA calculations, whereas, the corresponding results for $\zeta$ are shown in the lower panels. 
The dip in $\zeta$ appears in all three cases due to the cross-over of $P^{SNM}(\rho)$/$\rho$ to $P^{PNM}(\rho)$/$\rho$ as explained above. 
In the case of the first two EOSs, the dip appears isolated, as far as no direct Urca process takes place. However, in the case of the EOS 
for $E_s(\rho_0)$=36 MeV, in the rightmost panel, there is a direct Urca process superimposed in the region of density where
the cross-over density for the $P^{SNM}(\rho)$/$\rho$ and $P^{PNM}(\rho)$/$\rho$ curves for the EOS appears. Due to this cross-over effect, which takes place at a density about 1 fm$^{-3}$, the $\zeta$ curve 
shows a sudden fall in the Urca region followed by a sudden rise, and again falls back as the direct Urca switch-off density, which is also close to 1 fm$^{-3}$, is 
encountered. It may be mentioned here that the direct Urca threshold $L$ values for the EOSs corresponding to $E_s(\rho_0)$=30 
and 33 MeV lie close above their characteristic $L$ values, whereas in the case of $E_s(\rho_0)$=36 MeV the direct Urca threshold $L$ values lie close below the 
characteristic $L$ value for $E_s(\rho_0)$=36 MeV. 

%
%
%
%
%
 In the following we shall examine the implication of this sharp enhancement of the bulk viscosity 
due to
the appearance of direct Urca processes as $L$ increases, on the $r$-mode phenomenology of the pulsar NS, in particular, on the instability boundary and the spin-down features in newborn normal NSs. 

\subsection{Region of instability under $r$-mode oscillation}

We shall now examine the influence of the bulk viscosity, calculated consistently 
 to the EOSs of ANM for different values of $L$, on the $r$-mode instability boundary of pulsar NSs. 
It is well known that the bulk viscosity is the dominating restoring mechanism in the high temperature domain, $T\geq 10^{9}$K, of the instability arising out of continuous emission
of gravitational waves 
by pulsar NSs under $r$-mode oscillation.

The $r$-mode oscillations on the surface of rotating neutron stars are due to the surface currents 
to which the Coriolis force provides the real dynamics \cite{Andersson1998,Lindblom1998}. Out of the various $r$-modes, 
the lowest one $l$=2, $m$=2, is the most important one and the pulsar NS is subject to continuous emission of 
GWs by the Chandrasekhar-Friedman-Schutz mechanism \cite{Chandrasekhar1970,Friedman1978}. The instability due to the 
continuous emission of GWs under $r$-mode oscillation is counterbalanced by the various damping mechanisms in the 
volume of the NS and finally the oscillation is stabilized
with a frequency $\omega$. The time dependence of the $r$-mode oscillation is given by $e^{i\omega t-t/\tau}$, where 
the imaginary part time-scale $1/\tau$ is determined from the combined effects of gravitational radiation, 
different viscous mechanisms 
operating in the volume of the NS, etc. Here we shall restrict ourselves to the shear 
and bulk viscosity of the core as the damping forces, i.e., the minimal model, because our aim in this
work is focused to study the effect of the bulk viscosity 
on the $r$-mode instability when the slope parameter $L$ varies.
The 
imaginary part time-scale is now given by
\begin{eqnarray}
\frac{1}{\tau(\Omega,T)}=\frac{1}{\tau_{GR}(\Omega)}+\frac{1}{\tau_{BV}(\Omega,T)}+\frac{1}{\tau_{SV}(T)},
\label{eq.19}
\end{eqnarray}
where, $\Omega$ is the angular velocity of the rotating NS, $T$ is its temperature, 
$\frac{1}{\tau_{GR}}$ is  the gravitational radiation time-scale, and  $\frac{1}{\tau_{BV}}$ 
and $\frac{1}{\tau_{SV}}$ are the bulk and shear viscous time-scales of the core, respectively.
The $r$-mode is stable as long as $\frac{1}{\tau{(\Omega,T)}}$ ${\geq}$ 0. 
At given $T$, when $\Omega$ increases, the intensity of the
gravitational radiation 
also increases and a critical frequency, $\Omega_{c}$, is reached when the 
restoring viscous effects just counterbalances the instability due to the gravitational 
radiation, i.e., $\frac{1}{\tau{(\Omega_{c},T)}}$=0. Beyond $\Omega_{c}$ the $r$-mode becomes
 unstable and the NS enters into the region of instability. 
\begin{figure}[b]
\vspace{1.5cm}
	\begin{center}
	\includegraphics[width=0.9\columnwidth,angle=0]{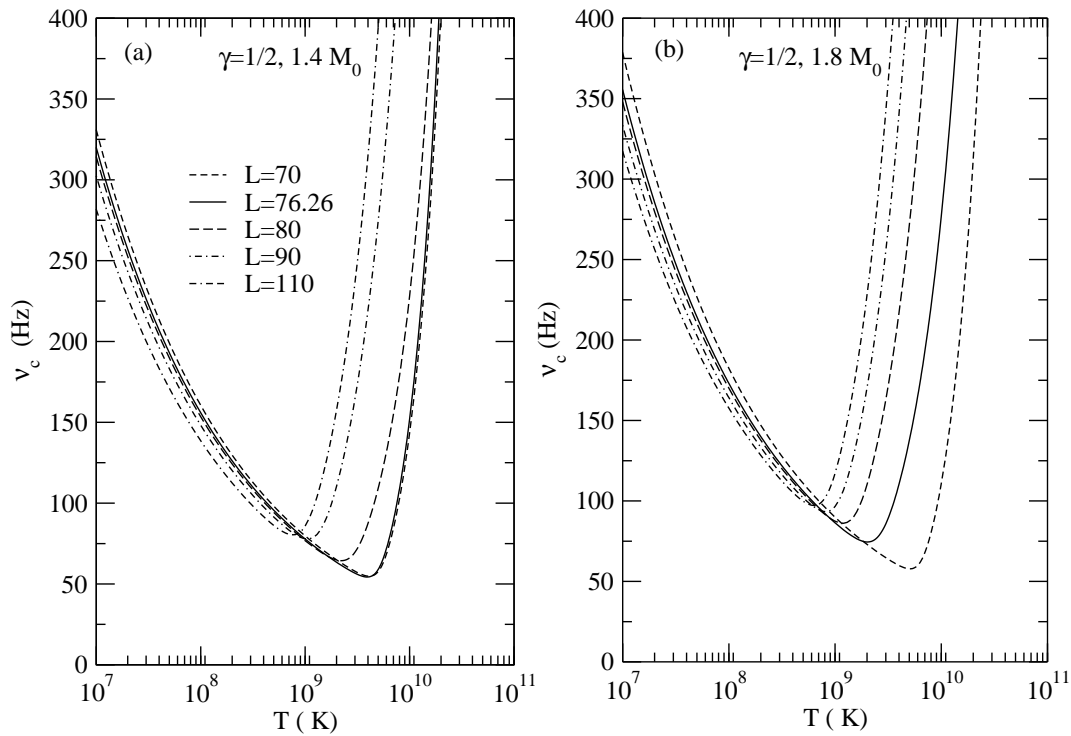}
		\end{center}
	
	\caption{The critical frequency, $\nu_{c}$, as a function of temperature $T$ for the EOSs 
of $\gamma$=1/2 having values of $L$=70, 76.26, 80, 90 and 110 MeV in pulsar NSs of mass 1.4 $M_\odot$ and 1.8 $M_\odot$
in panels (a) and (b), respectively.}\label{fig:1.4_1.8_12}
\end{figure}

The approximate expression for 
the bulk viscous time-scale valid for stars rotating with slow frequency is given by 
\cite{Vidana2012},
\begin{eqnarray}
&&\frac{1}{\tau_{BV}\left({\Omega,T}\right)}=\frac{4\pi R^{2l-2}}{690}
\left(\frac{\Omega}{\Omega_{0}}\right)^{4}\left(\int^{R}_{0} \rho_{m}(r^{\prime})r^{\prime (2l+2)}dr^{\prime}\right)^{-1}
\nonumber \\ &\times&\int^{R}_{0} \zeta\left(\frac{r^{\prime}}{R}\right)^{6}
\left[1+0.86\left(\frac{r^{\prime}}{R}\right)^{2}\right]r^{\prime 2} dr^{\prime} \quad (\mbox{\rm s}^{-1}),
\label{eq.20}
\end{eqnarray}
 where, $\rho_{m}(r^{\prime})=H(\rho(r^{\prime}))/c^2$ being the mass density at a distance $r^{\prime}$, 
$\Omega_0$=$(\pi G \bar{\rho})^{1/2}$ with $\bar{\rho}=(\frac{3 M}{4 \pi R^3})^{1/2}$ being the average density of
the NS of mass $M$ and radius $R$, which are obtained from the solution of the TOV equations. 
In many works \cite{Lindblom1998,Moustakidis2015}, including ours in Ref.~\cite{trr2018}, a 
general expression for the bulk viscosity, $\zeta$, derived from hydrodynamics was used. However, 
in the present work, the bulk viscous damping contribution is obtained from 
the Urca processes, which are consistent to the EOS used to compute the NS properties, as has been done by Vida\~na 
\cite{Vidana2012} and others \cite{Kolom2015}.
The gravitational and shear viscous time-scales are the same as used in 
our earlier work \cite{trr2018} and are reported in 
Eqs.~(\ref{eq.B.1})-(\ref{eq.B.4}) of Appendix B for the sake of completeness. 
\begin{figure}[ht]
\vspace{1.5cm}
	\begin{center}
		\includegraphics[width=0.9\columnwidth,angle=0]{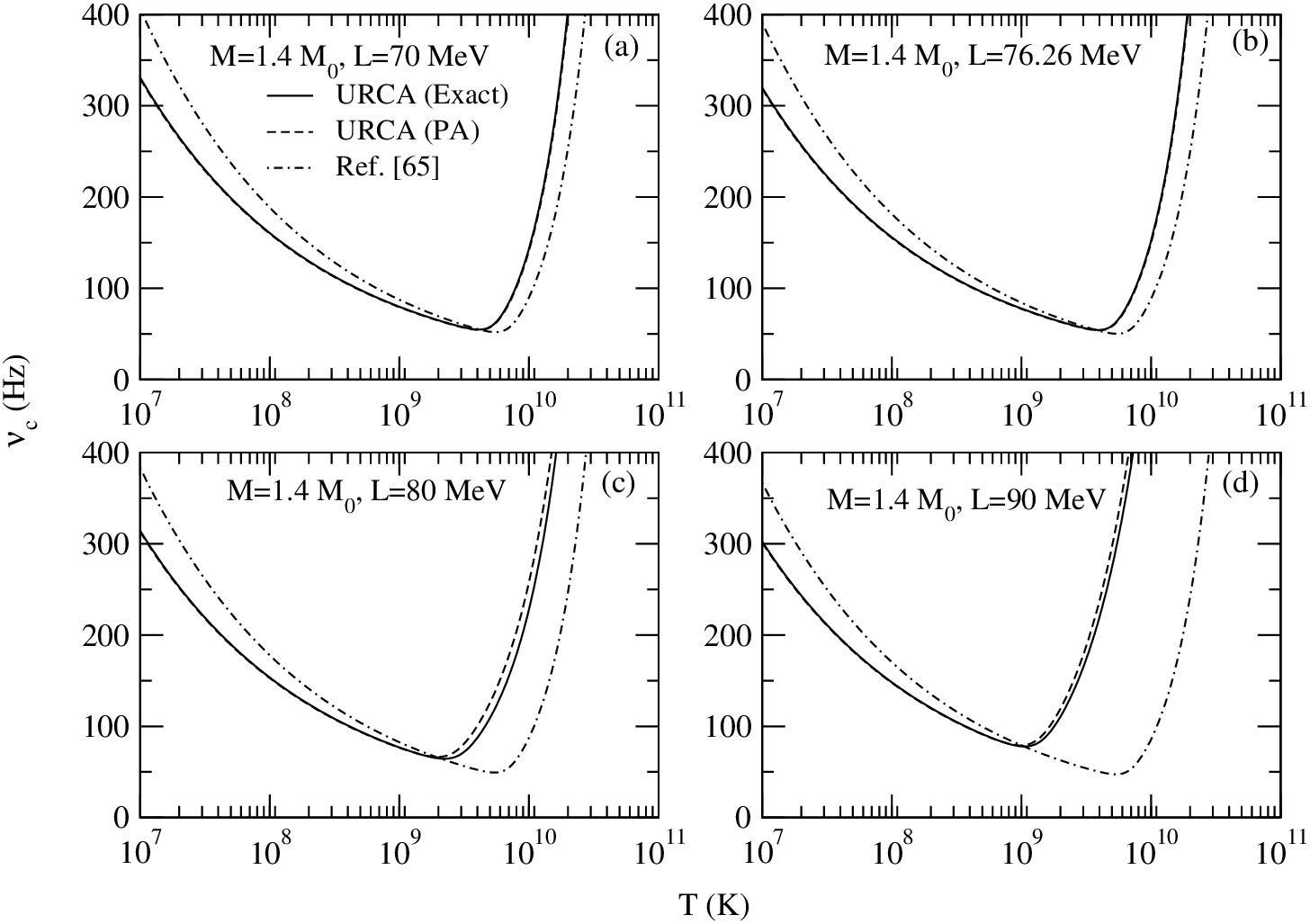}
		\end{center}
	\caption{Comparison of the instability boundary obtained 
using the total bulk viscosity $\zeta$ and the hydrodynamical approximation of Ref.
\cite{trr2018}.
The calculations have been performed for EOS of $\gamma$=1/2 with $L$=70, 76.26, 80 and 90 MeV for pulsar NS of 1.4 $M_{\odot}$ mass and using the exact EOS and its PA in ANM.
}
\label{fig:1.4_compare}
\end{figure}
\begin{figure}[ht]
\vspace{1.5cm}
	\begin{center}
			\includegraphics[width=0.9\columnwidth,angle=0]{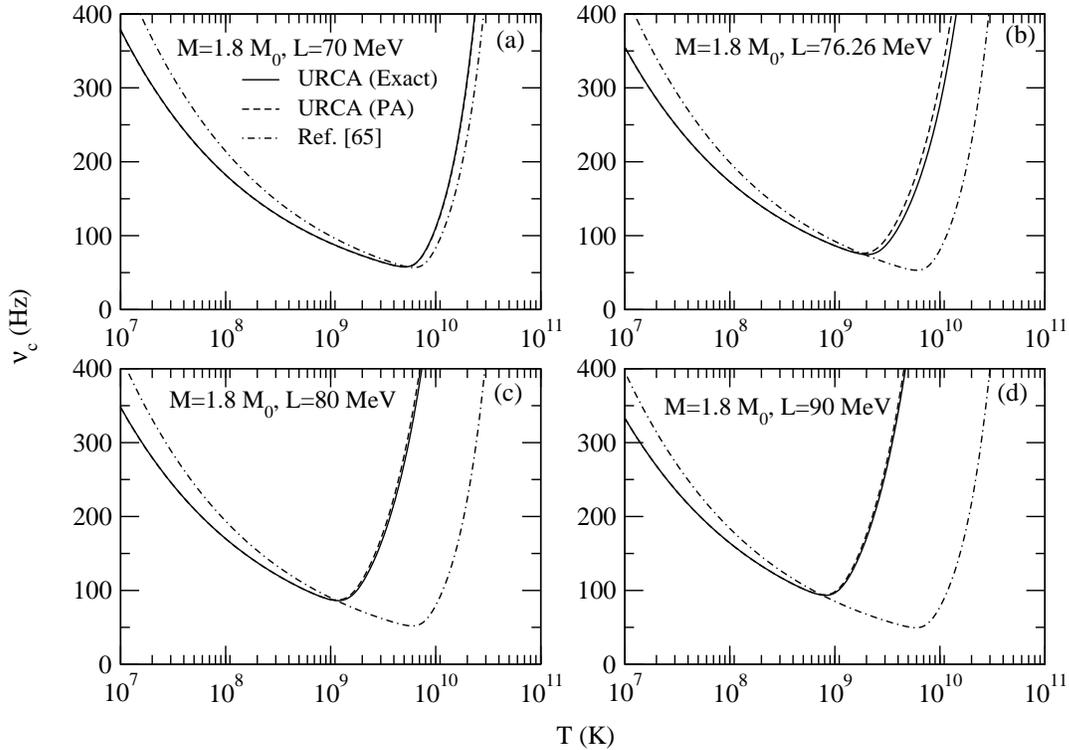}
		\end{center}
	\caption{ Comparison of the instability boundary obtained 
	using the total bulk viscosity $\zeta$ and the hydrodynamical approximation of Ref.\cite{trr2018}. The calculations have been performed for EOS of $\gamma$=1/2 with $L$=70, 76.26, 80 and 90 MeV for pulsar NS of 1.8 $M_{\odot}$ mass and using the exact EOS and its PA in ANM. 
}\label{fig:1.8_compare}
\end{figure} 
\begin{figure}[ht]
\vspace{1.5cm}
	\begin{center}
		\includegraphics[width=0.9\columnwidth,angle=0]{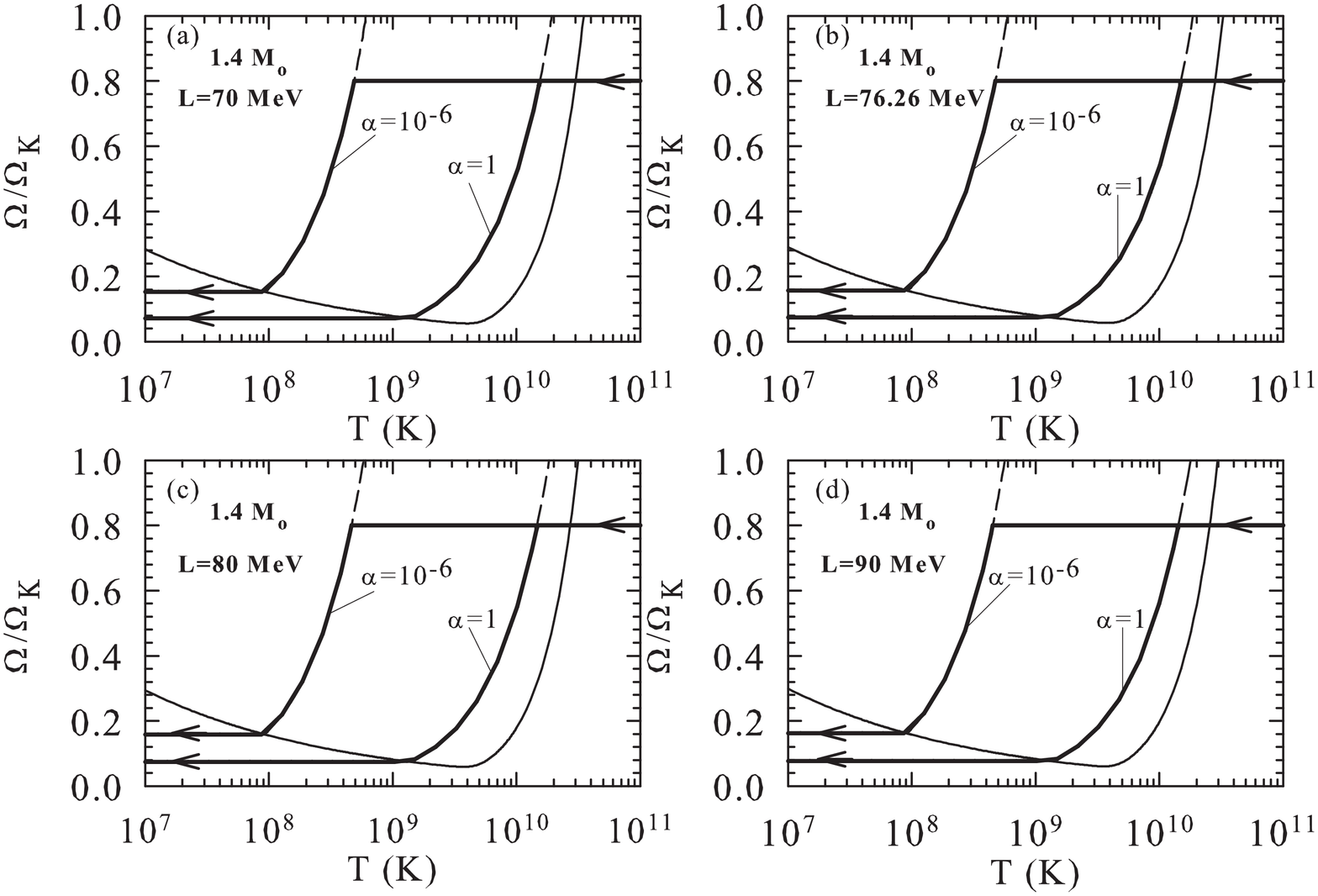}
		\end{center}
	\caption{Thermal steady state spin-down path from Eq.(\ref{eq.A10}) (thin long-dashed line) and from the numerical solution of Eq.(\ref{eq.A8}) with initial angular velocity $\Omega_{i}$=0.8$\Omega_{K}$ (thick continuous line) for two values of saturation amplitude $\alpha$=1 and 10$^{-6}$ inside the $r$-mode instability region of 1.4 $M_\odot$ pulsar NS.
The calculation is performed for four EOSs
of $\gamma$=1/2 having $L$=70, 76.26, 80 and 90 MeV, where the direct Urca has been suppressed. The value of the Keplerian velocity $\Omega_{K}$ are 5863.9, 5642.7, 5528.1 and 5295.6 Hz respectively for these EOSs.}
	\label{fig:MUomegahc_1.4}
\end{figure} 
\begin{figure}[ht]
\vspace{1.5cm}
	\begin{center}
		\includegraphics[width=0.9\columnwidth,angle=0]{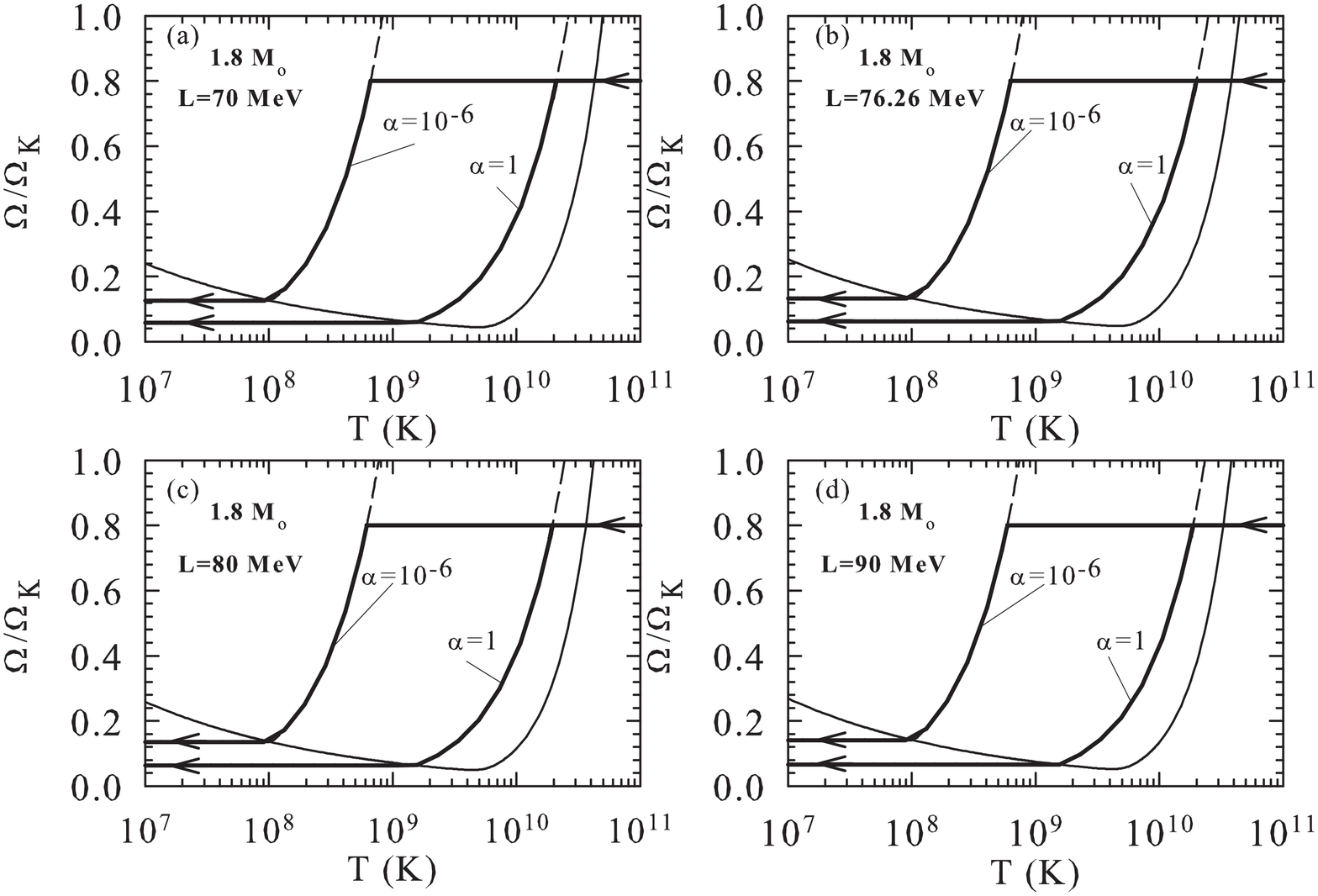}
		\end{center}
	\caption{ Thermal steady state spin-down path from Eq.(\ref{eq.A10}) (thin long-dashed line) and from the numerical solution of Eq.(\ref{eq.A8}) with initial angular velocity $\Omega_{i}$=0.8$\Omega_{K}$ (thick continuous line) for two values of saturation amplitude $\alpha$=1 and 10$^{-6}$ inside the $r$-mode instability region of 1.8 $M_\odot$ pulsar NS.
 The calculation is performed for four EOSs	
	of $\gamma$=1/2 having $L$=70, 76.26, 80 and 90 MeV, where the direct Urca has been suppressed. The value of the Keplerian velocity $\Omega_{K}$ are 7678.4, 7157.9, 6972.1 and 6603.0 Hz respectively for these EOSs.}
		\label{fig:MUomegahc_1.8}
\end{figure}

\begin{figure}[ht]
\vspace{1.5cm}
	\begin{center}
		\includegraphics[width=0.9\columnwidth,angle=0]{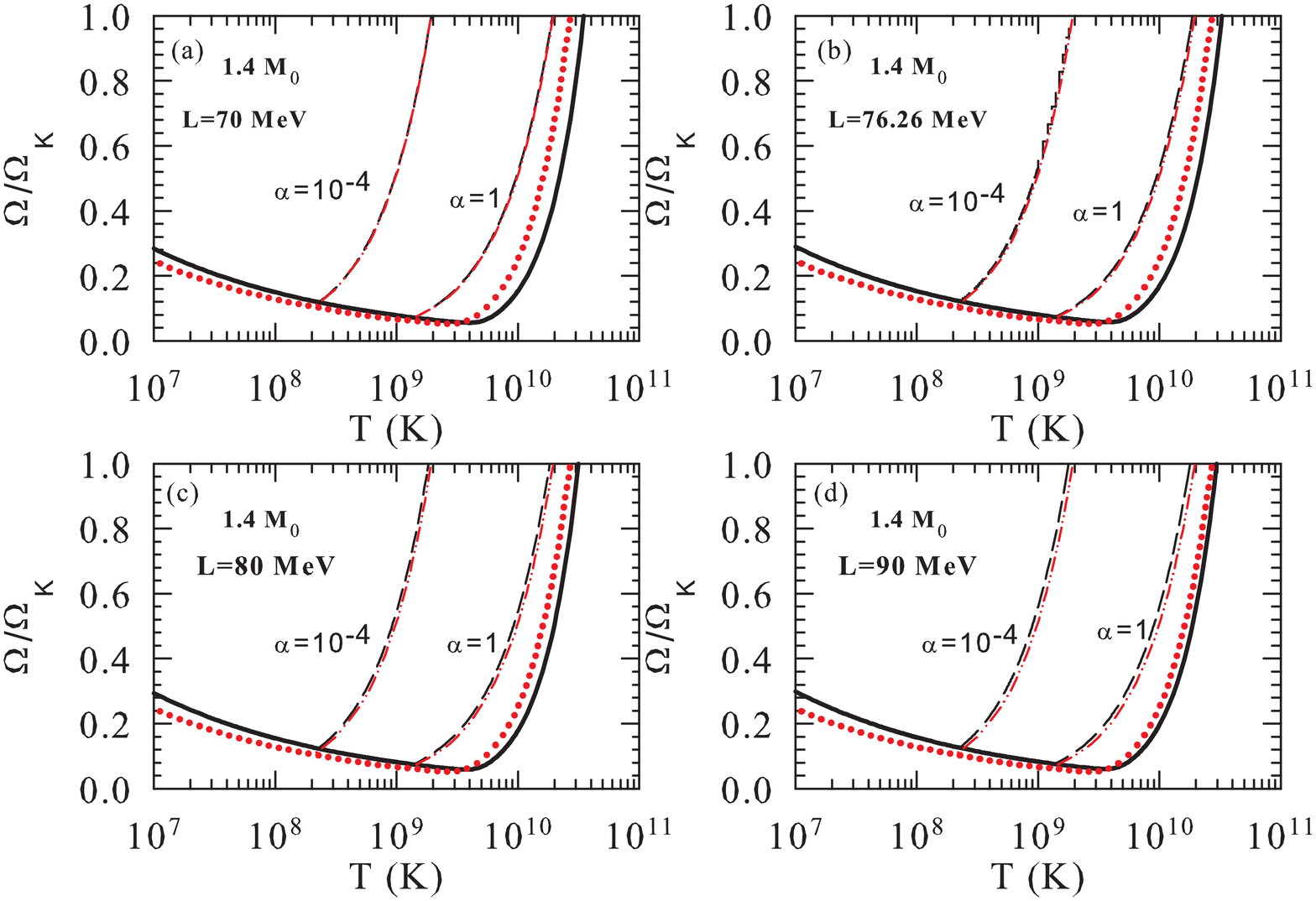}
		\end{center}
	\caption{(color online)Results of the instability boundary, steady state spin-down paths for the four EOSs of SEI having $\gamma$=1/2 and $L$=70, 76.26, 80 and 90 MeV in 1.4$M_\odot$ pulsar NS.
The results are compared with the predicitons obtained using the power law formulation of	
Alford et al. \cite{AMS2012,Alford2014} shown, in red, in the four panels. The thick dotted line in red is the instability boundary of Ref.\cite{Alford2014}.}\label{fig.omegahc_1.4_comparision}
\end{figure} 
The instability boundary is calculated by solving Eq.(\ref{eq.19}) for 
$\frac{1}{\tau{(\Omega_{c},T})}=0$, where 
the bulk viscous time-scale in Eq.~(\ref{eq.20}) has been evaluated by computing $\zeta$ 
taking into account the direct and modified Urca processes given by Eqs.~(\ref{eq.12}) and 
(\ref{eq.13})-(\ref{eq.16}), respectively. 
The calculations have been done for the 1.4 $M_\odot$ and 1.8 $M_\odot$ 
NS cases with
the EOSs of 
SEI corresponding to $\gamma$=1/2 for different $L$ values, namely, $L$=70, 76.26,
 80 and 90 MeV, 
using the exact EOS and its PA in ANM.
The results of the critical
 frequencies $\nu_{c}$ (=$\frac{\Omega_c}{2\pi}$) as a function of the temperature $T$ in 1.4 
$M_\odot$ and 1.8 $M_\odot$ NSs obtained in the exact case are shown in 
Figures~\ref{fig:1.4_1.8_12}(a) and \ref{fig:1.4_1.8_12}(b), respectively. 
From these figures, 
it follows that the instability
boundary in the high temperature region moves inward to the relatively 
low temperature values under the action
of Urca processes, which is a result similar to that found in Ref. \cite{Vidana2012}. This is due to the fact that as $L$ increases, the region of direct Urca covered 
in the volume of the NS also increases. In order to have a quantitative 
estimate of the influence of the Urca processes on the instability boundary,
we have 
compared the results obtained in the present work, computed using the EOS with $L$=70, 76.26, 80 and 90 MeV, with the calculation reported in the  Ref.\cite{trr2018} under the minimal model 
(without the viscous effect from the crust-core region).
In Ref.\cite{trr2018} a general expression
of $\zeta$, derived from hydrodynamical considerations (equation (11) of \cite{trr2018}), has been used.
 This comparison is shown in the different panels of 
Figures ~\ref{fig:1.4_compare} and \ref{fig:1.8_compare} for NSs of masses
1.4 $M_\odot$ and 1.8 $M_\odot$, respectively. In panels (a) of the two figures, corresponding to $L$=70 MeV, 
the high and low-temperature branches do not show much change. Whatever small changes noticed there are due to the 
modified Urca process as far as direct Urca processes are not predicted by this EOS. 
In panel (b) of Figure ~\ref{fig:1.4_compare}, which corresponds to the case of characteristic EOS having slope parameter 
$L$=76.26 MeV, a similar result is found for a 1.4 $M_{\odot}$ NS. In this case the direct Urca does not occur because the central density of the star is smaller than the threshold density for this process, as it can be seen from Table 2.
However, for this characteristic EOS, in case of a 1.8 $M_\odot$ NS, the direct Urca process takes place up to a 
distance of 6.19 km from the center. As a consequence, its instability boundary shows a marked inward shift as compared to the corresponding result of 1.4$M_\odot$ NS. 
For $L$$\geq$80 MeV the 
direct Urca process is allowed in both, 1.4 $M_\odot$ and 1.8 $M_\odot$ NSs, up to various distances from the center, as can be seen 
from Table 2. The impact of DU processes process in 1.4$M_\odot$ and 1.8$M_\odot$ mass NSs for the EOSs with $L$=80 and 90 MeV can be seen from the extent of inward 
shift of the high-temperature branches, which is shown in the panels (c) and (d) of Figure~\ref{fig:1.4_compare} and 
Figure~\ref{fig:1.8_compare}. The results for exact and PA are similar and only marginally small differences between the two 
calculations 
can be seen for the EOSs in the close vicinity of direct Urca threshold $L$-value. These small differences in the predictions under exact and PA vanishes rapidly 
when
occurrence of direct Urca in the volume of the NS increases as $L$ increases beyond the threshold $L$-value. 
%
%
%
%
%
%
%
%
%
%
In the work \cite{Vidana2012},  Vida{\~n}a examined the influence of the slope parameter $L$ by calculating the instability boundaries for a number of  model interactions, both microscopic as well as effective ones, having $L$ values ranging from 39.2 to 160.4 MeV. 
Although the work  
gives an overall influence of direct Urca, one cannot get a quantitative 
information about the impact of the slope parameter of the EOS, $L$, on the 
instability boundary for a given NS 
model, due to the fact that the
saturation 
properties and predictions in SNM of the interactions used in the study of Ref.\cite{Vidana2012} are not the same. Here, 
we make a
systematic analysis of the $L$ dependence of the $r$-mode instability features from the extent of occurrence of DU inside the normal fluid pulsar NS of given mass by varying the slope parameter 
$L$ 
of the EOS. 
To this end we use a set of EOSs of SEI with a given $\gamma$ value and whose saturation properties and predictions in SNM are the same, as discussed in subsection (2.3).
%
%
These EOSs differ only in their density dependence of the symmetry energy 
i.e, by the value of the slope parameter
 $L$, consequently predicting different properties in the isovector sector.
%
%
For an EOS of SEI with a given value of $\gamma$, the threshold value of $L$ for onset of direct Urca coincides with the critical $L$ value, discussed in sub-section (2.3), or lies very close to it. For a value of $L$ at and close around the threshold $L$-value, the direct Urca process may or may not be possible depending on the central density of the predicted NS of given mass. In general, for a $L$ value for which DU is allowed, the extent of occurrence inside the volume will depend on the mass and radius of the NS, as can be seen from the $R_{DU}$ values in
Table 2. For the EOS with a $L$ value 
that does not predict DU in NS of any mass, the instability boundary shows a very weak dependence on 
the microscopic details of the parameters involved, such as, mass distribution inside the NS, radius, etc.. This can be seen from the comparison of the curves for EOS $L$=70 MeV in the 1.4 $M_\odot$ and 1.8 $M_\odot$ NSs given in the two panels of Figure~\ref{fig:1.4_1.8_12}. This is in agreement with the findings in the earlier works of Lindblom et al. and Alford et al. \cite{Lindblom1998,AMS2012}. Under the present formulation, it can be said that the high-temperature branch of the instability boundary will 
not be much different in NSs of different masses if the $L$ value of the EOS is below the DU threshold value. However, there will be an inward shift in the high temperature branch if the 
slope parameter
$L$ of the EOS is above the threshold $L$ value for which DU processes are allowed. 
For a given EOS the magnitude 
of this shift will depend on the difference between the considered $L$ value 
and the corresponding threshold $L$ value for direct Urca processes. This shift also depends on the mass of the pulsar NS,
as can be seen from the two panels of Figure~\ref{fig:1.4_1.8_12}.
Extrapolating the insensitiveness 
to the details
of the EOS
in predicting the instability
boundary, Alford, Mahmoodifar and Schwenzer (AMS) \cite{AMS2012} have formulated a semi-analytical expression for describing the instability boundary through a power law approximation involving the dynamical variables. Semi-analytical expressions for the frequency and temperature at the minima of the 
instability boundary curve are also obtained under the AMS formulation,
which have values 59.05 Hz and 3.49$\times10^9$ K \cite{AMS2012}.
For the EOS of SEI with $L$=70 MeV, we obtain 54.80 Hz and $4.1\times10^9$K, which are in good agreement with the estimations of the AMS formulation.
It is to be noted here that in order to compare with the semi-analytical AMS predictions, we have 
considered only
the electron-electron ($e-e$) scattering in the shear viscosity
for having a compatible $T^{-5/3}$-dependence
in both the calculations. However, in rest of our calculations we have also considered the neutron-neutron ($n-n$) scattering which has a $T^{-2}$-dependence, alongwith the $e-e$ contribution in the shear viscosity. Using this power law formulation, Alford et al. have also obtained a semi-analytical expression for the spin-down path of a newly born hot NS inside the $r$-mode instability region \cite{Alford2014}. In the following, we shall study the spin-down feature using the present EOSs of SEI, where the influence of direct Urca on the path shall be explicitly examined.
%
%
%
\begin{figure}[ht]
\vspace{1.5cm}
	\begin{center}
		\includegraphics[width=0.9\columnwidth,angle=0]{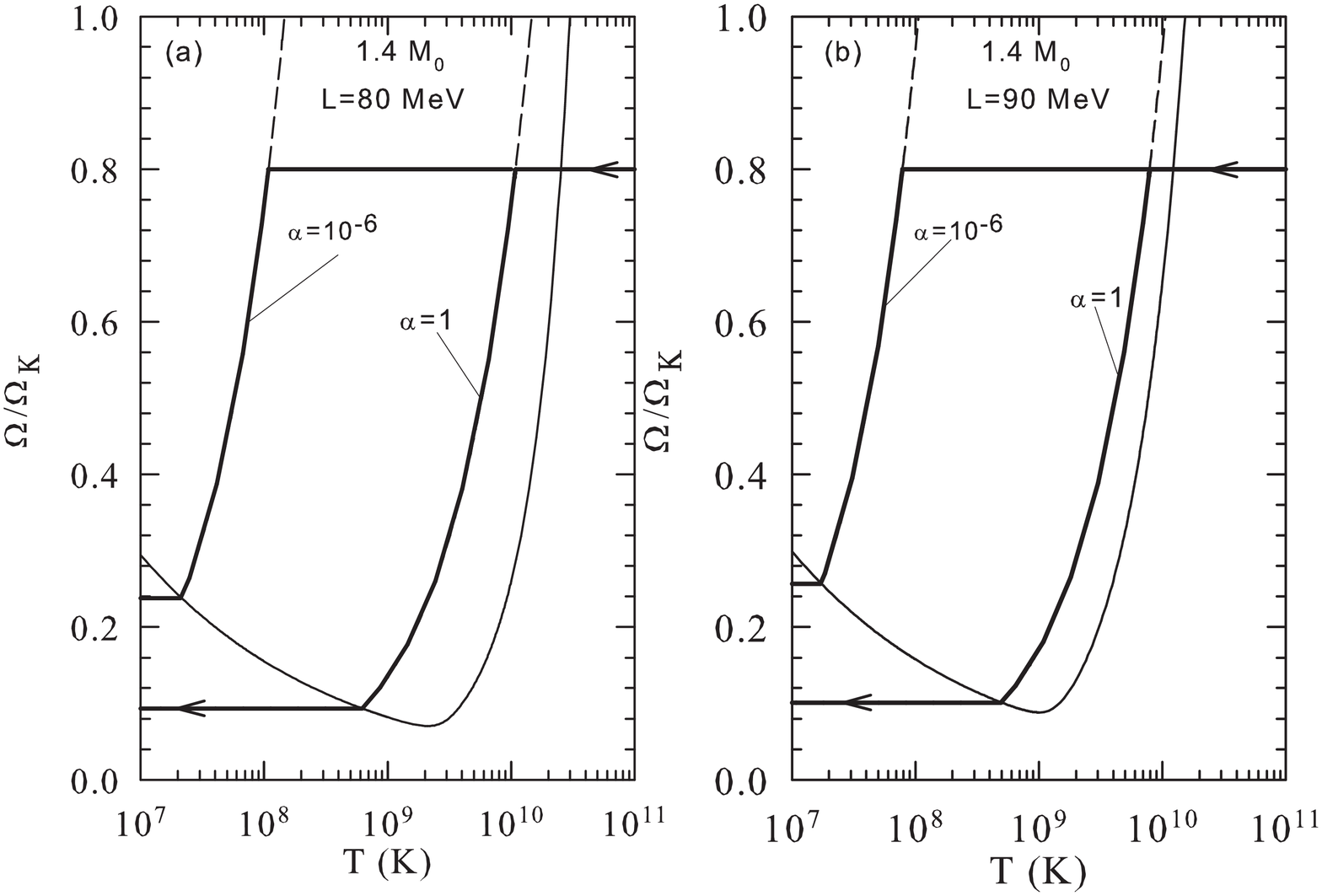}
		\end{center}
	\caption{Thermal steady state paths from Eq.(\ref{eq.A10}) (thin long-dashed line) and from the numerical solution of Eq.(\ref{eq.A8}) with initial rotational velocity $\Omega_{i}$=0.8$\Omega_{K}$ (continuous thick line) inside the $r$-mode instability region for 1.4$M_\odot$ pulsar NS.
The calculation has been done for two EOSs of
$\gamma$=1/2 and $L$=80 and 90 MeV taking into account the direct Urca contributions.	
	}
	\label{fig.DUomegahc_1.4}
\end{figure} 
\begin{figure}[ht]
\vspace{1.5cm}
	\begin{center}
		\includegraphics[width=0.9\columnwidth,angle=0]{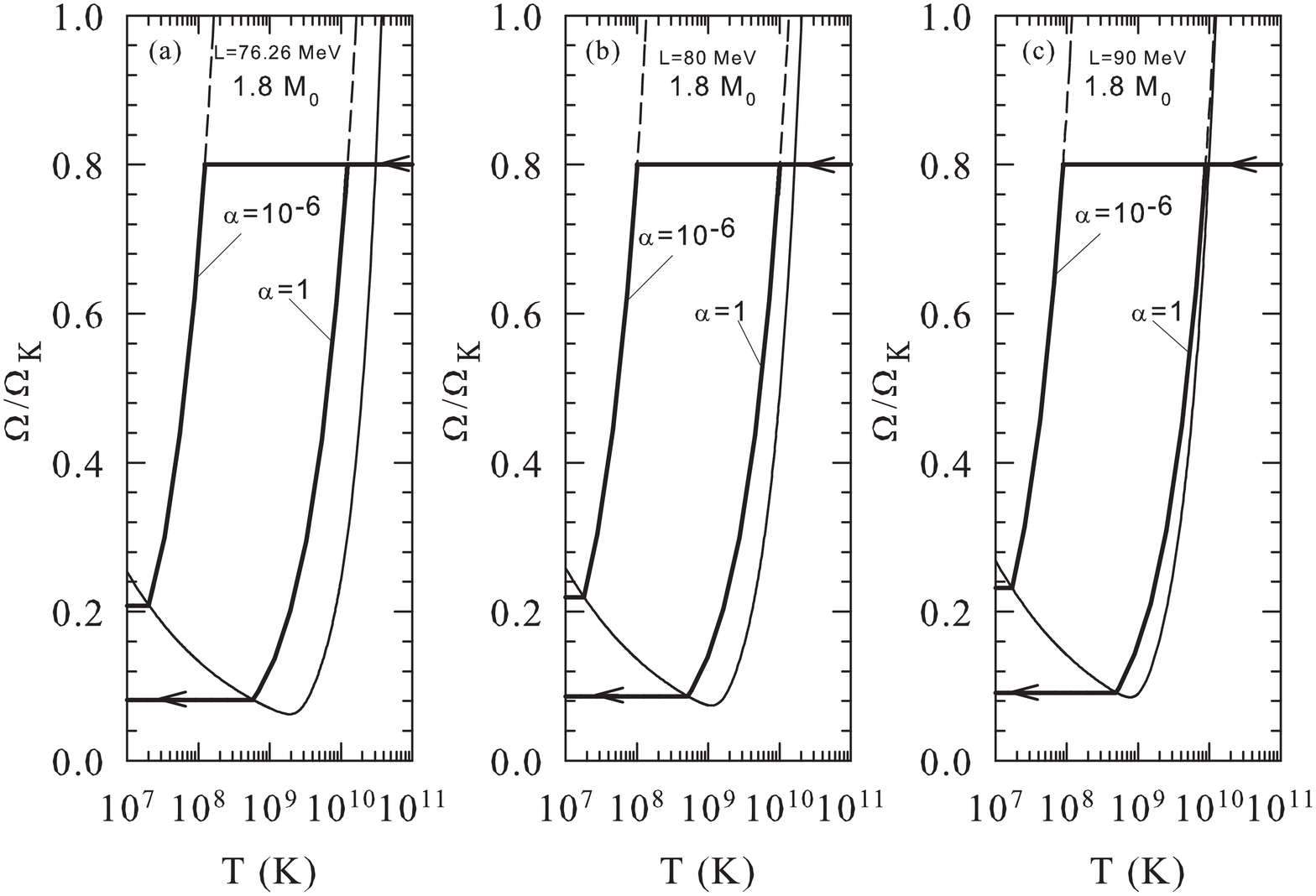}
		\end{center}
	\caption{ Thermal steady state paths from Eq.(\ref{eq.A10}) (thin long-dashed line) and from the numerical solution of Eq.(\ref{eq.A8}) with initial rotational velocity $\Omega_{i}$=0.8$\Omega_{K}$ (continuous thick line) inside the $r$-mode instability region for 1.8$M_\odot$ pulsar NS.
The calculation has been done for three EOSs of
$\gamma$=1/2 and $L$= 76.26, 80 and 90 MeV taking into account the direct Urca contributions.	
	}
	\label{fig.DUomegahc_1.8}
\end{figure}
\begin{figure}[ht]
\vspace{1.5cm}
	\begin{center}
		\includegraphics[width=0.9\columnwidth,angle=0]{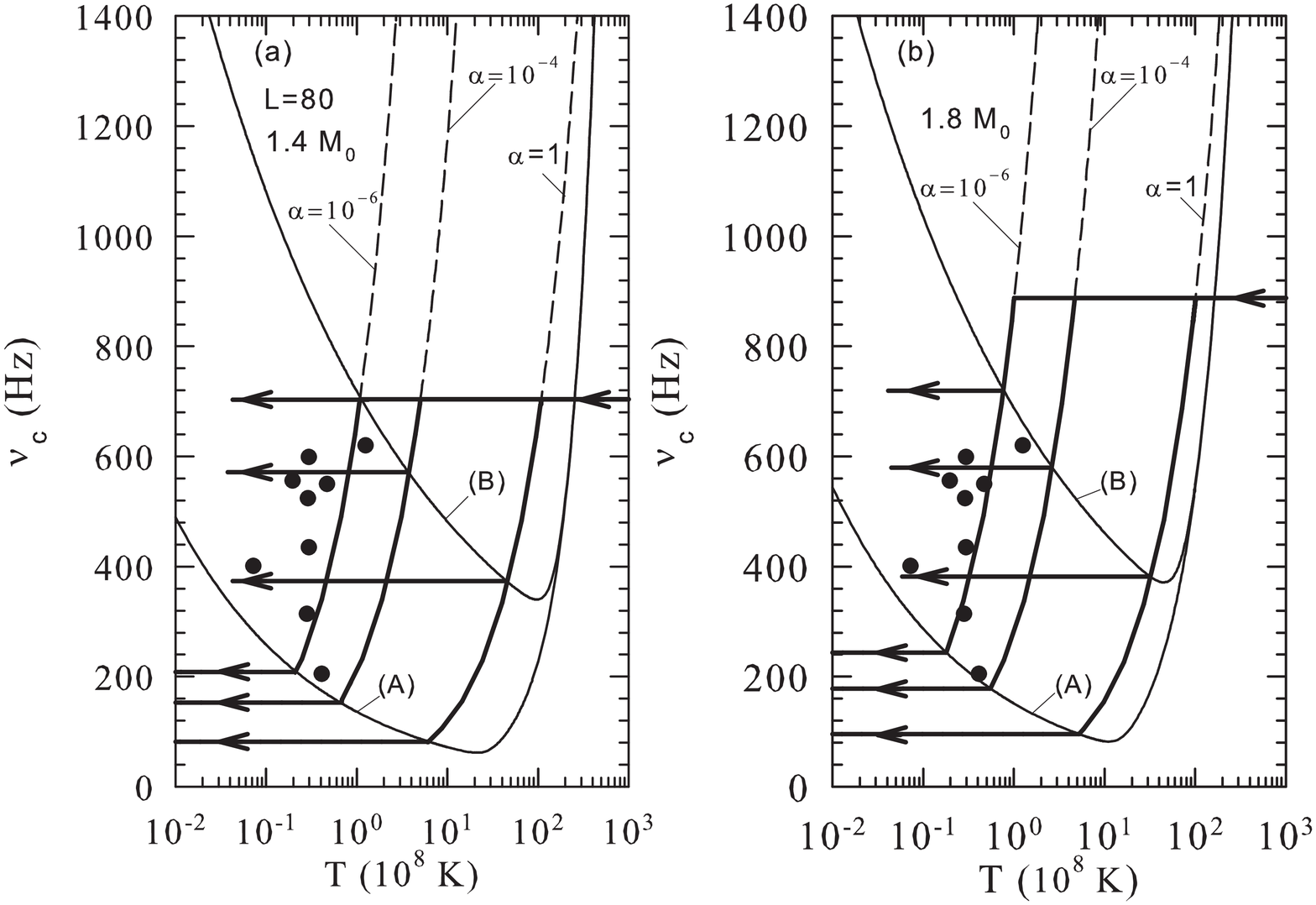}		
		\end{center}
	\caption{The instability boundary obtained under including the viscous layer damping contribution from the crust-core region, 
	 labelled (B), and excluding this effect (minimal model), labelled (A).
	The calculations are performed for the EOS with $\gamma$=1/2 and $L$=80 MeV.
	The position of the NSs of LMXBs and MSRPs are also indicated taking the data 
	from 
	Ref. \cite{Watts08,Heinke2007,Heinke2009,Tomsick2004}
	The spin-down paths inside the region of $r$-mode instability 
	instability are shown
	for three representative values of 
	the $r$-mode amplitude
	$\alpha$=1, 10$^{-4}$ and 10$^{-6}$. For details see the text.}	\label{graph_a1.4_1.8}
\end{figure}
\subsection{Spin-down of a newly born hot neutron star under $r$-mode oscillation}

A newly born NS in type-II supernovae has a temperature of the order of $10^{11}$K and a rotational frequency close to its Kepler limit. As the star cools due to neutrino emission and reaches the instability boundary in the high temperature branch, the $r$-mode instability sets in and the amplitude $\alpha$ goes on increasing until saturation is reached due to non-linear effects. Amongst the various known non-linear damping mechanisms \cite{Arras2003,Bondarescu2009,Bondarescu2013,Lindblom2001,Kastaun2011,Wu2001,Haskell2013}, it is not yet settled which of them actually saturates the mode. Hence, Owen et al. \cite{owen1998} proposed an approach, subsequently adopted by others as well \cite{AMS2012}, where a strong non-linear mechanism operates saturating the mode amplitude without going into any particular mechanism. Once the saturation of the $r$-mode amplitude is reached, the pulsar NS starts spinning down due to the continuous emission of gravitational waves in a thermal steady state, 
in which the
$r$-mode heating equals to the radiative cooling. At mode saturation the energy pumped into the $r$-mode from the rotational angular momentum reservoir is dissipated, due to the damping mechanism, in the core. The thermal steady state spin-down feature is a rigorous result, as has been shown by Alford et al. \cite{AMS2012}. The spin-down occurs till the star reaches the instability boundary at the low temperature 
branch. At this point the spin-down
ends, the $r$-mode decays and the gravitational emission stops. This spin-down of the young pulsar NS
exists during a period of time, which
ranges over from few years to several million years depending on the value of the saturation amplitude. 

The evolution equations, obtained from the energy and angular momentum conservation laws \cite{owen1998,Levin1999,HoLai2000},
 are given by
\begin{eqnarray}
\frac{d \alpha}{dt}=-\alpha\left(\frac{1}{\tau_{GR}}+\frac{1}{\tau_V}\left(\frac{1-Q\alpha^2}{1+Q\alpha^2}\right)\right),
\label{eq.A1}
\end{eqnarray}
\begin{eqnarray}
\frac{d \Omega}{dt}=-\frac{2\Omega Q\alpha^2}{\tau_V}\frac{1}{1+Q\alpha^2},
\label{eq.A2}
\end{eqnarray}
\begin{eqnarray}
\frac{d T}{dt}=-\frac{1}{C_V}\left(L_{\nu}-P_V\right),
\label{eq.A3}
\end{eqnarray}
 where $Q=\frac{3\widetilde{J}}{2\widetilde{I}}$. The dimensionless quantities are defined in \cite{owen1998} as 
$\widetilde{J}=\frac{1}{MR^{4}}\int_{0}^{R}\rho {r^{\prime}}^6 dr^{\prime}$ 
and $\widetilde{I}=\frac{I}{MR^2}=\frac{1}{MR^2}\int_{0}^{R}\rho {r^{\prime}}^4 dr^{\prime}$.
In these definition $I$ is
the moment of inertia of the star having mass $M$ and radius $R$. In Eq.(\ref{eq.A3}) $P_V=P_{SV}+P_{BV}$ is the viscous dissipation power under shear (SV) and bulk (BV) viscosities, 
which can be written in terms of the corresponding time-scales  as $1/\tau_{V}=1/\tau_{SV}+1/\tau_{BV}$. 
In this Eq.(\ref{eq.A3}), $C_V$ and $L_\nu$ are the specific heat and total neutrino luminosity of the star, respectively. At saturation $\frac{d\alpha}{dt}$=0 and Eq.~(\ref{eq.A1}) gives  
\begin{eqnarray}
\frac{1}{\tau_V}=\frac{1}{\tau_{GR}} \left(\frac{1+Q\alpha^2}{1-Q\alpha^2}\right),
\label{eq.A4}
\end{eqnarray}
The spin-down rate in Eq.~(\ref{eq.A2}) at saturation becomes,
\begin{eqnarray}
\frac{d \Omega}{dt}=-\frac{2\Omega Q\alpha^2}{|\tau_{GR}|}\frac{1}{1-Q\alpha^2}\approx-\frac{2\Omega Q\alpha^2}{|\tau_{GR}|},
\label{eq.A5}
\end{eqnarray}
and the time evolution of temperature in Eq.~(\ref{eq.A3}) 
can be written as
\begin{eqnarray}
\frac{d T}{dt}=-\frac{1}{C_V} \left(L_{\nu}+P_{G}\left(\frac{1+Q\alpha^2}{1-Q\alpha^2}\right)\right)
\approx-\frac{1}{C_V} \left(L_{\nu}+P_{G}\right),
\label{eq.A6}
\end{eqnarray}
where, we have used the fact that the power radiated by GWs, $P_G$, is equal to the 
dissipative power, $P_V$, due to the different viscous effects considered.
In obtaining Eqs.~(\ref{eq.A5}) and (\ref{eq.A6}) we have used the fact that $Q\alpha^2\textless $1. The value of $Q$ for all the four EOSs of $L$=70, 76.26, 80 and 90 MeV is $\sim$0.0934, which differs
in the fifth decimal place in these considered EOSs.
The maximum limiting value of $\alpha$ is 1 \cite {Alford2014}.
The neutrino luminosity is given by \cite{AMS2012,Mahmoodifar2013},
\begin{eqnarray}
L_{\nu}=\left(\frac{4\pi R_{DU}^3 \Lambda_{QCD}^{3}\widetilde{L}_{DU}}{\hbar^4 c^3 \Lambda_{EW}^4}\right)
(k_B T)^{6}
+       \left(\frac{4\pi   R^3    \Lambda_{QCD}    \widetilde{L}_{MU}}{\hbar^4 c^3 \Lambda_{EW}^4}\right)
(k_B T)^{8}
,
\label{eq.A81}
\end{eqnarray}
where, $T$ is the core temperature, $R_{DU}$ is the distance in the volume of the core over which direct Urca occurs, $\widetilde{L}_{DU}$ and $\widetilde{L}_{MU}$ are the
luminosity parameters for the direct and modified Urca processes, respectively. In Eq.(\ref{eq.A81})
$\Lambda_{QCD}$ and $\Lambda_{EW}$
are the characteristic strong and electroweak scales used to make these quantities dimensionless. Their typical values are $\Lambda_{QCD}=$1 GeV, $\Lambda_{EW}=$100 GeV.

We shall now compute the steady state spin-down path inside the region of $r$-mode instability using the constant amplitude  
approximation, as was 
done by Owen et al. \cite{owen1998}. The temperature and frequency dependence of the amplitude during spin-down 
was
discussed by Alford et al. in Ref. \cite{Alford2014} using a semi-analytical expression under a power law approximation. 
As the spin-down occurs in a thermally steady state where the cooling equals to the dissipative heating, $L_{\nu}+P_G$=0 \cite{Bondarescu2009}. From this condition, 
one can obtain the thermal steady state spin-down angular velocity as a function of the temperature $\Omega(T)$ as,
\begin{eqnarray}
\Omega(T)= \left(\frac{3^8 5^2 c^4}{2^{15} \hbar^4} 
\frac{\widetilde{L_i} \Lambda^{9-\theta}_{QCD}(k_B T)^{\theta}}{\widetilde{J}^2 \Lambda^{4}_{EW} G M^2 {R_i}^3 \alpha^2}\right)^{(1/8)}.
\label{eq.A10}
\end{eqnarray}
where $\theta=$ 8 (6), $\widetilde{L_i}=$ $\widetilde{L}_{{MU (DU)}}$ and $R_i=$ $R$ ($R_{DU}$) for MU (DU).

A young NS under the $r$-mode instability
will follow a path as described by Eq.~(\ref{eq.A10}) till it leaves the instability region. 
The thermal 
steady state path has been calculated from Eq.$~(\ref{eq.A10})$ for two representative  values of the saturation amplitude, $\alpha$=1 and $10^{-6}$.
To this end we use the EOSs
corresponding to $L$=70, 76.26, 80 and 90 MeV in both 1.4 $M_\odot$ and 1.8 $M_\odot$ NSs. First, we have computed the path by switching off the direct Urca processes, i.e., by including only the shear viscosity and modified Urca contributions. These results are shown in each of the four panels of Figures 
\ref{fig:MUomegahc_1.4} and \ref{fig:MUomegahc_1.8}
for 1.4 $M_\odot$ and 1.8 $M_\odot$ NSs, respectively. We also obtain the spin-down path in these 1.4 $M_\odot$ and 1.8 $M_\odot$ pulsar NSs from the solution of the evolution equation~(\ref{eq.A2}), which can be written as
\begin{eqnarray}
\Omega (t) =\left[\Omega_i^{-6}+6C^{\prime}(t-t_i)\right]^{-1/6},
\label{eq.A8}
\end{eqnarray}
where, $C^{\prime}=\frac{2^{18} \pi G M R^4 \widetilde{J} \alpha^2 Q}{3^8 5^2 c^7 (1-\alpha^2 Q)}$ and $\Omega_i$ is the initial frequency with which the pulsar has entered the instability region.  
In the late-time the spin-down strongly slows down and the initial condition i.e. $\Omega_i^{-6}$ in Eq. (\ref{eq.A8}) can be neglected \cite{Alford2014}. Also by approximating $1-\alpha^2 Q \approx 1$ in the denominator of $C^{\prime}$ and $t-t_i \approx t$, 
the time evolution of the temperature can be obtained from equations (\ref{eq.A8}) and (\ref{eq.A10}) as
\begin{eqnarray}
T(t) \approx
\left[\left(
\frac{3^6 5^2 c^7}{2^{18} \pi \alpha^2} \frac{\widetilde{I}}{\widetilde{J}^2 GM R^4 t}
\right)^{4/3}
\left(
\frac{2^{15}}{3^8 5^2} \left(\frac{\hbar}{c}\right)^4\frac{\widetilde{J}^2 G \Lambda_{EW}^4 M^2 R^3 \alpha^2}{{\widetilde{L}} \Lambda_{QCD}^{9-\theta} k_B^{\theta}}
\right)
\right]^{(\frac{1}{\theta})}.
\label{eq.A9}
\end{eqnarray}
The spin-down period is calculated from the condition $\Omega(t)=\Omega_f$, which is the frequency at the low temperature branch of the instability curve at which the pulsar NS emerges out of the instability region. This spin-down time period, $t_{sd}$, is obtained as,
\begin{eqnarray}
t_{sd}= \frac{\Omega_f^{-6}-\Omega_i^{-6}}{6 C^{\prime}}.
\label{eq.A11}
\end{eqnarray}

 The spin-down path is calculated from the solution of (\ref{eq.A8})  together with Eq.~(\ref{eq.A9}) with an initial rotational velocity 0.8$\Omega_K$, where $\Omega_K$ is the Kepler velocity of the NS. The results for the four considered EOSs in 1.4 $M_\odot$ and 1.8 $M_\odot$ pulsar NSs are shown in the different panels of Figures \ref{fig:MUomegahc_1.4} and \ref{fig:MUomegahc_1.8}
 by solid thick lines. Both the results of the spin-down path obtained from the steady state condition in Eq.(\ref{eq.A10}) and from the numerical solution of Eq.(\ref{eq.A8}) are in perfect agreement. The results for the final velocity, $\Omega_f$, where the pulsar leaves the instability region
has a higher value at a lower temperature 
for the smaller value of the saturation amplitude, as can be seen from these figures. Moreover, for a given saturation amplitude, the results of $\Omega_f$ are almost independent of NS mass and the EOS used to obtain it. This can be seen in the different panels of figures \ref{fig:MUomegahc_1.4} and \ref{fig:MUomegahc_1.8}, where the angular velocity $\Omega$ as a function of the temperature calculated with EOSs having different slope are shown for NSs of 1.4 $M_\odot$ and 1.8 $M_\odot$. 
This finding supports the power law formulation of Alford et al. \cite{AMS2012}, which is 
based on the assumption that the $r$-mode phenomenology has a weak dependence on the details of the EOS and the mass distribution in the NS. In Figure 15 we compare the results of the instability boundary and spin-down path obtained from the present calculation for the EOSs of $L$=70, 76.26, 80 and 90 MeV in a NS of 1.4 $M_\odot$ mass without direct Urca contribution together with the predictions of the power law formulation \cite{AMS2012,Alford2014}.
%
The SEI results and 
the power
law predictions are found to be in good agreement.
%
%
The spin-down rate and the duration for the pulsar NS to be inside the $r$-mode instability region, $t_{sd}$, strongly depends on the value of  the saturation amplitude. For $\alpha$=1, $t_{sd}$ is of the order of few years, whereas, for $\alpha$=$10^{-6}$ it is of the order $10^{10}$ years for all the four EOSs.

In order to examine the influence of direct Urca processes on the spin-down phenomenology, we now take into account the direct Urca in the 1.4 $M_\odot$ and 1.8 $M_\odot$ NSs for the EOSs wherever it is allowed and compute again the spin-down path. There is no occurrence of direct Urca for EOSs with $L$=70 and 76.26 MeV in 
the case of a
NS of 1.4$M_\odot$ and for the EOS with $L$=70 MeV for a NS of 1.8$M_{\odot}$.
In these cases the instability boundary, spin-down path, frequency at the exit point and duration inside the instability region are the same as in figures \ref{fig:MUomegahc_1.4} and \ref{fig:MUomegahc_1.8} for these EOSs in the respective pulsar NSs.
However,
for the EOSs, $L$=80 and 90 MeV in 1.4 $M_\odot$ NS and  $L$=76.26, 80 and 90 MeV in 1.8 $M_\odot$ NS, the high temperature branch shifts to the lower temperature side by different extents, which depends on the magnitude of occurrence of direct Urca inside the volume of the core. The spin-down paths inside the instability boundaries for the EOSs $L$=80 and 90 MeV with 1.4 $M_\odot$ and  for $L$=76.26, 80 and 90 MeV in
1.8 $M_\odot$ 
cases are shown in Figures \ref{fig.DUomegahc_1.4} and \ref{fig.DUomegahc_1.8}, respectively,
for 
saturation amplitudes $\alpha$= 1 and $10^{-6}$. Comparing 
these results with the ones displayed in the
Figures. \ref{fig:MUomegahc_1.4} and \ref{fig:MUomegahc_1.8},
we see that,
if direct Urca processes are taken into account, the path inside the region of instability shortens moving towards a relatively higher value of the final frequency, $\Omega_f$, at lower temperature, $T_f$, on the low-temperature 
branch. 
We can also see that, if direct Urca processes are included, the spin-down period, $t_{sd}$, decreases.
The results of the final frequency $\Omega_f$, temperature $T_f$ and spin-down period, $t_{sd}$ in 1.4 $M_\odot$ and 1.8 $M_\odot$ NSs 
are given in Table \ref{table3} for both,
without and with direct Urca contributions, when they are allowed, for four EOSs with
$L$=70, 76.26, 80 and 90 MeV. From the comparison of the results of without and with direct Urca processes, it can be seen that the spin-down features differ sizably depending on the extent of occurrence of direct Urca inside the bulk.
The magnitude of occurrence of direct Urca is, in turn, decided from the size of the pulsar NS as well as by the $L$-value of the EOS. The results of $\Omega_{f}$, $T_{f}$ and $t_{sd}$ for a 1.4$M_\odot$ (1.8$M_\odot$) NSs 
for the characteristic EOS having $L$=76.26 MeV 
without direct Urca contributions are 
66.3Hz (71.0 Hz), 13.8 K (15.5 K) and 5.48 yrs (3.51 yrs) for $\alpha$=1.
However, for the 1.8 $M_{\odot}$ NS case, again with the characteristic $L$ value EOS, direct Urca is allowed over a distance of 6.19 km inside the core, which results into the spin-down period, $t_{sd}$=0.69 yrs and $\Omega_{f}$=93.1 Hz, $T_{f}$=5.88K. An overall comparison of the results reported in Table \ref{table3} for different EOSs shows that the spin-down time $t_{sd}$ decreases by $\sim$ 4 to 5 times for $\alpha$=1 and $\sim$ 
10 to 15 times for $\alpha$=10$^{-6}$ in a 1.4$M_{\odot}$ NS due to direct Urca processes,
compared with the case without direct Urca, when the $L$-value of the EOS changes from $L$=80 to 90 MeV. The results for the EOSs allowing direct Urca given here can be considered as qualitative, 
as far as we have used the value of the luminosities  $\widetilde{L}$ from Ref.\cite{Alford2014}.
\begin{table}[ht]
\begin{center}
\caption{
Angular velocity $\Omega_{f}$ in units of Kepler velocity $\Omega_{K}$ ,
angular frequency $\nu_{f}$ in Hz at the point of exit from $r$-mode instability and 
the temperature $T_{f}$ in units of 10$^8$ K. The results are calculated for two values of saturation amplitude $\alpha$=1 and 10$^{-6}$ using the EOSs $L$=70, 76.26, 80 and 90 MeV in 1.4$M_\odot$ and 1.8$M_\odot$ NSs. The results for $L$-values for which two sets of values are given, the upper line corresponds to the results without including direct Urca processes, whereas,
the second line corresponds to the results when the allowed direct Urca processes are included.
}
\label{table3}
\renewcommand{\tabcolsep}{0.0991cm}
\renewcommand{\arraystretch}{1.2}
\begin{tabular}{|c|c|c|c|c|c|c|c|c|c|}\hline
\multicolumn{1}{|c|}{L}    & \multicolumn{1}{c|}{ $\alpha$}              &
\multicolumn{2}{c|}{${\Omega_f}/{\Omega_K}$}&\multicolumn{2}{c|}{ $\nu_f$}&
\multicolumn{2}{c|}{$T_f$}&\multicolumn{2}{c|}{$t_{sd}$}\\ 
\multicolumn{1}{|c|}{MeV}  & \multicolumn{1}{c|}{ }              &
\multicolumn{2}{c|}{}&\multicolumn{2}{c|}{$Hz$ }&
\multicolumn{2}{c|}{($10^8$ K)}&\multicolumn{2}{c|}{$ (yr)$}\\\hline 
\multicolumn{1}{|c|}{ }  & \multicolumn{1}{c|}{  }              &
\multicolumn{1}{c|}{(1.4 $M_{\odot}$)}&\multicolumn{1}{c|}{(1.8 $M_{\odot}$)}&
\multicolumn{1}{c|}{(1.4 $M_{\odot}$)}&\multicolumn{1}{c|}{(1.8 $M_{\odot}$)}&
\multicolumn{1}{c|}{(1.4 $M_{\odot}$)}&\multicolumn{1}{c|}{(1.8 $M_{\odot}$)}&
\multicolumn{1}{c|}{(1.4 $M_{\odot}$)}&\multicolumn{1}{c|}{(1.8 $M_{\odot}$)}\\\hline  
70.00  & 1.00  	&   0.0721   & 0.0592   &   67.28 	&72.34 		& 13.91  	& 15.51  	&   5.56  &3.84		\\\hline
76.26  & 1.00  	&   0.0739   & 0.0624   &   66.33 	&71.03 		& 13.84  	& 15.54  	&   5.48  &3.51		\\
       & 			  &         	  & 0.0817   &         	&93.05  	&         & 5.88    &         &0.69   \\\hline	
80.00  & 1.00  	&   0.0749   & 0.0636   &   65.75 	&70.61 		& 13.77  	& 15.54  	&   5.53  &3.38		\\
       & 			  &   0.0935   & 0.0863   &   82.07 	&95.79 		& 6.20   	& 5.19   	&   1.46  &0.54    \\\hline	
90.00  & 1.00				&   0.0763   & 0.0662   &   64.35 	&69.57 		& 13.65  	& 15.51  	&   5.57  &3.23			\\
       &			  &   0.1010   & 0.0911	 &	 85.10	&95.76 		& 4.99   	& 4.91  	&		1.04 	&0.48\\\hline
70.00  & $10^{-6}$ 	& 0.1527 & 0.1254 & 142.55 &153.28 	& 0.9323 	& 1.04   	&6.78$\times 10^{10}$&4.70$\times 10^{10}$\\\hline
76.26  & $10^{-6}$ 	& 0.1565 & 0.1321 & 140.54 &150.50 	& 0.9275 	& 1.04   	&6.69$\times 10^{10}$&4.29$\times 10^{10}$\\
       & 						&        & 0.2077 &        &236.67		&         & 0.2041  &                    &2.84$\times 10^{9}$\\\hline
80.00  & $10^{-6}$ 	& 0.1587 & 0.1348 & 139.31 &149.61 	& 0.9224  & 1.04   	&6.74$\times 10^{10}$& 4.14$\times 10^{10}$\\
       & 						& 0.2378 & 0.2195 & 208.73 &243.62		& 0.2151 	& 0.1802 	& 5.95$\times 10^{9}$& 2.22$\times 10^{9}$\\\hline
90.00  & $10^{-6}$ 	& 0.1618 & 0.1403 & 136.34 &147.40 	& 0.9143  & 1.04   	&6.79$\times 10^{10}$&3.95$\times 10^{10}$\\
       & 						& 0.2568 & 0.2318  & 216.45  &243.55	& 0.1732 & 0.1705 	& 4.24$\times 10^{9}$& 1.94$\times 10^{9}$\\\hline 
\end{tabular}
\end{center}
\label{tab:table3}
\end{table}

In order to complete our $r$-mode study in a normal fluid pulsar NS, we now
enhance damping by including the contribution from the crust-core region. The time-scale, $1/\tau_{VE}$, for the viscous layer damping in the crust-core transition region is reported in Eq. (8) of our previous work \cite{trr2018} and also given in Eq. (\ref{eqB.5}) of Appendix-B for ready reference. The present calculation of the contribution from $1/\tau_{VE}$ 
differs from that in \cite{trr2018} because the crust-core tansition density, $\rho_t$ and pressure, $P_t$ are calculated using the dynamical method \cite{Claudia2019}, instead of the thermodynamical approximation used in \cite {trr2018}. 
The instability boundary is now calculated by including $1/\tau_{VE}$ in Eq.~(\ref{eq.19}) assuming a rigid coupling between core and crust. The instability boundary taking into account the damping
due to the viscous layer contribution is shown for the EOS $L$=80 MeV together with the minimal model instability boundary for the 1.4 $M_\odot$ and 1.8 $M_\odot$ mass NSs in the two panels (a) and (b) of Figure \ref{graph_a1.4_1.8}, respectively.
The position of the NSs belonging to Low-mass X-ray Binaries (LMXBs) and milli-second radio pulsars (MSRPs) 
\cite{Watts08,Heinke2007,Heinke2009,Tomsick2004}
are also shown in the same figure. The viscous layer damping in the crust-core region 
plays an important role raising
the low-temperature branch, so that all the pulsar NSs shown in the figure are predicted to be stable. The spin-down paths for new born NSs are also shown inside the $r$-mode instability region for three representative values of amplitude $\alpha$=1, 10$^{-4}$ and 10$^{-6}$. For these three values of $\alpha$ the final frequency $\Omega_f$,
when the viscous layer damping from the crust-core region is included,
are 373.8 (388.9), 570.0 (590.3) and 703.9 (730.9) Hz for 1.4$M_\odot$  (1.8$M_\odot$) NSs, respectively. The spin-down period of the pulsar NS is also drastically shortened. In the 1.8$M_\odot$ NS case, the values of the spin-down period 
of the saturation amplitude,
$\alpha$=1, 10$^{-4}$ and 10$^{-6}$ are  $t_{sd}$=0.12$\times$10$^{-3}$, 1.0$\times$10$^{3}$ and 2.1$\times$10$^{6}$ yr, respectively. These results can be compared with the corresponding values of the minimal model for $\alpha$=1 and 10$^{-6}$ given in Table~\ref{table3}. For $\alpha$=10$^{-6}$ in 1.4$M_\odot$ pulsar NS, it is found that the NS will cool without being stopped by $r$-mode
heating, as it can be seen in Figure 18. 
This is in contrast to the 
result found in Refs.\cite{{Kokkotas2016},{Alford2015}}, where for such a situation the required value of $\alpha$ is less than 10$^{-10}$. In such cases GWs are not expected from the source. 
In a more realistic 
scenario
there is a velocity gradient in the crust-core viscous region, which is taken into account through a slippage factor \cite{Glampekadis2006}. 
All newborn young pulsars having an age around thousand years are found to have a spin frequency as low as $\nu\leq$62Hz. The $r$-mode can be a possible spin-down mechanism
that causes so low frequency.
Similarly there are known pulsars in LMXBs whose core temperature is measured in the range 10$^8$K
\cite{Haskell2012,Mahmoodifar2013} 
and according to the minimal model these sources
lie inside the $r$-mode instability region. In either of the cases there shall be emission of gravitational waves if the $r$-mode is operative.
The detection of GW signals from such sources 
is highly necessary, as far as they can  provide
information on the composition and EOS of the pulsar NS.

\subsection{Constraints on the intensity of gravitational waves from isolated pulsar neutron stars}
\begin{table}[t]
\begin{center}
\caption{The amplitude of the strain tensor $h_0$ for the EOS $\gamma$=1/2 and $L=$ 76.26, 80, 90 MeV in the case of 1.4 
$M_{\odot}$ and 1.8 $M_{\odot}$ neutron stars. The data for the spin frequency $\nu_s$ and distance $r$ to the source are 
taken from Ref.~\cite{Watts08}, whereas, the effective temperature $T_{eff}$ at the surface of the star and the core temperature
$T_{core}$ inside the core of the star are taken from Refs.~\cite{Heinke2007,Heinke2009,Tomsick2004}.
For the EOS of $L=$ 76.26 MeV, $\alpha$ is calculated from Eq.(49) and for the $L=$ 80 MeV and $L=$ 90 MeV, $\alpha$ is calculated from Eq.(50).
\label{Table:h0}}
\renewcommand{\tabcolsep}{0.0991cm}
\renewcommand{\arraystretch}{1.2}
\begin{tabular}{|c|c|c|c|c|c|c|}\hline
$L$&Source	   & $\nu_s$     &  r     & $kT_{eff} $  &  \multicolumn{1}{|c|}{$\alpha$} &  \multicolumn{1}{c|}{$h_0$}\\\hline
(MeV)&           & (Hz)        & (kpc)      & (eV)        &  \multicolumn{2}{c|}{1.4 $M_{\odot}$}\\\hline
76.26&4U1608-522			&620	&4.1$\pm$0.4		&$170$& $6.96\times 10^{-8} $ 	&$4.39^{-0.39}_{+0.48}\times10^{-28}$\\
     &IGR J00291+5934	&599	&5.0$\pm$1.0		&$71$& $1.39\times 10^{-8} $    &$6.50^{-1.08}_{+1.63}\times10^{-29}$\\
     &MXB 1659-29			&556	&11.5$\pm$1.5	  &$55$& $1.13\times 10^{-8} $	  &$1.83^{-0.21}_{+0.27}\times10^{-29}$\\
     &Aql X-1					&550	&4.55$\pm$1.35 	&$94$& $3.43\times 10^{-8} $	  &$1.36^{-0.31}_{+0.58}\times10^{-28}$\\
     &KS 1731-260			&524	&7.2$\pm$1.00	  &$70$& $2.31\times 10^{-8} $    &$5.02^{-0.61}_{+0.81}\times10^{-29}$\\
     &XTE J1751-305		&435	&9.0$\pm$3.00		&$71$& $5.01\times 10^{-8} $	  &$4.98^{-1.25}_{+2.48}\times10^{-29}$\\
     &SAX J1808-3658	&401	&3.5$\pm$0.1		&$30$& $1.24\times 10^{-8} $	  &$2.48^{-0.07}_{+0.07}\times10^{-29}$\\
     &XTE J1814-338		&314	&6.7$\pm$2.9		&$69$& $1.74\times 10^{-7} $	  &$8.75^{-2.65}_{+6.65}\times10^{-29}$\\
     &NGC 6440				&205	&8.5$\pm$0.4		&$87$& $1.52\times 10^{-6} $	  &$1.68^{-0.08}_{+0.08}\times10^{-28}$\\\hline
%
$L$&Source	   & $\nu_s$     &  r     & $T_{core} $  &  \multicolumn{1}{|c|}{$\alpha$} &  \multicolumn{1}{c|}{$h_0$}\\\hline
 (MeV) &         & (Hz)        & (kpc)      & (K)        &  \multicolumn{2}{c|}{1.4 $M_{\odot}$}\\\hline
80.00&4U1608-522			&620	&4.1$\pm$0.4		&$1.25\times10^8$& $6.61\times 10^{-6} $ 	  &$4.33^{-0.44}_{+0.42}\times10^{-26}$\\
		 &IGR J00291+5934	&599	&5.0$\pm$1.0		&$2.96\times10^7$& $1.01\times 10^{-7} $    & $4.89^{-0.82}_{+1.22}\times10^{-28}$\\
		 &MXB 1659-29			&556	&11.5$\pm$1.5	  &$1.96\times10^7$& $3.94\times 10^{-8} $	  &$6.65^{-0.77}_{+1.00}\times10^{-29}$\\
		 &Aql X-1					&550	&4.55$\pm$1.35 	&$4.70\times10^7$& $5.68\times 10^{-7} $	  &$2.34^{-0.53}_{+1.09}\times10^{-27}$\\
		 &KS 1731-260			&524	&7.2$\pm$1.00	  &$2.89\times10^7$& $1.60\times 10^{-7} $    &$3.61^{-0.44}_{+0.59}\times10^{-28}$\\
		 &XTE J1751-305		&435	&9.0$\pm$3.00		&$2.96\times10^7$& $3.62\times 10^{-7} $	  &$3.74^{-0.93}_{+1.87}\times10^{-28}$\\
		 &SAX J1808-3658	&401	&3.5$\pm$0.1		&$7.23\times10^6$& $7.31\times 10^{-9} $	  &$1.52^{-0.04}_{+0.05}\times10^{-29}$\\
		 &XTE J1814-338		&314	&6.7$\pm$2.9		&$2.82\times10^7$& $1.15\times 10^{-6} $	  &$6.02^{-1.82}_{+4.58}\times10^{-28}$\\
		 &NGC 6440				&205	&8.5$\pm$0.4		&$4.11\times10^7$& $1.97\times 10^{-5} $	  &$2.25^{-0.1}_{+0.11}\times10^{-27}$\\\hline
90.00&4U1608-522			&620	&4.1$\pm$0.4		&$1.25\times10^8$& $1.06\times 10^{-5} $	 &$7.53^{-0.67}_{+0.82}\times10^{-26}$\\
		 &IGR J00291+5934	&599	&5.0$\pm$1.0		&$2.96\times10^7$& $1.61\times 10^{-7} $   &$8.48^{-1.42}_{+2.12}\times10^{-28}$\\
		 &MXB 1659-29			&556	&11.5$\pm$1.5	  &$1.96\times10^7$& $6.29\times 10^{-8} $	 &$1.15^{-0.13}_{+0.17}\times10^{-28}$\\
		 &Aql X-1					&550	&4.55$\pm$1.35 	&$4.70\times10^7$& $9.06\times 10^{-7} $	 &$4.06^{-0.93}_{+1.71}\times10^{-27}$\\
		 &KS 1731-260			&524	&7.2$\pm$1.00	  &$2.89\times10^7$& $2.56\times 10^{-7} $   &$6.26^{-0.76}_{+1.01}\times10^{-28}$\\
		 &XTE J1751-305		&435	&9.0$\pm$3.00		&$2.96\times10^7$& $5.78\times 10^{-7} $   &$6.48^{-1.62}_{+3.25}\times10^{-28}$\\
		 &SAX J1808-3658	&401	&3.5$\pm$0.1		&$7.23\times10^6$& $1.17\times 10^{-8} $   &$2.63^{-0.07}_{+0.08}\times10^{-29}$\\
		 &XTE J1814-338		&314	&6.7$\pm$2.9		&$2.82\times10^7$& $1.84\times 10^{-6} $   &$1.04^{-0.31}_{+0.80}\times10^{-27}$\\
		 &NGC 6440				&205	&8.5$\pm$0.4		&$4.11\times10^7$& $3.14\times 10^{-5} $   &$3.90^{-0.18}_{+0.19}\times10^{-27}$\\\hline
\end{tabular}
\end{center}
\end{table}

The intensity of the GWs emitted by a NS under $r$-mode instability 
can be expressed
in terms of the lateral stress tensor, $h_0$, given by \cite {owen1998}
\begin{equation}
 h_0=\sqrt{\frac{8\pi}{5}}\frac{G}{c^5} \frac{1}{r} \alpha \omega^3 M R^3 \tilde{J},
\label{eq.h0}
\end{equation}
where, $r$ is the distance from the neutron star of mass $M$ and radius $R$ and the other quantities have their usual meaning as defined earlier. From Eq.~(\ref{eq.h0}) it becomes clear that the detectability
of GWs coming from $r-$mode
depends on the saturation amplitude $\alpha$, the frequency $\Omega$ and the distance $r$. We shall consider here 
LMXB pulsars, which are predicted
to lie inside the $r$-mode instability region and for which there are data available on their frequency, distance and core temperature. Assuming that the $r$-mode spin-down mechanism is dominant, these pulsars spin down along the thermal steady state path, which 
fulfill the condition,
$P_G$=$L_{\gamma}$+$L_{\nu}$, 
where $L_{\gamma}$ and $L_{\nu}$ are the photon and neutrino luminosities, respectively. Because of the standard cooling, the photon luminosity is given by 
\begin{equation}
L_{\gamma}=4 \pi R^{2 } \sigma T^{4}_{eff},
\label{eq:Lg}
\end{equation}
where $\sigma$ and $T_{eff}$ are the Stefan\-Boltzmann constant and surface temperature, respectively. 
The neutrino luminosity $L_{\nu}$ is 
obtained using 
Eq.~(\ref{eq.A81}) and $P_G$ is the power radiated as GW, which is given by
\begin{equation}
P_G=\frac{2^{17} \pi G  }{3^8 5^2 c^{7}} \hspace{0.2 cm}\alpha^2 \Omega^{8} \widetilde{J}^2 M^2 R^6. 
\label{eq:pg}
\end{equation}
We shall compute the strain amplitude, $h_0$, for the different EOSs of SEI in a 1.4$M_\odot$ pulsar NS. For the EOS with $L$=76.26 MeV, direct Urca does not occur and the thermal steady state condition becomes $P_{G}\approx$$L_{\gamma}$, since the neutrino luminosity for modified Urca processes can be neglected in comparison to the photon luminosity. For the pulsars considered 
here, it is found that $L_{\nu}$ is two order of magnitude
lower than the corresponding $L_{\gamma}$ value. A similar result has also been found using the APR EOS in Ref.
\cite{Mahmoodifar2013}.
Consequently,  the amplitude $\alpha$  in Eq.(\ref{eq:pg}) can be written as
\begin{equation}
 \alpha=\frac{5\hspace{0.1cm} 3^4 c^{7/2}}{2^8 \widetilde{J} M R^3 \Omega^4} \left[\frac{L_{\gamma}}{2 \pi G}\right]^{1/2}.
\label{eq:alp}
\end{equation}
The values of the amplitude of  the strain tensor $h_0$
for nine pulsars in the instability region are calculated using $\alpha$ given by Eq.(\ref{eq:alp}) and 
the results are reported in 
Table~\ref{Table:h0} 
for the EOS $L$=76.26 MeV. In the case of EOSs with $L$=80 and 90 MeV, the direct Urca processes are allowed in 1.4$M_\odot$ NSs to different extents. In this scenario the thermal steady state condition for the pulsar becomes $P_{G}\approx$$L_{\nu}$,
where both
photon and neutrino luminosities due to modified Urca processes are small as compared with the neutrino luminosity due to the direct Urca.
From this condition $\alpha$ is obtained as,
\begin{equation}\label{eq:alpha}
 \alpha=\frac{5\hspace{0.1cm} 3^4 c^{7/2}}{2^8 \widetilde{J} M R^3 \Omega^4} \left[\frac{L_{\nu}}{2 \pi G}\right]^{1/2}.
\end{equation}
With the help of the $\alpha$ calculated from 
Eq.(\ref{eq:alpha}) for the EOSs with $L$=80 and 90 MeV,
the GW strain amplitude, $h_0$, is
obtained for nine different
LMXB pulsars
in the 1.4$M_\odot$ 
 case
and the 
corresponding values
are given in Table~\ref{Table:h0}. From the results of $h_0$ of 
the nine pulsars in Table~\ref{Table:h0}, it can be inferred that the intensity of the GW increases with increase in the $L$ value. This fact is due to the comparatively larger coverage of direct Urca in the volume of the pulsar for higher $L$-value. For the EOS of $L$=76.26 MeV, where direct Urca is not allowed in the pulsar according to the SEI 
model, $h_0$ is found to be small as compared with the predictions of EOSs with larger value of the slope parameter.
In the case of a EOS with 
$L$=80 MeV, the maximum value for $h_0$ is predicted for the 4U1608-522 pulsar in the range 3.89-4.75$\times$10$^{-26}$, which increases by a factor of 1.7 as $L$ increases to 90 MeV. The overall range of $h_0$ for all these nine pulsar NSs lies between 10$^{-29}$ to 10$^{-26}$. Therefore, if the $r$-mode is the dominating spin-down mechanism, then the detection of the GWs of such intensities could be possible with the help of second and third generation detectors. 
This conclusion is in agreement with the findings of Kokkotas and Schwenzer \cite{Kokkotas2016}. 
\begin{figure}[t]
\vspace{1.5cm}
	\begin{center}
		\includegraphics[width=0.9\columnwidth,angle=0]{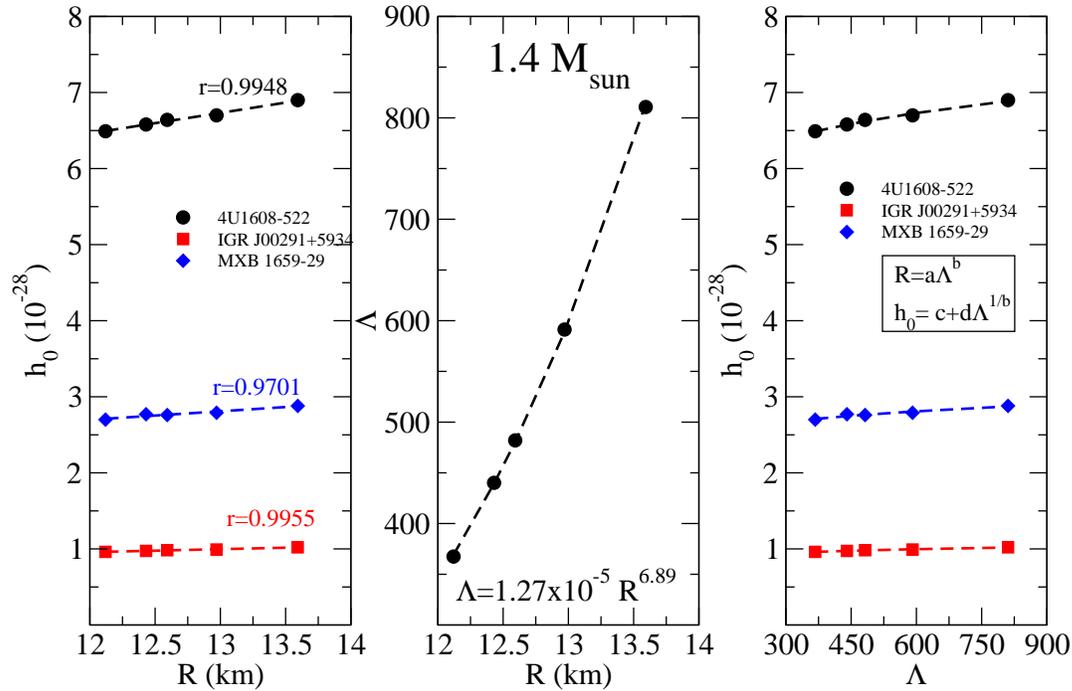}
		\end{center}
	\caption{Left panel: strain parameter $h_0$ against the radius $R$ of the neutron star for three representative neutron stars. The plot also includes the linear regression of the results, and their 
	correlation number $r$. Central panel: dimensionless mass-weighted tidal deformability $\Lambda$ against the radius $R$ of the neutron star. The figure includes the fit of the results, and its numerical expression. 
	Right panel: strain parameter $h_0$ against the dimensionless mass-weighted tidal deformability $\Lambda$ for three representative pulsar neutron stars. The figure includes the fit $h_0= c+d\Lambda^{1/b}$ resulting from the fits coming from the left and central panels. 
	All results are evaluated for a $1.4 M_\odot$ neutron star. \label{fig:h01}}
\end{figure}
From Eqs.~(\ref{eq.h0}), (\ref{eq:alp}) and (\ref{eq:alpha}), one can see that the amplitude of the strain tensor depends on two different types of factors. 
On the one hand, there are factors, such as the spin frequency, the distance to the source, the effective temperature and the frequency of the $r$-mode that are related to each particular pulsar considered in Table~\ref{Table:h0}. On the other hand, the remaining contributions to $h_0$ depend on the underlying EOS 
of the NS. It is easy to see from Eq.~(\ref{eq.h0}) that, globally, the latter contributions to $h_0$ are proportional to the radius of the star because $\alpha \sim 1/MR^2$. 
This dependence can be seen in the left hand side panel of Figure~\ref{fig:h01},
where this correlation is shown for the pulsars 4U1608-522, J00291+5934 and MXB 1659-29. In this panel we plot the strain tensor amplitude as a function of the radius of the NS, which is determined using the EOSs with $L$=70, 76.26, 80, 90 and 100 MeV.
The linear behaviour of this correlation can clearly be appreciated by the correlation
factors, which are close to 1. Therefore, it is expected that the amplitude of the strain tensor will
also show a dependence on the tidal deformability $\Lambda$, predicted by the same EOS, similar to that exhibited by the radius $R$ of the star. 
From Eq.~(\ref{eq:lambda}) one can see that $\Lambda \sim R^5$~\cite{Flanagan08, Hinderer08, Hinderer2010}. However, due to the more involved 
dependence of the Love number $k_2$ on the radius, the power will be 
different from $5$. The tidal deformability as a function of the radius 
of the star is displayed in the central panel of Figure~\ref{fig:h01} 
 computed with 
using SEI with different slope parameters. 
We see that these EOSs predict that the tidal deformability follows a power law of $\Lambda \sim R^{6.89}$ 
for the NS of $1.4 M_\odot$. 
From our previous analysis, it is expected that the amplitude of the strain tensor as a function of the tidal deformability behaves as 
$\sim \Lambda^{1/6.89}$ 
for NS of $1.4 M_\odot$. 
This plot is shown in the rightmost panel of Figure~\ref{fig:h01}. 
It should be pointed out, however, that the range of values of the tidal 
deformability, which are constrained by the $L$-values considered in this calculation, is samll
and, 
therefore, $\Lambda^{1/6.89}$ 
takes almost a constant value in the ranges of tidal deformabilities considered in the figures.
These correlations amongst the radius of the star, its tidal deformability and the amplitude of the strain tensor 
are almost insensitive to the mass
of the pulsar NS. 
For example, if one considers the 1.8 $M_{\odot}$ NS, one obtains
$\Lambda^{1/7.88}$ and $h_0 \sim \Lambda^{1/7.88}$. 

\section{Summary and Conclusion}\label{sec1.4}
We have examined the influence of the stiffness of the symmetry energy, characterized by the slope parameter $L$ of the EOS on the $r$-mode phenomenology in normal fluid pulsar NSs. In this context, we have studied how the spin-down 
features associated with newborn NSs under the $r$-mode instability are modified due to the variation of the $L$-value of the
 EOS. An essential ingredient for this study is the bulk viscosity, arising from the modified and the direct Urca processes, which is determined consistently with the underlying EOS. For our study we have constructed a set of EOSs 
using the simple effective interaction.
All these EOSs correspond to an incompressibiliy $K(\rho_0)$=246 MeV ($\gamma$=1/2), have the
same saturation properties and a 
symmetry energy at saturation $E_s(\rho_0)$=35 MeV, but different stiffness of the symmetry energy, as the slope 
parameter $L$ is the only variable quantity in our model. The set of EOSs thus obtained in the SEI model allows to study in a transparent way the role of the slope parameter
 of the EOS on the $r$-mode phenomenology in the pulsar NSs. In this
SEI model the upper limit of $L$ is restricted to a maximum value of 90 MeV, which is extracted from the analysis of the tidal 
deformability $\tilde{\Lambda}$ data of the GW170817 event. On the other hand, the maximum mass constraint $\sim$2$M_{\odot}$ 
puts the 
lower limit on the slope parameter value as $L\geq$ 76.26 MeV. This range of the slope parameter of the SEI EOS, namely $76 \lesssim L \lesssim 90$ MeV,
 conforms to the stiff-to-soft and soft-to-stiff density dependence inferred by Krastev and Li from an 
analysis of the GW170817 data with the MDI model assuming a NS of 1.4$M_{\odot}$.

Our investigations show that for the set of EOSs of SEI, 
there exists a lower minimum threshold value of $L$, referred to as critical $L$-value, for initiating the direct Urca process in the NSM. The threshold density for direct Urca, i.e., the density at which the direct Urca starts, depends on the EOS and is maximum for the one corresponding to the critical $L$-value. This threshold density decreases as the $L$-value of the EOS increases. At the threshold density of direct Urca, the bulk viscous coefficient $\zeta$ shows a sharp
increase and maintains the enhanced value over the range of density of occurrence of direct Urca. This enhancement of $\zeta$ as direct Urca processes occur has the consequence of shifting the high temperature branch of the $r$-mode instability boundary inward to the relatively lower temperature region and thereby decreasing the area under the instability boundary. In the present model the critical $L$ corresponds to a value around 76.26 MeV, which coincides with the lower limit of the range of $L$ considered in this work and for which the direct Urca threshold density is 
$\rho$=0.54 $fm^{-3}$. The central density of a typical 
1.4$M_{\odot}$ NS predicted by this critical $L$-value EOS is smaller than the threshold density and therefore the direct Urca reactions do not occur in these typical mass NSs for this critical EOS. Whereas, for NSs of higher mass, as for example
1.8$M_{\odot}$, the critical EOS predicts a central density higher than the threshold value
%
and therefore the direct Urca processes are predicted in the higher mass NSs. Thus within our model, around the lower limit $L$=76.26 MeV, direct Urca processes
 do not contribute to $\zeta$ in standard NS of 1.4$M_{\odot}$, but they have contribution in heavier NS, as for instance the one with 1.8$M_{\odot}$. As $L$ increases, for example $L=80$ MeV, the threshold density for direct Urca 
decreases below the central density of the 1.4$M_{\odot}$ NS predicted by this EOS and direct Urca processes contribute to $\zeta$ in typical as well as heavier mass NSs. The direct Urca contribution to $\zeta$ increases with further increase in the $L$, as the volume of the NS covered under direct Urca is larger due to the decrease in the threshold density. Thus, for a given pulsar mass, the magnitude of the shift of the higher temperature branch to the lower temperature side depends on the extent of occurrence of direct Urca in
the volume of the NS core, which is decided from the slope parameter of the EOS. On the other hand, for a given $L$ allowing direct Urca in both typical as well as heavy mass NSs, this shift will be larger for the heavier pulsars.

%

It is found that if the bulk viscosity is calculated taking into account the contribution of modified Urca processes 
only, then the $r$-mode instability region remains almost the same irrespective of the $L$-value of the EOS as well 
as the mass of the pulsar NS. The spin-down features of the pulsar, such as, frequency and temperature at the exit point of 
$r$-mode instability boundary and the period of instability are having very close values in this case. This is in 
conformity with the power law formulation of the $r$-mode instability worked out by Alford and collaborators. 
%
However,
EOSs with slope parameter $L$ larger than the critical value, allow direct Urca processes in the NSs if the central density is higher than the threshold value.
In such a case the $r$-mode instability boundary will critically depend on the value of $L$ and mass of the pulsar NS. The spin-down features will vary accordingly depending on the magnitude of $L$ of the EOS and mass of the pulsar.
In the present model, for a saturation amplitude $\alpha$=1 (10$^{-6}$) the time duration inside the instability region, $t_{sd}$, varies within the range 1.46 yr to 1.04 yr (5.95$\times$10$^{9}$ yr to 
4.24$\times$10$^{9}$ yr) as $L$ changes from 80 MeV to 90 MeV in case of a newborn NS of mass
1.4$M_{\odot}$ produced under the initial conditions of frequency $\Omega$=0.8$\times\Omega_K$ and temperature $T$=10$^{11}$K. For the same $\alpha$ and the range of variation in $L$, $t_{sd}$ will lie within 0.54 yr and 0.48 yr (2.22$\times$10$^{9}$ yr to 1.94$\times$10$^{9}$ yr) if the newborn NS produced under the same initial conditions has mass 1.8$M_{\odot}$. 
This variation in the area under the instability 
boundary depending on the value of $L$ of the EOS and mass of the pulsar produces widely different predictions for $\Omega_f$, $T_f$ and $t_{sd}$ that cannot be explained 
under the  power law formulation in its current form.

%

The intensity of the gravitational waves emitted during the $r$-mode spin-down process has been calculated for different LMXB pulsars in the standard mass model of 1.4$M_\odot$. The strain tensor amplitudes are obtained in the range 10$^{-26}$ to 10$^{-29}$, which are in 
harmony with the results available in earlier literature. These results show that if the $r$-mode dominates the GW emission, then their 
detection would be possible with the incoming second and third generation detectors. Once such facilities for detection of gravitational 
waves in the range of the $r$-mode emission are developed, information about the mass, age and distance of the pulsar NSs may be extremely 
useful for narrowing down the prevailing uncertainty in the $L$ value and the EOS of isospin rich dense nuclear matter.

\section*{Acknowledgments}
The authors thank the referees for their valuable suggestions that could improve the manuscript. Valuable discussions with Isaac Vida\~na are greatly appreciated. C.G, M.C. and X.V. acknowledge support from Grant FIS2017-87534-P from MINECO and FEDER,
and Project MDM-2014-0369 of ICCUB (Unidad de 
Excelencia Mar\'{\i}a de Maeztu) from MINECO. C.G. also acknowledges Grant BES-2015-074210 from MINECO. 
%
\section*{Appendix A}
\renewcommand{\theequation}{A. \arabic{equation}}

As mentioned in section 2, asymmetric nuclear matter computed with the SEI 
in equation (\ref{eq.17}) depends on the parameters $\gamma$, $b$, 
$\varepsilon_{0}^{l}$, $\varepsilon_{0}^{ul}$, 
$\varepsilon_{\gamma}^{l}$,$\varepsilon_{\gamma}^{ul}$, $\varepsilon_{ex}^{l}$,
 $\varepsilon_{ex}^{ul}$ and $\alpha$. 
The new parameters are connected to the parameters of the interaction
through the following relations:
\begin{eqnarray}
\varepsilon_{0}^{l}=\rho_0\left[\frac{t_0}{2}\left(1-x_0\right)
+\left(W+\frac{B}{2}-H-\frac{M}{2}\right)\pi^{3/2}\alpha^3\right] \nonumber \\
\nonumber \\
\varepsilon_{0}^{ul}=\rho_0\left[\frac{t_0}{2}\left(2+x_0\right)
+\left(W+\frac{B}{2}\right)\pi^{3/2}\alpha^3\right] \nonumber \\
\nonumber \\
\varepsilon_{\gamma}^{l}=\frac{t_3}{12}\rho_0^{\gamma+1}(1-x_3),
\varepsilon_{\gamma}^{ul}=\frac{t_3}{12}\rho_0^{\gamma+1}(2+x_3) \nonumber \\
\nonumber \\
\varepsilon_{ex}^{l}=\rho_0\left(M+\frac{H}{2}-B-\frac{W}{2}\right)
\pi^{3/2}\alpha^3 \nonumber \\
\nonumber \\
\varepsilon_{ex}^{ul}=\rho_0\left(M+\frac{H}{2}\right)
\pi^{3/2}\alpha^3.
\label{eq.A.1}
\end{eqnarray}
The energy density in asymmetric nuclear matter resulting from SEI having Gaussian form reads 
\begin{eqnarray}
H(\rho_n,\rho_p)&=&\frac{3\hslash^2}{10m}\left(k_n^2\rho_n+k_p^2\rho_p\right)
+\frac{\varepsilon_{0}^{l}}{2\rho_0}\left(\rho_n^2+\rho_p^2\right)
+\frac{\varepsilon_{0}^{ul}}{\rho_0}\rho_n\rho_p \nonumber \\
&&+\left[\frac{\varepsilon_{\gamma}^{l}}{2\rho_0^{\gamma+1}}\left(\rho_n^2+\rho_p^2\right)
+\frac{\varepsilon_{\gamma}^{ul}}{\rho_0^{\gamma+1}}\rho_n\rho_p\right]
\left(\frac{\rho({\bf R})}{1+b\rho({\bf R})}\right)^{\gamma} \nonumber \\
&&+\frac{\varepsilon_{ex}^{l}}{2\rho_0}(\rho_n^2 J(k_n) + \rho_p^2 J(k_p))\nonumber \\
&&+\frac{\varepsilon_{ex}^{ul}}{2\rho_0}\frac{1}{\pi^2}\left[\rho_n\int_0^{k_p}I(k,k_n)k^2dk
+\rho_p\int_0^{k_n}I(k,k_p)k^2dk\right] 
\label{eq.A.2}
\end{eqnarray}
where the functions $J(k_i)$ and $I(k,k_i)$ with $k_i=(3 \pi^2 \rho_i)^{1/3}$
($i=n,p$) are given by
\begin{eqnarray}
J(k_i)= \frac{3\Lambda^3}{2{k_i^3}} \bigg[\frac{\Lambda^3}{8{k_i^3}}
-\frac{3 \Lambda}{4{k_i}}
 -\left(\frac{\Lambda^3}{8{k_i^3}}-\frac{\Lambda}{4{k_i}}\right) e^{-4{k_i^2}/\Lambda^2}
+\frac{\sqrt{\pi}}{2} \textrm{erf}\left(2k_i/\Lambda\right) \bigg]
\label{eq.A.3}
\end{eqnarray}
and
\begin{eqnarray}
I(k,k_{i}) &=& \frac{3 \Lambda^3}{8k_{i}^{3}}
\bigg[\frac{\Lambda}{k}
\left(e^{-\left(\frac{k+k_{i}}{\Lambda}\right)^2}
-e^{-\left(\frac{k-k_{i}}{\Lambda}\right)^2}\right) \nonumber \\
&+&
\sqrt{\pi}\left(\textrm{erf}{\bigg(\frac{k+k_{i}}
{\Lambda}\bigg)}-\textrm{erf}{\bigg(\frac{k-k_{i}}
{\Lambda}\bigg)}\right) \bigg] ,
\label{eq.A.4}
\end{eqnarray}
with $\Lambda=2/\alpha$.
The neutron single-particle energy is

\begin{eqnarray}
\nonumber \varepsilon^{n}(k,\rho_{n},\rho_{p})
=\frac{\hbar^{2}k^{2}}{2m}+\frac{\varepsilon_{0}^{l}}{\rho_{0}}\rho_{n}+
 \frac{\varepsilon_{0}^{ul}}{\rho_{0}}\rho_{p}+ \left(\frac{\varepsilon_{\gamma}^{l}}{\rho_{0}^{\gamma+1}}\rho_{n} + \frac{\varepsilon_{\gamma}^{ul}}{\rho_{0}^{\gamma+1}}\rho_{p} \right)\left(\frac{\rho}{1+b\rho}\right)^{\gamma}      \\ \nonumber 
+\frac{\varepsilon_{ex}^{l}\rho_{n}}{\rho_{0}}\left[\frac{3\Lambda^{4}}{8kk_{n}^{3}}\left(e^{-\left(\frac{k+k_{n}}{\Lambda}\right)^{2}}-e^{-\left(\frac{k-k_{n}}{\Lambda}\right)^{2}}\right)+ \frac{3\Lambda^{3}}{4k_{n}^{3}}\int_{\frac{k-k_{n}}{\Lambda}}^{\frac{k+k_{n}}{\Lambda}}e^{-t^{2}dt}\right]\\ \nonumber
+\frac{\varepsilon_{ex}^{ul}\rho_{p}}{\rho_{0}}\left[\frac{3\Lambda^{4}}{8kk_{p}^{3}}\left(e^{-\left(\frac{k+k_{p}}{\Lambda}\right)^{2}}-e^{-\left(\frac{k-k_{p}}{\Lambda}\right)^{2}}\right)+ \frac{3\Lambda^{3}}{4k_{p}^{3}}\int_{\frac{k-k_{p}}{\Lambda}}^{\frac{k+k_{p}}{\Lambda}}e^{-t^{2}dt}\right]\\
+
\left[\frac{\varepsilon_{\gamma}^{l}}{2\rho^{\gamma+1}_{0}}(\rho_{n}^{2}+\rho_{p}^{2})+\frac{\varepsilon_{\gamma}^{ul}}{\rho^{\gamma+1}_{0}}\rho_{n}\rho_{p}\right]\frac{\gamma \rho^{\gamma-1}}{(1+b\rho)^{\gamma+1}}
\label{eq.A.5}
\end{eqnarray}
The expression for the proton single-particle energy can be written from the above expression
by interchanging the indexes $n$ and $p$.
The neutron (proton) chemical potential $\mu_n$ $(\mu_p)$ can be obtained from the neutron (proton) single-particle energy by replacing $k$ by $k_{n}$ ($k_{p}$), where $k_{n}$ ($k_{p}$) is the neutron (proton) Fermi momentum.

The neutron effective mass is given by

\begin{eqnarray}
\left(\frac{m}{m^*}(k,\rho_n,\rho_p)\right)_n&=&
1+
\frac{\varepsilon_{ex}^{l}\rho_{n}}{\rho_{0}}\left(\frac{m}{\hbar^2}\right)
[
\frac{3\Lambda^{2}}{4kk_{n}^{3}}\left(1-\frac{\Lambda^{2}}{2k^2}\right)
\left(e^{-\left(\frac{k+k_{n}}{\Lambda}\right)^{2}}-
e^{-\left(\frac{k-k_{n}}{\Lambda}\right)^{2}}\right)\nonumber \\
&-&
\frac{3\Lambda^{3}}{4k^2k_{n}^{3}}
\left(\left(\frac{k+k_{n}}{\Lambda}\right)e^{-\left(\frac{k+k_{n}}{\Lambda}\right)^{2}}-
\left(\frac{k-k_{n}}{\Lambda}\right)e^{-\left(\frac{k-k_{n}}{\Lambda}\right)^{2}}\right)
]
\nonumber \\
&+&
\frac{\varepsilon_{ex}^{ul}\rho_{p}}{\rho_{0}}\left(\frac{m}{\hbar^2}\right)
[
\frac{3\Lambda^{2}}{4kk_{p}^{3}}\left(1-\frac{\Lambda^{2}}{2k^2}\right)
\left(e^{-\left(\frac{k+k_{p}}{\Lambda}\right)^{2}}-
e^{-\left(\frac{k-k_{p}}{\Lambda}\right)^{2}}\right)\nonumber \\
&-&
\frac{3\Lambda^{3}}{4k^2k_{p}^{3}}
\left(\left(\frac{k+k_{p}}{\Lambda}\right)e^{-\left(\frac{k+k_{p}}{\Lambda}\right)^{2}}-
\left(\frac{k-k_{p}}{\Lambda}\right)e^{-\left(\frac{k-k_{p}}{\Lambda}\right)^{2}}\right)
]\nonumber \\
\label{eq.A.6}
\end{eqnarray}
The proton effective mass $\left(\frac{m}{m^{*}}(k,\rho_{n},\rho_{p})\right)_{p}$ can be written from the above expression
by interchanging the indexes $n$ and $p$.

The expression for the energy density in SNM is
\begin{eqnarray}
 H(\rho)&=&\frac{3\hbar^{2}}{10m}k_{f}^{2}\rho + \frac{\varepsilon_{0}}{2}\frac{\rho^{2}}{\rho_{0}}
+\frac{\varepsilon_{\gamma}}{2\rho_{0}^{\gamma+1}}\rho^{2}\left(\frac{\rho}{1+b\rho}\right)^{\gamma}\\
&+&
\frac{\varepsilon_{ex}}{2\rho_{0}}\rho^{2}\left[\frac{3\Lambda^{6}}{16k_{f}^{6}}
-\frac{9\Lambda^{4}}{8k_{f}^{4}}+\left(\frac{3\Lambda^{4}}{8k_{f}^{4}}-
\frac{3\Lambda^{6}}{16k_{f}^{6}}\right)e^{\frac{-4k_{f}^{2}}{\Lambda^{2}}}+
\frac{3\Lambda^{3}}{2k_{f}^{3}}\int_{0}^{\frac{2k_{f}}{\Lambda}}e^{-t^{2}}dt\right] , \nonumber
\label{eq.A.7}
\end{eqnarray}
with $k_{f}=(3 \pi^2 \rho/2)^{1/3}$ being the Fermi momentum in SNM,
whereas, the expression for the energy density in PNM is
\begin{eqnarray}
H(\rho)&=&\frac{3\hbar^{2}}{10m}k_{n}^{2}\rho + \frac{\varepsilon_{0}^{l}}{2}\frac{\rho^{2}}{\rho_{0}}+
\frac{\varepsilon_{\gamma}^{l}}{2\rho_{0}^{\gamma+1}}\rho^{2}\left(\frac{\rho}{1+b\rho}\right)^{\gamma}\\&+&
\frac{\varepsilon_{ex}^{l}}{2\rho_{0}}\rho^{2}\left[\frac{3\Lambda^{6}}{16k_{n}^{6}}-
\frac{9\Lambda^{4}}{8k_{n}^{4}}+\left(\frac{3\Lambda^{4}}{8k_{n}^{4}}-
\frac{3\Lambda^{6}}{16k_{n}^{6}}\right)e^{\frac{-4k_{n}^{2}}{\Lambda^{2}}}+
\frac{3\Lambda^{3}}{2k_{n}^{3}}\int_{0}^{\frac{2k_{n}}{\Lambda}}e^{-t^{2}}dt\right] . \nonumber 
\label{eq.A.8}
\end{eqnarray}


\section*{Appendix B}
\renewcommand{\theequation}{B.\arabic{equation}}
The expression for the time scale is 
\begin{equation}
\frac{1}{\tau_{i}}=-\frac{P_i}{E_m}
\label{eq.B.01}
\end{equation}
where,
\begin{equation}
E_m=\frac{1}{2} \alpha^2 \Omega^2 M R^2 \widetilde{J},
\label{eq.B.02}
\end{equation}
is the energy of the $r$-mode and $P_i$, $i$=GR, BV, SV and VE, is power radiated under gravitational radiation, bulk viscosity,
shear viscosity and viscous dissipation at the
boundary layer of the crust-core, respectively.

The analytical expression for the gravitational radiation time scale is 
\begin{equation}
\frac{1}{\tau_{GR}}=\frac{-32 \pi G \Omega^{2l+2}}{c^{2l+3}} \frac{(l-1)^{2l}}{[(2l+1)!!]^2}\left(\frac{l+2}{l+1}\right)^{(2l+2)} 
\int^{R}_{0}\rho_{m}(r^{\prime})r^{\prime (2l+2)} dr^{\prime} \hspace{0.5cm}\left( s^{-1}\right), 
\label{eq.B.1}
\end{equation}
where $G$ is the gravitational constant
and $\rho_{m}(r^{\prime})=H(\rho(r^{\prime}))/c^2$ being the mass density at a distance $r^{\prime}$. 

The shear viscous dissipation time-scale $1/\tau_{SV}$ is 
\begin{equation}
\frac{1}{\tau_{SV}}=(l-1) (2l+1) \left(\int^{R}_{0}\rho_{m}(r^{\prime})r^{\prime (2l+2)} dr^{\prime}\right)^{-1}\int^{R}_{0}\eta \hspace{0.1cm} r^{\prime (2l)} dr^{\prime} 
\hspace{0.5cm}\left( s^{-1}\right), 
\label{eq.B.2}
\end{equation}
where $\eta$ is the shear viscosity and $R$ is the radius of the NS.
The effects of the viscosity 
come from the electron-electron (ee) and neutron-neutron (nn) scattering. The respective viscosities $\eta^{ee(nn)}$ are given 
by 
\begin{equation}
\eta^{nn}=347 \left(\frac{\rho_m}{g \hspace{0.1cm} cm^{-3}}\right)^{(9/4)} \left(\frac{K}{T}\right)^{2} \hspace{0.5cm} 
\left({g \hspace{0.1cm} cm^{-1} \hspace{0.1cm} s^{-1}}\right), 
\label{eq.B.3}
\end{equation}
\begin{equation}
\eta^{ee}= 4.26 \times 10^{-26}\left(Y_p \rho \right) \left(\frac{K}{T}\right)^{5/3} \hspace{0.5cm} 
\left({g \hspace{0.1cm} cm^{-1} \hspace{0.1cm} s^{-1}}\right), 
\label{eq.B.4}
\end{equation}
where the temperature $T$ is in Kelvin.
The time scale for viscous layer dissipation in crust-core region is,
\begin{equation}
\frac{1}{\tau_{VE}}=\left[\frac{1}{2\Omega} \frac{2^{l+3/2}(l+1)!}{l(2l+1)!!I_{l}}\sqrt{\frac{2\Omega R_{c}^{2} \rho_{c}}{\eta_c}}
\int^{R_{c}}_{0} \frac{\rho_m(r^{\prime})}{\rho_{c}}\left(\frac{r^{\prime}}{R_{c}}\right)^{2l+2} \frac{dr^{\prime}}{R_c}\right]^{-1} \hspace{0.5cm}\left( s^{-1}\right), 
\label{eqB.5}
\end{equation}
 where{ $\rho_c$ is the mass density corresponding to the crust-core transition density}. $I_{l}$ in equation ({\ref{eqB.5}}) 
has the value $I_{2}=0.80411$ for $l=2$ \cite{Lindblom2000}. 
%

The various time-scales for the $l=2$ $r$-mode result into the following analytical expressions:
\begin{equation}
\frac{1}{{\tau}_{GR}}=-1.3705 \times 10^{-33} \left[I(R_x)_{,6}\right] \left(\frac{\Omega}{{\rm Hz}}\right)^6 \hspace{0.5cm}
\left( s^{-1}\right),
\label{eqb.11}
\end{equation}
\begin{equation}
\frac{1}{{\tau}_{VE}}=\frac{2.8230 \times 10^{-18} \left[\sqrt{\eta_c \rho_c}\right]}{\left[I(R_x)_{,6}\right]}
\left(\frac{R_c}{{\rm km}}\right)^6 
\left(\frac{\Omega}{{\rm Hz}}\right)^{1/2}
\hspace{0.5cm}
\left( s^{-1}\right),
\label{eqb.12}
\end{equation}
\begin{equation}
\frac{1}{{\tau}_{SV}}=
\frac{2.8086\times 10^{-22}}{\left[I(R_x)_{,6}\right]}
\left[
\left(\frac{K}{T}\right)^2 I^{nn}_{SV}(R_x)_{,4}
+
\left(\frac{K}{T}\right)^{5/3} I^{ee}_{SV}(R_x)_{,4}
\right]\hspace{0.5cm}
\left( s^{-1}\right),
\label{eqb.13}
\end{equation}
\begin{eqnarray}
\frac{1}{{\tau}_{BV}}&=&\frac{5.808\times 10^{-39}}{\left[I(R_x)_{,6}\right]} 
\left(\frac{M_{\odot}}{M}\right)^2
\left(\frac{\Omega}{{\rm Hz}}\right)^{2}\nonumber
\\ 
&&
\times \Bigg[ 
\left[\left(\frac{R}{{\rm km}}\right)^2 I^{DU}_{BV}(R_x)_{,8}+ 0.86\hspace{0.1cm} I^{DU}_{BV}(R_x)_{,10}\right]
\left(\frac{T}{10^9 K}\right)^{4}\nonumber
\\ 
&&
+
\left[\left(\frac{R}{{\rm km}}\right)^2 I^{MU}_{BV}(R_x)_{,8}+ 0.86\hspace{0.1cm} I^{MU}_{BV}(R_x)_{,10}\right]
\left(\frac{T}{10^9 K}\right)^{6}
\Bigg]
\hspace{0.1cm}\left( s^{-1}\right),
\label{eqb.14}
\end{eqnarray}
The 
various $I$-functions appearing in the above equations (\ref{eqb.11})-(\ref{eqb.14}) read
  \begin{eqnarray}
I(R_x)_{,6}=\int_{0}^{R_x} \left[\frac{H(\rho(r))}{{\rm MeV fm}^{-3}}\right]\left(\frac{r}{{\rm km}}\right)^6 d\left(\frac{r}{{\rm km}}\right),
\label{eqb.15}
  \end{eqnarray}
	  \begin{eqnarray}
I^{nn}_{SV}(R_x)_{,4}=\int_{0}^{R_x} 
\left(\eta^{\prime}\right)^{nn}
\left(\frac{r}{{\rm km}}\right)^4 d\left(\frac{r}{{\rm km}}\right),
\label{eqb.16}
  \end{eqnarray}
	  \begin{eqnarray}
I^{ee}_{SV}(R_x)_{,4}=\int_{0}^{R_x} 
\left(\eta^{\prime}\right)^{ee}
\left(\frac{r}{{\rm km}}\right)^4 d\left(\frac{r}{{\rm km}}\right),
\label{eqb.17}
  \end{eqnarray}
   \begin{eqnarray}
I^{DU}_{BV}(R_x)_{,8}=\int_{0}^{R_x} 
\left(\zeta^{\prime}\right)^{DU}
\left(\frac{r}{{\rm km}}\right)^8 d\left(\frac{r}{{\rm km}}\right),
\label{eqb.18}
  \end{eqnarray}
	   \begin{eqnarray}
I^{DU}_{BV}(R_x)_{,10}=\int_{0}^{R_x} 
\left(\zeta^{\prime}\right)^{DU}
\left(\frac{r}{{\rm km}}\right)^{10} d\left(\frac{r}{{\rm km}}\right),
\label{eqb.19}
  \end{eqnarray}
   \begin{eqnarray}
I^{MU}_{BV}(R_x)_{,8}=\int_{0}^{R_x} 
\left(\zeta^{\prime}\right)^{MU}
\left(\frac{r}{{\rm km}}\right)^8 d\left(\frac{r}{{\rm km}}\right),
\label{eqb.20}
  \end{eqnarray}
	   \begin{eqnarray}
I^{MU}_{BV}(R_x)_{,10}=\int_{0}^{R_x} 
\left(\zeta^{\prime}\right)^{MU}
\left(\frac{r}{{\rm km}}\right)^{10} d\left(\frac{r}{{\rm km}}\right),
\label{eqb.21}
  \end{eqnarray}
	where the upper limit $R_x$ of the integrals is $R$, the radius of the fluid star, 
if the effect of the crust is not considered and $R_{c}$, the core radius, if the crust is explicitly taken into account and\\
${\eta^{nn}} = \left(\eta^{\prime}\right)^{nn} \left(\frac{K}{T}\right)^{2} (g \hspace{0.1 cm} cm^{-1}\hspace{0.1 cm} s^{-1})$,\\
$\eta^{ee}=\left(\eta^{\prime}\right)^{ee}\left(\frac{K}{T}\right)^{5/3} (g \hspace{0.1 cm} cm^{-1}\hspace{0.1 cm} s^{-1})$,\\
$\zeta^{DU}=\left(\zeta^{\prime}\right)^{DU} \left(\frac{T}{K}\right)^6\left(\frac{10^4}{\omega}\right)^2 (g \hspace{0.1 cm} cm^{-1}\hspace{0.1 cm} s^{-1})$,\\
$\zeta^{MU}=\left(\zeta^{\prime}\right)^{MU}\left(\frac{T}{K}\right)^4\left(\frac{10^4}{\omega}\right)^2 (g \hspace{0.1 cm} cm^{-1}\hspace{0.1 cm} s^{-1})$.

The various $I$-functions in Eqs. (\ref{eqb.15})-(\ref{eqb.21}) are calculated for both the minimal model
($R_x$ = $R$) and by including the viscous layer contributions from the crust-core region ($R_x$ = $R_c$)
for 1.4 $M_{\odot}$ and  1.8 $M_{\odot}$ NSs for the EOSs with $L$= 70 MeV, 76.26 MeV, 80 MeV and 90 MeV and the results are given in Table 5.

\begin{table}[ht]
\begin{center}
\caption{The values of the $I(R_x)$-functions in equations (\ref{eqb.15})-(\ref{eqb.21}) calculated 
for $R_x= R$ (total radius of the star) and for $R_x= R_c$ (radius of the stellar core). Results 
are shown for stellar masses of 1.4 $M_{\odot}$ (upper line) and 1.8 $M_{\odot}$ (lower line) for each of the four EOSs
of $\gamma=1/2$ and $L$=70 MeV, 76.26 MeV, 80 MeV and 90 MeV.} 
\renewcommand{\tabcolsep}{0.04cm}
\renewcommand{\arraystretch}{1.0}
\begin{tabular}{|c|c|c|c|c|c|c|c|}\hline
\multicolumn{1}{|c}{$L$}&
\multicolumn{1}{|c}{$I(R)_{,6}$}&
\multicolumn{1}{|c}{$I^{nn}_{SV}(R)_{,4}$}&
\multicolumn{1}{|c}{$I^{ee}_{SV}(R)_{,4}$}&
\multicolumn{1}{|c}{$I^{DU}_{BV}(R)_{,8}$}&
\multicolumn{1}{|c}{$I^{DU}_{BV}(R)_{,10}$}&
\multicolumn{1}{|c}{$I^{MU}_{BV}(R)_{,8}$}&
\multicolumn{1}{|c|}{$I^{MU}_{BV}(R)_{,10}$}
\\\hline
%
%
  70.00 &5.8515 $\times 10^{8}$&9.9156 $\times 10^{39}$&1.6532$\times 10^{37}$
	&0&0&4.4407$\times 10^{-28}$&4.4642$\times 10^{-26}$\\
   & 5.3212 $\times 10^{8}$&1.9536 $\times 10^{40}$&2.5330$\times 10^{37}$
	&0&0&2.5000$\times 10^{-28}$&2.3289$\times 10^{-26}$\\\hline
	76.26 &6.4689 $\times 10^{8}$&9.4968 $\times 10^{39}$&1.8397$\times 10^{37}$
	&0&0&6.6247$\times 10^{-28}$&6.6872$\times 10^{-26}$\\
 &6.4849 $\times 10^{8}$&1.8160 $\times 10^{40}$&3.3822$\times 10^{37}$
	&1.7538$\times 10^{-6}$& 5.6088$\times 10^{-5}$&5.2067$\times 10^{-28}$&4.9764$\times 10^{-26}$\\\hline
	80.00 &6.7851 $\times 10^{8}$&9.2014 $\times 10^{39}$&1.9019$\times 10^{37}$
	&3.9113$\times 10^{-7}$&7.5206$\times 10^{-6}$&8.0429$\times 10^{-28}$&8.1393$\times 10^{-26}$\\
	 &6.9666 $\times 10^{8}$&1.7687 $\times 10^{40}$&3.7722$\times 10^{37}$
	&2.7698$\times 10^{-5}$&1.3452$\times 10^{-3}$&7.0663$\times 10^{-28}$&6.7883$\times 10^{-26}$\\\hline
	90.00 &7.6668 $\times 10^{8}$&8.9338 $\times 10^{39}$&2.10260 $\times 10^{37}$
	&1.8843$\times 10^{-5}$&7.4327$\times 10^{-4}$& 1.2386$\times 10^{-27}$& 1.2647$\times 10^{-25}$\\
	&7.9988 $\times 10^{8}$&1.6579 $\times 10^{40}$&4.5264 $\times 10^{37}$
	&2.0122$\times 10^{-4}$&1.1611$\times 10^{-2}$& 1.2987$\times 10^{-27}$& 1.2641$\times 10^{-25}$\\\hline
\multicolumn{1}{|c}{$L$}&
\multicolumn{1}{|c}{$I(R_c)_{,6}$}&
\multicolumn{1}{|c}{$I^{nn}_{SV}(R_c)_{,4}$}&
\multicolumn{1}{|c}{$I^{ee}_{SV}(R_c)_{,4}$}&
\multicolumn{1}{|c}{$I^{DU}_{BV}(R_c)_{,8}$}&
\multicolumn{1}{|c}{$I^{DU}_{BV}(R_c)_{,10}$}&
\multicolumn{1}{|c}{$I^{MU}_{BV}(R_c)_{,8}$}&
\multicolumn{1}{|c|}{$I^{MU}_{BV}(R_c)_{,10}$}
\\\hline
%
%
  70.00 & 5.8378 $\times 10^{8}$&9.9125 $\times 10^{39}$&1.6530$\times 10^{37}$
	&0&0&4.4252$\times 10^{-28}$&4.4451$\times 10^{-26}$\\
  & 5.3114 $\times 10^{8}$&1.9533 $\times 10^{40}$&2.5328$\times 10^{37}$
	&0&0&2.4901$\times 10^{-28}$&2.3179$\times 10^{-26}$\\\hline
	76.26 &6.4536 $\times 10^{8}$&9.4946 $\times 10^{39}$&1.8396$\times 10^{37}$
	&0&0&6.6094$\times 10^{-28}$&6.6671$\times 10^{-26}$\\
		 &6.4743 $\times 10^{8}$&1.8158 $\times 10^{40}$&3.3821$\times 10^{37}$
	&1.7538$\times 10^{-6}$& 5.6088$\times 10^{-5}$&5.1956$\times 10^{-28}$&4.9629$\times 10^{-26}$\\\hline
	80.00 &6.7710 $\times 10^{8}$&9.1993 $\times 10^{39}$&1.9018$\times 10^{37}$
	&3.9113$\times 10^{-7}$&7.5206$\times 10^{-6}$&8.0270$\times 10^{-28}$&8.1179$\times 10^{-26}$\\
	&6.9553 $\times 10^{8}$&1.7685 $\times 10^{40}$&3.7722$\times 10^{37}$
	&2.7698$\times 10^{-5}$&1.3452$\times 10^{-3}$&7.0544$\times 10^{-28}$&6.7735$\times 10^{-26}$\\\hline
	90.00 &7.6499 $\times 10^{8}$&8.9315 $\times 10^{39}$&2.1025 $\times 10^{37}$
	&1.8843$\times 10^{-5}$&7.4327$\times 10^{-4}$& 1.2367$\times 10^{-27}$& 1.2621$\times 10^{-25}$\\
	 &7.9850 $\times 10^{8}$&1.6577 $\times 10^{40}$&4.5264 $\times 10^{37}$
	&2.0122$\times 10^{-4}$&1.1611$\times 10^{-2}$& 1.2972$\times 10^{-27}$& 1.2622$\times 10^{-25}$\\\hline
\end{tabular}
\end{center}
\end{table}
\section*{References}
\end{document}